\documentclass[rmp,aps,twocolumn,nofootinbib]{revtex4}

\usepackage{graphicx}
\usepackage{amssymb}
\usepackage{amsbsy}
\usepackage{mathbbol}
\usepackage{psfrag}

\def\kB{{k_{\rm B}}}

\begin{document}


\title{The numerical renormalization group method for
     quantum impurity systems}

\author{Ralf Bulla}

\email{Ralf.Bulla@Physik.Uni-Augsburg.De}

\affiliation{Theoretische Physik III, Elektronische Korrelationen und
Magnetismus, Institut f\"ur Physik, Universit\"at Augsburg,
86135 Augsburg, Germany}

\author{Theo Costi}
\email{t.costi@fz-juelich.de}
\affiliation{Institut f\"ur Festk\"orperforschung,
Forschungszentrum J\"ulich, 
52425 J\"ulich, Germany}

\author{Thomas Pruschke}
\email{pruschke@theorie.physik.uni-goettingen.de}
\affiliation{Institut f\"ur Theoretische Physik, Universit\"at G\"ottingen, 
37077 G\"ottingen, Germany}

\date{\today}

\begin{abstract}
In the beginning of the 1970's, Wilson developed the concept
of a fully non-perturbative renormalization group transformation.
Applied to the Kondo problem, this numerical 
renormalization group method (NRG) gave for the first
time the full crossover from the high-temperature phase of a 
free spin to the low-temperature phase of a completely screened
spin. The NRG has been later generalized to a variety
of quantum impurity problems.
The purpose of this review is to give
a brief introduction to the NRG method including some guidelines
of how to calculate physical quantities, and to survey the development of
the NRG method and its various applications over
the last 30 years.
These applications include variants of the original Kondo problem
such as the non-Fermi liquid behavior in the two-channel Kondo model,
dissipative quantum systems such as the spin-boson model, and lattice
systems in the framework of the dynamical mean field theory.
\end{abstract}

\maketitle

\tableofcontents

\section{Introduction}
\label{sec:intro}

The last decades saw a steadily increasing interest in
a wide range of physical systems involving quantum impurities.
The expression `quantum impurity system' is used in a
very general sense here, namely a 
small system (the impurity) with only a few degrees
of freedom coupled to a large system (the environment or bath)
with very many degrees
of freedom, and where both subsystems have to be treated quantum
mechanically. 

The use of the terminology `impurity' is due 
to historical reasons. In the Kondo
problem, the small system is a magnetic impurity,
such as an iron ion, interacting with the conduction electrons of a 
nonmagnetic metal such as gold \cite{Hewson:1993}.
Other realizations are for example artificial impurities such as
quantum dots hosting only a small number of electrons. Here,
the environment is formed by the electrons in the leads.
The term
`quantum impurity systems' can also be used for what are traditionally
called `dissipative systems'. As an example, let the impurity 
correspond to a spin degree of freedom and the 
environment be built up by a bosonic bath; this describes
the well-known spin-boson model experimentally relevant,
for example, for
dissipative two-level systems like tunneling centers in glasses
\cite{Leggett:1987}.

Any theoretical method for the investigation of quantum impurity
systems has to face a number of serious obstacles.
First of all,
because the environment typically consists of a
(quasi-)continuum of quantum-mechanical degrees of freedom,
one has to consider a wide range of
energies -- from a high-energy cut-off (which
can be of the order of several eV) down to
arbitrarily small excitation energies. On the other hand, because the 
impurity degrees of freedom usually form an interacting
quantum-mechanical system, their coupling to a continuum of
excitations with arbitrarily small energies can result in infrared
divergencies in perturbational treatments. A well-known example for
this difficulty is the Kondo problem: Its physics is governed by an
energy scale, the Kondo temperature $T_{\rm K}$, which depends
non-analytically on the spin-exchange coupling $J$ between the
impurity and the conduction band of the host, $\ln T_{\rm K} \propto
1/J$ (see \textcite{Hewson:1993} for a detailed description of the
limitations of the perturbational approach for the Kondo model and the
single-impurity Anderson model). 
One is thus faced with the task to perform non-perturbative
calculations for an interacting quantum-mechanical
many-body system with a continuum of excitations covering a broad
spectrum of energies.

A very efficient way to treat systems with such a broad and continuous
spectrum of energies is the renormalization group approach. It allows,
in general, to go in a certain sequence of renormalization group
steps from high energies, such as the bandwidth, to low energies,
such as the Kondo temperature. General introductions of renormalization
group concepts have been given in 
\textcite{Goldenfeld:1992,Ma:1976,Salmhofer:1999}
(see also the original papers: \textcite{Wilson:1974} and 
\textcite{Wilson:1975b}).
Here we focus on a
specific implementation of the renormalization
group idea: Wilson's numerical renormalization group method \cite{Wilson:1975},
referred to as `NRG' in the remainder of the review. This approach is
different from most renormalization group methods as it is 
non-perturbative in all system parameters; however, the price one has to
pay is that the renormalization group steps have to be
performed numerically.

The general strategy of the NRG is the following (more details
are given in Sec.~\ref{sec:nrg-intro}). 
As specific example, let us consider the Kondo model which describes
a magnetic impurity with spin $\vec{S}$ coupled to the electrons
of a conduction band, assumed to be non-interacting, via an interaction of the form $J\vec{S}\cdot\vec{s}$,
with $\vec{s}$ the spin of the electrons at the impurity site. The
NRG starts with a logarithmic discretization of the conduction band
in intervals 
$[\Lambda^{-(n+1)}\omega_{\rm c},\Lambda^{-n}\omega_{\rm c}]$ and
$[-\Lambda^{-n}\omega_{\rm c},-\Lambda^{-(n+1)}\omega_{\rm c}]$
($n=0,1,2,\ldots$). We shall call $\Lambda>1$ the NRG discretization parameter.
After a sequence of transformations, the discretized model
is mapped onto a semi-infinite chain with the impurity spin 
representing the first site of the chain. The Kondo model
in the semi-infinite chain form is diagonalized iteratively,
starting from the impurity site and successively adding degrees
of freedom to the chain. Due to the logarithmic 
discretization, the hopping parameters between neighboring
sites fall off exponentially, i.e.\ going along the chain
corresponds to accessing decreasing energy scales in the calculation.
  
In this way, Wilson achieved a non-perturbative description
of the full crossover from a free impurity spin at high temperatures
to a screened spin at low temperatures \cite{Wilson:1975}, 
thus solving the so-called Kondo problem as discussed in detail in
\textcite{Hewson:1993}. After this first application more than
30 years ago, the NRG has been successfully generalized and applied to
a much wider range of quantum impurity problems. The first extension
was the investigation of the single-impurity Anderson model \cite{Anderson:1961}, which extends
the Kondo model by including charge fluctuations at the impurity site. 
\textcite{Krishnamurthy:1980a,Krishnamurthy:1980b}
described all the technical details, the analysis of 
fixed points, and the calculation of static quantities for this model.

In the following, the development of the NRG 
concentrated on the analysis of more complicated impurity
models, involving either more environment or impurity degrees
of freedom. For example, in the two-channel Kondo model
the impurity spin couples to two conduction bands. This model,
which has a non-Fermi liquid fixed point with associated
non-Fermi liquid properties, has been first investigated with the NRG by \textcite{Cragg:1980}.
The numerical calculations for such a two-channel model are, however, much more cumbersome
because the Hilbert space grows by a factor 16
in each step of the iterative diagonalization, instead
of the factor 4 in the single-channel case. 
\textcite{Pang:1991} and \textcite{Affleck:1992} later
analyzed the stability of the non-Fermi liquid fixed point
with respect to various perturbations such as channel
anisotropy. 

The two-impurity Kondo model as paradigm for
the competition of local Kondo screening and non-local magnetic
correlations was studied with NRG by
\textcite{Jones:1987,Jones:1988,Sakai:1990,Sakai:1992,Sakai:1992b,Silva:1996}.
Here, the focus was on the question, if the two regimes are connected
by a quantum-phase transition or rather by a smooth crossover.
Later on, such studies were extended to the two-channel situation, too \cite{Ingersent:1992}.

Originally, the NRG was used to determine thermodynamic properties of
quantum impurity systems.
The calculation of dynamic quantities with the NRG started
with the $T=0$ absorption and photoemission spectra of the x-ray 
Hamiltonian \cite{Oliveira:1981,Oliveira:1985}, 
followed by the $T=0$ single-particle
spectral function for the orbitally non-degenerate and degenerate
Anderson model 
\cite{Frota:1986,Brito:1990,Sakai:1989,Costi:1990,Costi:1992b}. 
The resulting spectral functions are obtained on all energy scales,
with a resolution proportional to the frequency as discussed
in detail in Sec.~\ref{subsec:dynamics}. 
Calculation of finite-temperature spectral functions
$A(\omega,T)$ are more problematic since all excitations can, in
principle, contribute. 
Nevertheless, the NRG has been
shown to give accurate results for $A(\omega,T)$, too, which also
allows to calculate transport properties
\cite{Costi:1992b,Costi:1994b,Suzuki:1996}. 
A subsequent development is the introduction of the concept of the
reduced density matrix, which allows to calculate dynamic quantities
in equilibrium in the presence of external fields \cite{Hofstetter:2000b}.
The calculation of non-equilibrium transient dynamics requires a multiple-shell
NRG procedure \cite{Costi:1997a} and has been accomplished
with the aid of a complete basis set and the reduced density matrix \cite{Anders:2005c}.
The first applications of this approach 
show very promising results, both for fermionic and bosonic systems
\cite{Anders:2005c,Anders:2006b,Anders:2006c}. Another recent generalization
of the NRG approach is to quantum impurities coupled to a bosonic bath 
(bosonic NRG, see \textcite{Bulla:2005}; for early related approaches see
\textcite{Evangelou:1982}). The bosonic NRG has already been
successfully applied to the sub-Ohmic spin-boson model which
shows a line of quantum critical points separating localized and
delocalized phases \cite{Bulla:2003b}.

Additional motivation to further improve the NRG method came from
the development of the dynamical mean-field theory 
(DMFT) \cite{Metzner:1989,Georges:1996}
in which a lattice model of correlated electrons, such as 
the Hubbard model, is mapped onto a single-impurity Anderson model
with the impurity coupled to a bath whose structure has to be determined
self-consistently. This requires the NRG to handle impurity models
with an arbitrary density of states of the conduction electrons
and to calculate directly 
the impurity self-energy \cite{Bulla:1998}. The first applications of the NRG
within the DMFT framework concentrated on the Mott transition
in the Hubbard model and accurate results could be obtained for
both $T=0$ \cite{Sakai:1994,Bulla:1999} and finite temperatures 
\cite{Bulla:2001}. Within DMFT, the NRG
has been applied to the periodic Anderson model 
\cite{Pruschke:2000}, the Kondo
lattice model \cite{Costi:2002},
 multi-band Hubbard models \cite{Pruschke:2005}, the ferromagnetic Kondo lattice model
with interactions in the band \cite{Liebsch:2006}, and 
to lattice models with a coupling to local phonon modes such as the Holstein
model \cite{Meyer:2002} and the Hubbard-Holstein model 
\cite{Koller:2004a}.

The observation that
the coupling between electronic degrees of freedom in quantum dots
and the surrounding leads can give rise to Kondo-like features in
the transport characteristics, has also led to a resurgence of
interest in quantum impurity systems, both experimentally
and theoretically. An important feature of quantum dot systems
is their flexibility and a number of different set-ups have
been realized so far, and investigated theoretically by various
methods including the NRG. Applications of the NRG in this field
include the standard Kondo effect 
\cite{Izumida:1998,Gerland:2000,Costi:2001,Borda:2005},
coupled quantum dots 
\cite{Hofstetter:2002,Hofstetter:2004,Borda:2003,Cornaglia:2005c,Galpin:2006a},
quantum dots in a superconductor \cite{Choi:2004}, and
quantum dots coupled to ferromagnetic leads \cite{Martinek:2003,Choi:2004b}.

From this brief overview one can see that the range of
applicability of the NRG has widened considerably since
Wilson's original paper, covering physical phenomena such
as the  Mott transition, quantum dot physics, local criticality, 
dissipative quantum systems, etc. Further applications are still
lying ahead and various optimizations of the technique itself are
still being developed -- we shall come back to this point in the
summary section.

This paper is the first review of the NRG approach (since Wilson's
original paper on the Kondo problem) which attempts to cover
both the technical details and all the various applications. In this
way, the reader should get an overview over the field, learn about
the current status of the individual applications, and (hopefully)
come up with ideas for further calculations. This review can
only be a start for a deeper understanding of the NRG. The
following shorter reviews on selected topics are also helpful:
section 4 in \textcite{Hewson:1993} contains a pedagogical
introduction to the NRG as applied to the Kondo problem,
\textcite{Gonzalez:1998} discuss the soft-gap Anderson and Kondo models,
\textcite{Costi:1999} gives a general overview of the key concepts,
including the application to the anisotropic Kondo model,
\textcite{Bulla:2005} present a detailed introduction to
the bosonic NRG,
and, finally, the two papers on the first calculations for
the single-impurity Anderson model 
by \textcite{Krishnamurthy:1980a,Krishnamurthy:1980b}
are still valuable reading for an overview of
the method and the details of the analysis of fixed points.
%
%

The review is organized as follows:
in Sec.~\ref{sec:nrg-intro} we start with an introduction to
the basic concepts of the NRG approach. The single-impurity
Anderson model serves as an example here, but the strategy applies 
to quantum impurity systems quite generally. At the end
of this section, we discuss the flow of the many-particle
eigenenergies and the appearance of fixed points in this flow.
This analysis already gives important insights into the
physics of a given model, but the calculation of physical
quantities needs some extra care, as described in Sec.~\ref{sec:nrg-calc}.
This section is divided into Sec.~\ref{subsec:thermodynamics}
(thermodynamic and static quantities, such as entropy, specific
heat and susceptibilities) and Sec.~\ref{subsec:dynamics}
(dynamic quantities, both in equilibrium and non-equilibrium).

The following two sections deal with the various
applications of the NRG and we distinguish here between
quantum impurity systems (Sec.~\ref{sec:imp}) and lattice models
within DMFT (Sec.~\ref{sec:dmft}). Section \ref{sec:imp}
covers most of the work using the NRG which has been published
so far. We shall present results for systems which show
the standard Kondo effect (Sec.~\ref{subsec:kondo}, this
also includes most of the NRG-results on quantum dots), the
two-channel Kondo problem (Sec.~\ref{subsec:two-channel}),
models displaying impurity quantum phase transitions 
(Sec.~\ref{subsec:qpt}), quantum impurity systems with orbital
degrees of freedom (Sec.~\ref{subsec:orbital}), and, finally,
impurities coupled to bosonic degrees of freedom (Sec.~\ref{subsec:bosons}).
The section on lattice models within DMFT (Sec.~\ref{sec:dmft}) covers
calculations for the Hubbard model (Sec.~\ref{subsec:hubbard}), 
the periodic Anderson
and Kondo lattice models (Sec.~\ref{subsec:pam}), and lattice models with
coupling to phonons (Sec.~\ref{subsec:phonons}).

In the summary we shall discuss open problems as well as possible
directions for future developments of the NRG approach.

Let us finish the introduction with a
few remarks on the selection of the material presented and the references:
Due to the flexibility of the NRG, the review covers a broad
range of physical phenomena, in particular in 
Secs.~\ref{sec:imp} and \ref{sec:dmft}. We shall, however, only
give very brief introductions to these phenomena and refer
the reader to the references given in the individual subsections, in
particular reviews or seminal books.  
%
%
Furthermore, due to lack of space, we shall mostly not 
review the results from other theoretical
approaches which have been applied to quantum impurity systems,
such as Bethe ansatz, quantum Monte Carlo, resolvent perturbation theory,
local-moment approach, etc, unless these appear crucial for an understanding
of relevant NRG results. Comparisons between the results from
NRG and these approaches are, in most cases, included in the
papers cited here (see also \textcite{Hewson:1993}). 
This means that we shall focus, almost completely,
on references which use the NRG.

\section{Introduction to the Numerical
         Renormalization Group Method}
\label{sec:nrg-intro}
The NRG method can be
applied to systems of the following form: a quantum mechanical
impurity with a small number of degrees of freedom
(so that it can be diagonalized exactly) coupling to a bath
of fermions or bosons, usually with continuous excitation
spectrum.
There is no restriction on the structure of the impurity part of
the Hamiltonian; it might contain, for example, a Coulomb
repulsion of arbitrarily large strength. The bath, however,
is required to consist
of non-interacting fermions or bosons, otherwise the various mappings
described below cannot be performed.

Whenever we discuss, in this review, models of a different kind,
such as the Hubbard model, these models will be mapped
onto impurity models of the above type. For the Hubbard
model and other lattice models of correlated electrons this
is achieved via the dynamical mean-field theory, see Sec.~\ref{sec:dmft}.

\begin{figure}[htb]
\centerline{
  \includegraphics*[width=3.1in]{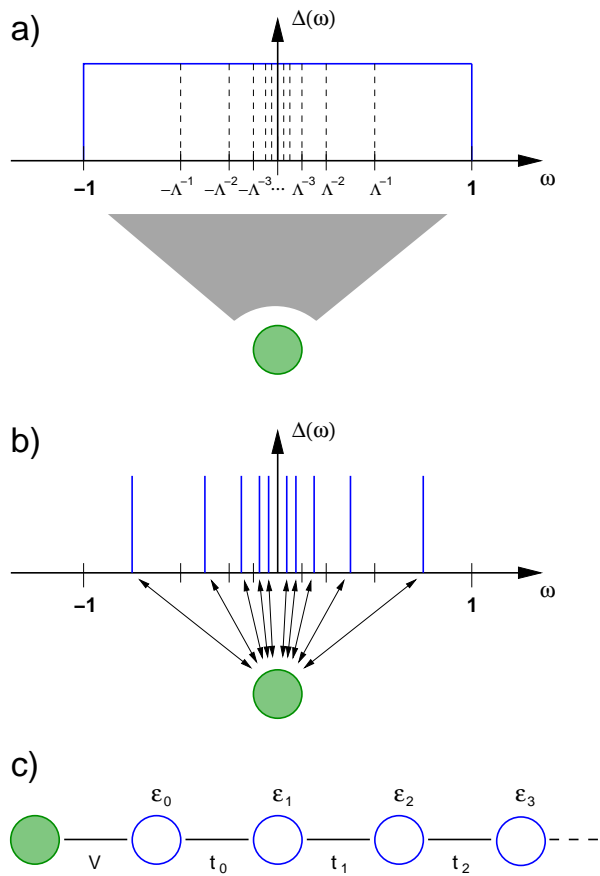}}
\caption{
Initial steps of the NRG illustrated for the single-impurity Anderson
model in which an impurity (filled circle) couples to a 
continuous conduction band via the hybridization function $\Delta(\omega)$;
a) a logarithmic set of intervals is introduced through the NRG discretization
parameter $\Lambda$; b) the continuous spectrum within each of these
intervals is approximated by a single state; c) the resulting discretized
model is mapped onto a semi-infinite chain where the impurity couples
to the first conduction electron site via the hybridization $V$;
the parameters of the tight-binding model 
(see Eq.~(\ref{eq:H_si})) are $\varepsilon_n$
and $t_n$.
}
\label{fig:mapping}
\end{figure}

Before we start with the technical details of the NRG approach,
let us give a brief overview of the general strategy. For basically
all NRG applications, one proceeds as follows:
\begin{itemize}
  \item[a)] Division of the energy support of the bath spectral function
            into a set of logarithmic intervals.
  \item[b)] Reduction of the continuous spectrum to a
            discrete set of states (logarithmic discretization).
  \item[c)] Mapping of the discretized model onto a semi-infinite chain.
  \item[d)] Iterative diagonalization of this chain.
  \item[e)] Further analysis of the many-particle energies, matrix elements,
            etc., calculated during the iterative diagonalization. This
            yields information on fixed points, static and dynamic properties
            of the quantum impurity model.
\end{itemize}
Parts a),b) and c) of this strategy are sketched in Fig.~\ref{fig:mapping},
where we consider a constant bath spectral function within the
interval $[-1,1]$. The NRG discretization parameter $\Lambda$
defines a set of discretization points, $\pm\Lambda^{-n}$, 
$n=0,1,2,\ldots$, and a corresponding set of intervals. The continuous
spectrum in each of these intervals (Fig.~\ref{fig:mapping}a)
is approximated by a single state (Fig.~\ref{fig:mapping}b).
The resulting discretized model is mapped onto a semi-infinite
chain with the impurity (filled circle) corresponding to the
first site of this chain. Due to the logarithmic discretization,
the hopping matrix elements decrease exponentially with distance
from the impurity, $t_n\propto\Lambda^{-n/2}$.

While the various steps leading to the semi-infinite chain are fairly
straightforward from a mathematical point of view, the 
philosophy behind this strategy is probably 
not so obvious.
%

Quite generally, a numerical diagonalization of Hamiltonian matrices
allows to take into account the various impurity-related terms in the
Hamiltonian, such as a local Coulomb repulsion,
non-perturbatively. Apparently, the actual implementation of such a
numerical diagonalization scheme requires some sort of discretization of the
original model, which has a continuum of bath states. There are, however,
many ways to discretize such a system, so let us try to explain why the
logarithmic discretization is the most suitable one here. As it turns
out, quantum impurity models are very often characterized by energy
scales orders of magnitudes smaller than the bare energy scales of the
model Hamiltonian. If the ratio of these energy scales is, for example,
of the order of $10^5$, a linear discretization would require energy intervals
of size at most $10^{-6}$ to properly resolve the lowest scale in the
system. Since for a finite system the splitting of energies is roughly
inversely proportional to the system size, one would need of the order
of $10^6$ sites, which renders an exact diagonalization impossible. 

Apparently, the logarithmic discretization reduces this problem
in that the low-energy resolution now depends exponentially on
the number of sites in the discretized
model. Of course, the accuracy of such an approach has to be
checked by suitable extrapolations of the discretization parameters,
in particular a $\Lambda\to 1$ extrapolation, which recovers the
original continuum model. 
Very often it turns out that 
already for $\Lambda$ of the order of $2$ the
results are accurate to within a few percent and a $\Lambda\to 1$ extrapolation
indeed reproduces exact results, if these are available.

However, this argument in favor of the logarithmic discretization does
neither explain the need for a mapping to a chain 
Hamiltonian as in Fig.~\ref{fig:mapping}c, nor how the 
problem of an exponentially growing Hilbert space with increasing
chain length is resolved. As far as the first point is concerned,
an iterative diagonalization for the discretized model with a star
geometry as in Fig.~\ref{fig:mapping}b has
been implemented for the spin-boson model \cite{Bulla:2005}.
For reasons which are not yet completely clear,
such a `star-NRG' is only partly successful.
Let us just mention here that for a fermionic model such
as the single-impurity Anderson model, the iterative
diagonalization of the model in the semi-infinite chain form
is much more convenient since one site of the chain
can be added in each step without violating particle-hole symmetry
(for a detailed discussion of this point see \textcite{Bulla:2005}).

The quantum impurity model in the semi-infinite chain form is
solved by iterative diagonalization, which means that in each
step of the iterative scheme one site of the chain is added to the
system and the Hamiltonian matrices of the enlarged cluster are 
diagonalized numerically. As already pointed out, without taking further
steps to reduce the size of the Hilbert space this procedure would have to end for chain sizes of
$\approx10$. Here the renormalization group concept enters
the procedure through the dependence of the hopping matrix elements
on the chain length, $t_n\propto\Lambda^{-n/2}$. Adding
one site to the chain corresponds to decreasing the relevant 
energy scale by a factor $\sqrt{\Lambda}$. Furthermore, because the
coupling $t_n$ to the newly added site falls off exponentially, only
states of the shorter chain within a comparatively small energy window
will actually contribute to the states of the chain with the
additional site. This observation allows to introduce a
very simple truncation scheme: after each step only the 
lowest lying $N_{\rm s}$ many-particle states are retained
and used to build up the Hamiltonian matrices of the
next iteration step, thus keeping the size of the Hilbert space fixed
as one walks along the chain.

All these technical steps will be discussed in detail in the following.
Let us briefly remark on the general set-up of this section.
We would of course like to keep this section as general as possible, because
it should serve as an introduction to the NRG technique, whose
application to a variety of problems is then the subject of the
remainder of this review. 
This quest for generality is, however, contrasted by the large variety
of possible impurity-bath interactions. 
Instead of presenting
explicit formulae for all possible quantum impurity models, we 
restrict ourselves to the single-impurity Anderson model as a specific
-- and most important -- example here. 
The original introductions to the technique for
the Kondo model
\cite{Wilson:1975} and the single-impurity Anderson model 
\cite{Krishnamurthy:1980a,Krishnamurthy:1980b}
were restricted to a constant bath density of states
(or better, a constant hybridization function  $\Delta(\omega)$
as defined below).
Here, we consider a general frequency dependent
$\Delta(\omega)$ from the outset.
This generalization is essential for various applications of
the NRG (the soft-gap models, see Sec.~\ref{subsubsec:loc-crit},
and lattice models within DMFT, see Sec.~\ref{sec:dmft})
where the physics is largely determined by the frequency
dependence of $\Delta(\omega)$.
If the hybridization function is non-zero
for positive frequencies only, the manipulations of the bath
degrees of freedom equally hold for a bosonic bath, see
\textcite{Bulla:2005}.

In this section we cover the steps a), b), c) and d) of the
list given above. Concerning the analysis of the data
[step e) in the list], we discuss the flow of the many-particle 
spectra and all related issues here. The calculation of static
and dynamic quantities will be described in Sec.~\ref{sec:nrg-calc}.

\subsection{Structure of the Hamiltonian}

The Hamiltonian of a general quantum impurity model consists
of three parts, the impurity $H_{\rm imp}$, the bath
$H_{\rm bath}$, and the impurity-bath interaction $H_{\rm imp-bath}$:
\begin{equation}
   H = H_{\rm imp} + H_{\rm bath} + H_{\rm imp-bath} \ .
\end{equation}
For the single-impurity Anderson model (SIAM) \cite{Anderson:1961}
with the Hamiltonian $H=H_{\rm SIAM}$, these three terms are given by:
\begin{eqnarray}
  H_{\rm imp} &=&   \sum_{\sigma} \varepsilon_{\rm f} f^\dagger_{\sigma}
                             f_{\sigma}
                 + U  f^\dagger_{\uparrow} f_{\uparrow}
                       f^\dagger_{\downarrow} f_{\downarrow} \ ,
                \nonumber \\
   H_{\rm bath}   &=&    
      \sum_{k \sigma} \varepsilon_k c^\dagger_{k\sigma} c_{k\sigma} \ ,
            \nonumber \\
   H_{\rm imp-bath}  &=&    
       \sum_{k \sigma} V_k
      \Big( f^\dagger_{\sigma} c_{k \sigma}
                                    +   c^\dagger_{k\sigma} f_{\sigma} \Big)
 \ .
\label{eq:siam}
\end{eqnarray}
In this Hamiltonian, the fermionic operators $c_{k\sigma}^{(\dagger)}$
correspond to  band states with spin $\sigma$ and energy $\varepsilon_k$,
and the $f_{\sigma}^{(\dagger)}$ to the impurity states with
energy  $\varepsilon_{\rm f}$. The
Coulomb interaction between two electrons occupying the impurity site
is parametrized by $U$ and 
both subsystems are coupled via a $k$-dependent hybridization
$V_k$.

The influence of the bath on the impurity is completely determined by
the 
so-called hybridization function $\Delta(\omega)$:
\begin{equation}
\Delta(\omega)= \pi \sum_k V_k^2 \delta(\omega - \varepsilon_k) \ .
\end{equation}
Thus, if we are only interested in the impurity contributions to the physics
of the SIAM, we can rewrite the Hamiltonian in a variety of ways,
provided the manipulations involved do not change the form of $\Delta(\omega)$.
Without loss of generality, we assume that the support of
$\Delta(\omega)$ completely lies within the interval $[-D,D]$, with
$D>0$ chosen suitably. Henceforth, we will use $D=1$ as energy unit.

One such possible reformulation is given by the following Hamiltonian:
\begin{eqnarray}
  H &=&  H_{\rm imp}
           + \sum_{ \sigma}\int_{-1}^1 {\rm d} \varepsilon \, g(\varepsilon)
            a^\dagger_{\varepsilon \sigma} a_{\varepsilon \sigma}
           \nonumber \\
           &+&  \sum_{ \sigma} \int_{-1}^1 {\rm d} \varepsilon \,
                   h(\varepsilon) \Big( f^\dagger_{\sigma}
                   a_{\varepsilon \sigma}  +
                     a^\dagger_{\varepsilon \sigma} f_{\sigma} \Big) .
\label{eq:siam_int}
\end{eqnarray}
Here we introduced a one-dimensional energy representation for the 
conduction band with band cut-offs at energies $\pm 1$, a dispersion
$g(\varepsilon)$ and a hybridization $h(\varepsilon)$. 
The band operators fulfill the standard fermionic commutation relations: 
$\left[a_{\varepsilon\sigma}^{\dagger},
a_{\varepsilon'\sigma'}\right]_+=\delta(\varepsilon-\varepsilon')
\delta_{\sigma\sigma'}$.
It has been shown in \textcite{Bulla:1997a} that the two functions
$g(\varepsilon)$ and $h(\varepsilon)$ are related to the
hybridization function $\Delta(\omega)$ via
\begin{equation}
   \Delta(\omega) = 
   \pi\frac{{\rm d} \varepsilon (\omega)}{{\rm d}\omega }
                   h[\varepsilon(\omega)]^2 \ , 
\label{eq:diffeq}
\end{equation}   
where $\varepsilon(\omega)$ is the inverse function to $g(\varepsilon)$,
$g[\varepsilon(\omega)] = \omega$.
For a given $\Delta(\omega)$ there are obviously many possibilities to
divide the $\omega$-dependence between $\varepsilon(\omega)$ and
$h(\varepsilon(\omega))$. This feature will turn out to be
useful later.

For a constant $\Delta(\omega)\!=\!\Delta_0$ within the interval
$[-1,1]$, Eq.~(\ref{eq:diffeq}) can be fulfilled by choosing 
$\varepsilon(\omega)= \omega$ (this corresponds
to $g(\varepsilon) = \varepsilon$) and 
$h^2(\varepsilon)= \Delta_0/\pi$.

\subsection{Logarithmic discretization}

The Hamiltonian in the integral representation 
Eq.~(\ref{eq:siam_int}) is a convenient
starting point for the logarithmic discretization of the conduction
band. As shown in Fig.~\ref{fig:mapping}a, the parameter
$\Lambda>1$ defines a set of intervals with discretization
points
\begin{equation}\label{eq:discretization_points}
   x_n = \pm \Lambda^{-n} \ , \ \ n=0,1,2,\ldots \ .
\end{equation}
The width of the intervals is given by
\begin{equation}
   d_n =  \Lambda^{-n}(1-\Lambda^{-1}) \ .
\end{equation}
Within each interval we now introduce a complete set of orthonormal functions
\begin{equation}
   \psi_{np}^\pm (\varepsilon) = \left\{ \begin{array}{ll}
                          \frac{1}{\sqrt{d_n}} e^{\pm i\omega_n p \varepsilon}
                          & \mbox{for}\ x_{n+1} < \pm\varepsilon <
                            x_n \\
                          0 & \mbox{outside this interval} \ .
                                         \end{array}  \right. 
\end{equation}
The index $p$ takes all integer values between $-\infty$
and $+\infty$, and the fundamental frequencies for each interval are given by
$\omega_n = 2\pi/d_n$. The next step is to expand the conduction
electron operators $a_{\varepsilon \sigma}$ in this basis, i.e.
\begin{equation}
    a_{\varepsilon \sigma} = \sum_{np} \Big[ a_{n p \sigma}
       \psi^+_{np} (\varepsilon) + b_{n p \sigma}
       \psi^-_{np} (\varepsilon)  \Big] \ ,
\end{equation}
which corresponds to a Fourier expansion in each of the intervals.
The inverse transformation reads
\begin{eqnarray}
   a_{np\sigma} &=& \int_{-1}^1 {\rm d} \varepsilon \, \big[ \psi^+_{np}
                     (\varepsilon) \big]^* a_{\varepsilon \sigma} \ ,
        \nonumber \\
   b_{np\sigma} &=& \int_{-1}^1 {\rm d} \varepsilon \, \big[ \psi^-_{np}
                     (\varepsilon) \big]^* a_{\varepsilon \sigma} \ .
\end{eqnarray}
The operators $a^{(\dagger)}_{np\sigma}$ and
$b^{(\dagger)}_{np\sigma}$ defined in this way fulfill
the usual fermionic commutation relations.
The Hamiltonian Eq.~(\ref{eq:siam_int}) is now expressed in terms
of these discrete operators.

In particular, the transformed hybridization
term (first part only) is
\begin{eqnarray}
  \int_{-1}^1 {\rm d} \varepsilon \,  h(\varepsilon)
        f^\dagger_{\sigma}  a_{\varepsilon \sigma} =&
f^\dagger_{\sigma}& \sum_{np} 
      \left[ a_{np\sigma} \int^{+,n}{\rm d} \varepsilon\,  h(\varepsilon) 
                \psi_{np}^+ (\varepsilon)  \right. \nonumber \\
   & + & \! \left.   b_{np\sigma} \int^{-,n}{\rm d} \varepsilon\,  h(\varepsilon) 
                \psi_{np}^- (\varepsilon) \right]  ,
\label{eq:hyb_map}
\end{eqnarray}
where we have defined
\begin{equation}
  \int^{+,n} {\rm d} \varepsilon \equiv
    \int_{x_{n+1}}^{x_n} {\rm d} \varepsilon \ \ , \ \ 
  \int^{-,n} {\rm d} \varepsilon \equiv
    \int_{-x_n}^{-x_{n+1}} {\rm d} \varepsilon \   .
\end{equation}
For a constant $h(\varepsilon)=h$, the integrals in Eq.~(\ref{eq:hyb_map})
filter out the $p=0$ component only
\begin{equation}
   \int^{\pm,n} {\rm d} \varepsilon\,  h \psi_{np}^\pm (\varepsilon)
 = \sqrt{d_n} h \delta_{p,0} \ .
\end{equation}
In other words, the impurity couples only to the $p=0$ components
of the conduction band states. It will become clear soon, that this point
was essential in Wilson's original line of arguments, so we would like
to maintain 
this feature ($h(\varepsilon)$ being constant in each interval
of the logarithmic discretization) also for a general, non-constant
$\Delta(\omega)$. Note that this restriction for the function $h(\varepsilon)$
does not lead to additional approximations for a non-constant
$\Delta(\omega)$ as one can shift all the remaining $\varepsilon$-dependence
to the dispersion $g(\varepsilon)$, see Eq.~(\ref{eq:diffeq}).

As discussed in \textcite{Chen:1995a} in the context of the soft-gap model
(see also \textcite{Chen:1995c}),
one can even set $h(\varepsilon)=h$ for all $\varepsilon$. Here we
follow the proposal of \textcite{Bulla:1997a}, that is we introduce a step
function for $h(\varepsilon)$
\begin{equation}
   h(\varepsilon) = h_n^\pm \ \ , \ \ x_{n+1} < \pm \varepsilon < x_n \ ,
\end{equation}
with $h_n^\pm$ given by the average of the hybridization function
$\Delta(\omega)$ within the respective intervals,
\begin{equation}
     {h_n^\pm}^2 = \frac{1}{d_n} \int^{\pm,n} {\rm d}\varepsilon\, 
                   \frac{1}{\pi} \Delta(\varepsilon) \ .
\end{equation}
This leads to the following form of the hybridization term
\begin{equation}
    \int_{-1}^1 {\rm d} \varepsilon \,  h(\varepsilon)
        f^\dagger_{\sigma}  a_{\varepsilon \sigma} =
        \frac{1}{\sqrt{\pi}} f^\dagger_{\sigma} \sum_n
        \left[ \gamma_n^+ a_{n0\sigma} + \gamma_n^- b_{n0\sigma} \right] \ ,
\end{equation}
with ${\gamma_n^\pm}^2 = \int^{\pm,n} {\rm d}\varepsilon\,\Delta(\varepsilon)$.

Next, we turn to the conduction electron term, which transforms into
\begin{eqnarray}
   &&  \int_{-1}^1 {\rm d}\varepsilon\,  g(\varepsilon) 
         a^\dagger_{\varepsilon \sigma} a_{\varepsilon \sigma} =
          \sum_{np} 
         \Big(\xi_n^+  a^\dagger_{np\sigma} a_{np\sigma}
                +\xi_n^- b^\dagger_{np\sigma} b_{np\sigma} \Big)
  \nonumber \\
  && + \sum_{n,p\ne p^\prime}
    \Big(\alpha_n^+ (p,p^\prime)  a^\dagger_{np\sigma} a_{np^\prime \sigma}
          - \alpha_n^- (p,p^\prime) b^\dagger_{np\sigma} b_{np^\prime\sigma} \Big) \ .
  \nonumber \\
 \label{eq:p_ne-pp}
\end{eqnarray}
The first term on the right hand side of Eq.~(\ref{eq:p_ne-pp}) is diagonal
in the index $p$. The discrete set of energies $\xi_n^{\pm}$ can be
expressed as \cite{Bulla:1997a}
\begin{equation}
  \xi_n^\pm = \frac{\int^{\pm,n} {\rm d} \varepsilon \Delta(\varepsilon) 
         \varepsilon}{
                   \int^{\pm,n} {\rm d} \varepsilon \Delta(\varepsilon) }
         \ \ 
    \left[ 
          = 
           \frac{1}{2}\Lambda^{-n}(1+\Lambda^{-1}) \right]
   \label{eq:xin}         \ ,
\end{equation}
where we added the result for a constant $\Delta(\varepsilon)$ in
brackets.
The coupling of the conduction band states with different $p,p^\prime$ (the second
term) recovers the continuum (no approximation has been made so far, 
Eq.~(\ref{eq:p_ne-pp}) is still exact). For the case of a linear dispersion,
$g(\varepsilon)=\varepsilon$, the prefactors $\alpha_n^\pm (p,p^\prime)$
are the same for both sides of the discretization and
take the following form
\begin{equation}\label{eq:p_ne-pp_expl}
  \alpha_n^\pm (p,p^\prime) = \frac{1-\Lambda^{-1}}{2\pi i}
           \frac{\Lambda^{-n}}{p^\prime - p}
          \exp\bigg[ \frac{2 \pi i (p^\prime - p)}{1 -\Lambda^{-1}}
                       \bigg]   \ .
\end{equation}
The actual discretization of the Hamiltonian is now achieved by dropping
the terms with $p\ne0$ in the expression for the
conduction band Eq.~(\ref{eq:p_ne-pp}).
This is, of course, an approximation, the quality of which is not clear from
the outset. To motivate this step we can argue that 
(i) the $p\ne0$ states couple only indirectly to the impurity
(via their coupling to the $p=0$ states in 
Eq.~(\ref{eq:p_ne-pp})) and 
(ii) the coupling between the  $p=0$ and $p\ne0$ states has a prefactor
$(1-\Lambda^{-1})$ which vanishes in the limit $\Lambda\to 1$. 
In this sense one can view the couplings to the states with $p\ne0$
as small parameters and consider the
restriction to $p=0$ as zeroth order step
in a perturbation expansion with respect to the coefficients
$a_n^\pm(p,p')$
\cite{Wilson:1975}. As it turns
out, the accuracy of the results obtained from the $p=0$ states only 
is surprisingly good even for values of $\Lambda$
as large as $\Lambda=2$, so that in all NRG
calculations the $p\ne0$ states have never been considered so far.

Finally, after dropping the $p\ne0$ terms and relabeling the
operators $a_{n0\sigma}\equiv a_{n\sigma}$, etc., we arrive at the discretized
Hamiltonian as depicted by Fig.~\ref{fig:mapping}b 
\begin{eqnarray}
  H &=&  H_{\rm imp}
           + \sum_{n\sigma} \left[
                  \xi_n^+
                 a^\dagger_{n\sigma} a_{n\sigma} +
                   \xi_n^-
                 b^\dagger_{n\sigma} b_{n\sigma} \right] \nonumber \\
         &+&  \frac{1}{\sqrt{\pi}}\sum_{\sigma} f^\dagger_{\sigma} \sum_n
        \left( \gamma_n^+ a_{n\sigma} + \gamma_n^- b_{n\sigma} \right) \nonumber \\
         &+& \frac{1}{\sqrt{\pi}} \sum_{\sigma} \left[ \sum_n
        \left( \gamma_n^+ a_{n\sigma}^\dagger + \gamma_n^- b_{n\sigma}^\dagger 
           \right) \right]         f_{\sigma}  \ .
        \label{eq:Hdisc}
\end{eqnarray}
Before we continue with the mapping of the Hamiltonian 
Eq.~(\ref{eq:Hdisc}) onto a semi-infinite chain, Fig.~\ref{fig:mapping}c,
let us make a few remarks on alternative discretizations of the continuous bath
spectral function.

The above procedure obviously applies for general asymmetric $\Delta(\omega)$,
also for different upper and lower cutoffs, $D_{\rm u}$ and $D_{\rm
  l}$. A special case is 
$D_{\rm l}=0$, which occurs for a bosonic bath, see
\textcite{Bulla:2005}; here the logarithmic discretization is performed
for positive frequencies only, and the operators 
$b_{n\sigma}^{(\dagger)}$ in Eq.~(\ref{eq:Hdisc}) are no longer present.

In Sec.~\ref{sec:nrg-calc} we shall discuss that the discreteness of the model
Eq.~(\ref{eq:Hdisc}) can be (in some cases) problematic for the
calculation of physical quantities. As it is not possible in the actual
calculations to recover the continuum by taking the limit $\Lambda\to 1$
(or by including the $p\ne 0$ terms), it has been suggested to
average over various discretizations for fixed $\Lambda$
\cite{Frota:1986,Yoshida:1990,Oliveira:1994}. The discretization
points are then modified as
\begin{equation}
  x_n = \left\{
   \begin{array}{lcl}
     1 & : & n=0 \\
     \Lambda^{-(n+Z)} & : & n\ge 1 \ ,
   \end{array} 
   \right. 
\end{equation}
where $Z$ covers the interval $[0,1)$. This `$Z$-trick' is,
indeed, successful as it removes certain artificial oscillations 
(see Sec.~\ref{subsubsec:Zaverage}), but it should be stressed here
that the continuum limit
introduced by integrating over $Z$ is not
the same as the true continuum limit $\Lambda\to 1$.

Another shortcoming of the discretized model is that the hybridization
function $\Delta(\omega)$ is systematically underestimated. It is
therefore convenient to multiply $\Delta(\omega)$ with the
correction factor
\begin{equation}
   A_\Lambda = \frac{1}{2} \ln \Lambda \frac{\Lambda +1}{\Lambda-1} \ ,
\end{equation}
which accelerates the convergence to the continuum limit.
For a recent derivation of this correction factor
see \textcite{Campo:2005}, where it was also shown that by a suitable 
modification of the discretization procedure, the factor $A_\Lambda$
can be taken into account from the outset.

\subsection{Mapping on a semi-infinite chain}
\label{subsec:semi-infinite}

According to Fig.~\ref{fig:mapping}b and c, the next step is to transform
the discretized Hamiltonian Eq.~(\ref{eq:Hdisc}) into a semi-infinite
chain form with the first site of the chain (filled circle in
Fig.~\ref{fig:mapping}c) representing the impurity degrees of freedom.
In the chain Hamiltonian, the impurity directly couples only
to one conduction electron degree of freedom with operators
$c^{(\dagger)}_{0\sigma}$, the form of which can be directly read off from
the second and third line in Eq.~(\ref{eq:Hdisc}). With the definition
\begin{equation}
    c_{0\sigma} = \frac{1}{\sqrt{\xi_0}} \sum_n \left[
          \gamma_n^+  a_{n\sigma}   + \gamma_n^- b_{n\sigma} \right],
\label{eq:c0}
\end{equation}
in which the normalization constant is given by
\begin{equation}
   \xi_0 = \sum_n \left(  (\gamma_n^+)^2 +(\gamma_n^-)^2 \right) 
       = \int_{-1}^1 {\rm d} \varepsilon \Delta(\varepsilon) \ , 
\end{equation} 
the hybridization term can be written as
\begin{equation}
 \frac{1}{\sqrt{\pi}} f^\dagger_{\sigma} \sum_n
        \left( \gamma_n^+ a_{n\sigma} + \gamma_n^- b_{n\sigma} \right) 
= \sqrt{\frac{\xi_0}{\pi}} 
            f^\dagger_{\sigma}c_{0\sigma} \ ,
\label{eq:hyb}
\end{equation} 
(similarly for the hermitian conjugate term).
Note that for a $k$-independent hybridization, 
$V_k=V$ in Eq.~(\ref{eq:siam}), the coupling in Eq.~(\ref{eq:hyb})
reduces to $\sqrt{\xi_0/\pi}=V$.

The operators $c^{(\dagger)}_{0\sigma}$ represent the first site of the 
conduction electron part of the semi-infinite chain. These
operators are of course not orthogonal to the operators
$a_{n\sigma}^{(\dagger)}$, $b_{n\sigma}^{(\dagger)}$. Constructing
a new set of mutually orthogonal operators $c^{(\dagger)}_{n\sigma}$
from
$c^{(\dagger)}_{0\sigma}$ and $a_{n\sigma}^{(\dagger)}$, $b_{n\sigma}^{(\dagger)}$
by a standard Gram-Schmidt procedure leads to
the desired chain Hamiltonian, which takes the form
\begin{eqnarray}
  &H& =  H_{\rm imp}
            + \sqrt{\frac{\xi_0}{\pi}}\sum_{\sigma} \left[
            f^\dagger_{\sigma}c_{0\sigma} + 
            c^\dagger_{0\sigma}f_{\sigma}  \right]  \nonumber \\
           & +& \sum_{\sigma n=0}^\infty \left[ 
               \varepsilon_n c_{n\sigma}^\dagger c_{n\sigma}
               + t_n \left( c_{n\sigma}^\dagger c_{n+1\sigma}
                  + c_{n+1\sigma}^\dagger c_{n\sigma}\right)\right] \ ,
           \nonumber \\
\label{eq:H_si}
\end{eqnarray}
with the operators $c^{(\dagger)}_{n\sigma}$ corresponding to the $n$th
site of the conduction electron part of the chain. The parameters
of the chain are the on-site energies $\varepsilon_n$ and the
hopping matrix elements $t_n$. The operators $c^{(\dagger)}_{n\sigma}$
in Eq.~(\ref{eq:H_si}) and the operators 
$\{ a^{(\dagger)}_{n\sigma},b^{(\dagger)}_{n\sigma} \}$ in 
Eq.~(\ref{eq:Hdisc})
are related via an orthogonal transformation
\begin{eqnarray}
   a_{n\sigma} = \sum_{m=0}^\infty u_{mn} c_{m\sigma} \ \ , \ \
   b_{n\sigma} = \sum_{m=0}^\infty v_{mn} c_{m\sigma} \ ,
\nonumber \\
   c_{n\sigma} = \sum_{m=0}^\infty \left[
                 u_{nm} a_{m\sigma} + v_{nm} b_{m\sigma} \right] \ .
\end{eqnarray}
From the definition of $c_{0\sigma}$ in Eq.~(\ref{eq:c0}) we can read off
the coefficients $u_{0m}$ and $u_{0m}$
\begin{equation}
   u_{0m} = \frac{\gamma_m^+}{\sqrt{\xi_0}} \ \ , \ \
   v_{0m} = \frac{\gamma_m^-}{\sqrt{\xi_0}} \ .
\end{equation}
For the remaining coefficients $u_{nm}$, $v_{nm}$, as well as for
the parameters $\varepsilon_n$, $t_n$, one can derive recursion
relations following the scheme described in detail in,
for example, Appendix A
of \textcite{Bulla:2005}. The starting point here is the equivalence
of the free conduction electron parts
\begin{eqnarray}
&& \sum_{n\sigma} \left[
                  \xi_n^+
                 a^\dagger_{n\sigma} a_{n\sigma} +
                   \xi_n^-
                 b^\dagger_{n\sigma} b_{n\sigma} \right] = \nonumber \\
&& \hspace*{-1cm}\sum_{\sigma n=0}^\infty \left[ 
               \varepsilon_n c_{n\sigma}^\dagger c_{n\sigma}
               + t_n \left( c_{n\sigma}^\dagger c_{n+1\sigma}
                  + c_{n+1\sigma}^\dagger c_{n\sigma}\right)\right]  .
\end{eqnarray}
The recursion relations are initialized by the equations
\begin{eqnarray}
   \varepsilon_0 &=& \frac{1}{\xi_0} 
                     \int_{-1}^1  {\rm d} \varepsilon \Delta(\varepsilon) 
                      \varepsilon  \ , \nonumber \\
   t_0^2 &=& \frac{1}{\xi_0} \sum_m \left[  
                    ( \xi_m^+ -  \varepsilon_0)^2 (\gamma_m^+)^2 +
                    ( \xi_m^- -  \varepsilon_0)^2 (\gamma_m^-)^2 \right] 
                      \ , \nonumber \\
   u_{1m} &=& \frac{1}{t_0} ( \xi_m^+ -  \varepsilon_0)  u_{0m}\ , \nonumber \\
   v_{1m} &=& \frac{1}{t_0} ( \xi_m^- -  \varepsilon_0)  v_{0m}\ .
\end{eqnarray}
For $n\ge 1$, the recursion relations read
\begin{eqnarray}
   \varepsilon_n &=& \sum_m  \left( \xi_m^+ u_{nm}^2 + \xi_m^- v_{nm}^2
                             \right)
                              \ , \nonumber \\
   t_n^2 &=& \sum_m  \left[ (\xi_m^+)^2 u_{nm}^2 + (\xi_m^-)^2 v_{nm}^2
                             \right]  - t_{n-1}^2 - \varepsilon_n^2
                      \ , \nonumber \\
   u_{n+1,m} &=& \frac{1}{t_n} \left[ 
                  ( \xi_m^+ -  \varepsilon_n)  u_{nm} - 
                  t_{n-1} u_{n-1,m} \right]   \ , \nonumber \\
   v_{n+1,m} &=& \frac{1}{t_n} \left[ 
                  ( \xi_m^- -  \varepsilon_n)  v_{nm} - 
                  t_{n-1} v_{n-1,m} \right] 
\ .
\end{eqnarray}
Note that for a particle-hole symmetric hybridization
function, $\Delta(\omega)=\Delta(-\omega)$, the on-site energies
$\varepsilon_n$ are zero for all $n$.

For a general hybridization function, the recursion relations have to
be solved numerically. Although these relations are fairly easy to
implement, it turns out that the iterative solution breaks down
typically after about 20-30 steps. The source of this instability 
is the wide range of values for the parameters entering the
recursion relations (for instance for the discretized energies
$\xi_m^\pm$). In most cases this problem can be overcome by using
arbitrary precision routines for the numerical calculations.
Furthermore, it is helpful to enforce the normalization of the vectors
$u_{nm}$ and $v_{nm}$ after each step.

Analytical solutions for the recursion relations have so far been given
only for few special cases. Wilson derived a formula for the
$t_n$ for a constant density of states of the conduction electrons
in the Kondo version of the impurity model \cite{Wilson:1975};
this corresponds to a constant hybridization function $\Delta(\omega)$
in the interval $[-1,1]$. Here we have $\varepsilon_n=0$
for all $n$ and the expression for the $t_n$ reads
\begin{equation}
   t_n = 
         \frac{\left( 1+ \Lambda^{-1} \right) 
               \left(1-\Lambda^{-n-1}\right) }{2\sqrt{1-\Lambda^{-2n-1}}
           \sqrt{1-\Lambda^{-2n-3}}}\, \Lambda^{-n/2} \ .
\end{equation}
(Similar expressions have been given for the soft-gap model, see
\textcite{Bulla:1997a}.)
In the limit of large $n$ this reduces to
\begin{equation}
   t_n \longrightarrow
         \frac{1}{2}\left( 1+ \Lambda^{-1} \right) 
              \, \Lambda^{-n/2} \ .
  \label{eq:tn-inf}
\end{equation}
The fact that the $t_n$ fall off exponentially with the distance from
the impurity is essential for the following discussion, so let
us briefly explain where this $n$-dependence comes from. Consider
the discretized model Eq.~(\ref{eq:Hdisc}) 
with a finite number $1+M/2$ ($M$ even) of conduction
electron states for both positive and negative energies
(the sum over n then goes from $0$ to $M/2$). This corresponds to
$2+M$ degrees of freedom which result in $2+M$ sites of the 
conduction electron part of the chain after
the mapping to the chain Hamiltonian. The lowest energies in the
discretized model Eq.~(\ref{eq:Hdisc}) are the energies $\xi_{M/2}^\pm$ which,
for a constant hybridization function,
are given by $\xi_{M/2}^\pm = \pm \frac{1}{2}\Lambda^{-M/2}(1+\Lambda^{-1})$, 
see Eq.~(\ref{eq:xin}).
This energy shows up in the chain Hamiltonian as the last hopping
matrix element $t_{M}$, so we have $t_{M}\sim \xi_{M/2}$
equivalent to Eq.~(\ref{eq:tn-inf}).

Equation (\ref{eq:H_si}) is a specific one-dimensional representation of the
single-impurity Anderson model Eq.~(\ref{eq:siam}) with the special feature
that the hopping matrix elements $t_n$ fall off exponentially.
As mentioned above, this representation is not exact since in
the course of its derivation, the $p\ne 0$ terms 
have been dropped. We should stress here that the dimensionality of 
the chain Hamiltonian is not related to that of the original model
which describes, for example, an impurity in a three-dimensional
host (apparently, this only holds for a non-interacting conduction
band). Nevertheless, the conduction electron
sites of the chain do have a physical meaning in the original
model as they can be viewed as a sequence of shells centered around
the impurity. The first site of the conduction electron chain
corresponds to the shell with the maximum of its wavefunction closest
to the impurity \cite{Hewson:1993,Wilson:1975}; this shell is coupled
to a shell further away from the impurity and so on.

\subsection{Iterative diagonalization}

The transformations described so far are necessary to map the
problem onto a form (the semi-infinite chain, Eq.~(\ref{eq:H_si}))
for which an iterative renormalization group (RG) procedure
can be defined. This is the point at which, finally, the RG
character of the approach enters.

The chain Hamiltonian Eq.~(\ref{eq:H_si}) can be viewed as a series 
of Hamiltonians $H_N$ ($N=0,1,2,\ldots$)
which approaches $H$ in the limit $N\to\infty$.
\begin{equation}
   H = \lim_{N\to\infty}  \Lambda^{-(N-1)/2} H_N \ ,
   \label{eq:H-HN}
\end{equation}
with
\begin{eqnarray}
 H_N =  \Lambda^{(N-1)/2}
          \bigg[   H_{\rm imp} + 
             \sqrt{\frac{\xi_0}{\pi}}\sum_{\sigma} \left(
            f^\dagger_{\sigma}c_{0\sigma} + 
            c^\dagger_{0\sigma}f_{\sigma}  \right) 
         \nonumber \\
              +  \sum_{\sigma n=0}^{N}  
               \varepsilon_n c_{n\sigma}^\dagger c_{n\sigma}
            + \sum_{\sigma n=0}^{N-1}
           t_n \left( c_{n\sigma}^\dagger c_{n+1\sigma}
                  + c_{n+1\sigma}^\dagger c_{n\sigma}\right) 
 \bigg] \ .
\nonumber \\
\label{eq:HN}
\end{eqnarray}
The factor $\Lambda^{(N-1)/2}$ in Eq.~(\ref{eq:HN}) (and, consequently,
the factor $\Lambda^{-(N-1)/2}$ in Eq.~(\ref{eq:H-HN}))
has been chosen to cancel the $N$-dependence of $t_{N-1}$, the
hopping matrix element between the last two sites of $H_N$.
Such a scaling is useful for the discussion of fixed points, as described
below. For a different $n$-dependence of $t_n$, as for the spin-boson
model \cite{Bulla:2005}, the scaling factor has to be changed accordingly.
(The $n$-dependence of $\varepsilon_n$ is, in most cases, irrelevant
for the overall scaling of the many-particle spectra.)

Two successive Hamiltonians are related by
\begin{eqnarray}
   H_{N+1} = \sqrt{\Lambda} H_N 
        + \Lambda^{N/2} \sum_\sigma  
        \varepsilon_{N+1} c_{N+1\sigma}^\dagger c_{N+1\sigma}
     \nonumber \\
+ \Lambda^{N/2} \sum_\sigma
              t_N
 \Big(
                c^\dagger_{N \sigma} c_{N+1 \sigma}
             +   c^\dagger_{N+1 \sigma} c_{N\sigma}  \Big) \ ,
  \label{eq:HN-rec}
\end{eqnarray}
and the starting point of the sequence of Hamiltonians is
given by
\begin{eqnarray}
   H_0 = \Lambda^{-1/2} \bigg[ H_{\rm imp} + 
           \sum_\sigma  
               \varepsilon_0 c_{0\sigma}^\dagger c_{0\sigma} 
         \nonumber \\
              +   \sqrt{\frac{\xi_0}{\pi}}\sum_{\sigma} \left(
            f^\dagger_{\sigma}c_{0\sigma} + 
            c^\dagger_{0\sigma}f_{\sigma}  \right) 
           \bigg]\ .
\end{eqnarray}
This Hamiltonian corresponds to a two-site cluster formed by the
impurity and the first conduction electron site. Note that in
the special case of the single-impurity Anderson model, one can
also choose $H_{-1}=\Lambda^{-1}H_{\rm imp}$ as the starting point
(with a proper renaming of parameters and operators) since
the hybridization term has the same structure as the hopping term
between the conduction electron sites.

The recursion relation Eq.~(\ref{eq:HN-rec}) can now  
be understood in terms of a renormalization group
transformation $R$:
\begin{equation}
   H_{N+1} = R \left( H_N \right) \ .
\end{equation}
In a standard RG transformation, the Hamiltonians are
specified by a set of parameters $\vec{K}$ and
the mapping $R$ transforms the Hamiltonian $H(\vec{K})$
into another Hamiltonian {\em of the same form}, $H(\vec{K^\prime})$,
with a new set of parameters $\vec{K^\prime}$. Such
a representation does not exist, in general, for the $H_N$
which are obtained in the course of the iterative diagonalization
to be described below. Instead, we characterize $H_N$, and
thereby also the RG flow, directly by the many-particle
energies $E_N(r)$
\begin{equation}
   H_N \vert r \rangle_N = E_N(r) \vert r \rangle_N \ \ , \ \
   r=1,\ldots,N_{\rm s} \ ,
   \label{eq:HN-EN}
\end{equation}
with the eigenstates $ \vert r \rangle_N $ and $N_{\rm s}$ the
dimension of $H_N$.
This is particularly useful in the crossover regime between
different fixed points, where a description in terms of an
effective Hamiltonian with certain renormalized parameters
is not possible. Only in the vicinity of the fixed points
(except for certain quantum critical points) one can go
back to an effective Hamiltonian description, as described
below.

\begin{figure}[!t]
\centerline{
  \includegraphics*[width=3.4in]{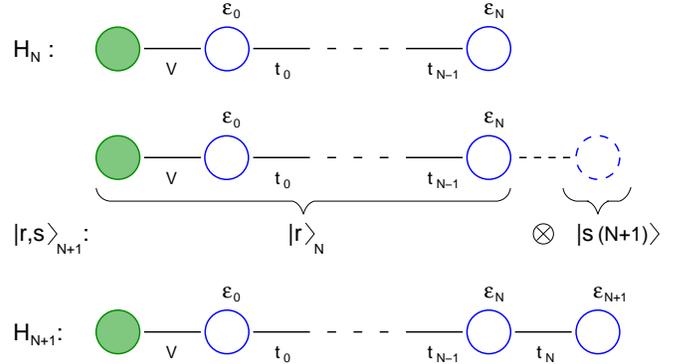}}
\caption{
In each step of the iterative diagonalization scheme one site of the
chain (with operators $c_{N+1}^{(\dagger)}$ 
and on-site energy $\varepsilon_{N+1}$) is added to the Hamiltonian
$H_N$. A basis $\vert r;s \rangle_{N+1}$ for the resulting
Hamiltonian, $H_{N+1}$, is formed by the eigenstates of $H_N$,
$\vert r \rangle_N$, and a basis of the added site, $\vert s(N+1) \rangle$.
}
\label{fig:it-diag1}
\end{figure}

Our primary aim now is to set up an iterative scheme
for the diagonalization of $H_N$, in order to discuss the
flow of the many-particle energies $E_N(r)$. Let us
assume that, for a given $N$, the Hamiltonian $H_N$ has
already been diagonalized, as in Eq.~(\ref{eq:HN-EN}).
We now construct a basis for $H_{N+1}$, as
sketched in Fig.~\ref{fig:it-diag1}:
\begin{equation}
   \vert r;s \rangle_{N+1} = \vert r \rangle_N \otimes \vert s(N+1) \rangle 
     \ .
   \label{eq:basis}
\end{equation}
The states $\vert r;s \rangle_{N+1}$ are product states consisting
of the eigenbasis of $H_N$ and a suitable basis $\vert s(N+1) \rangle$
for the added site (the new degree of freedom).
From the basis Eq.~(\ref{eq:basis}) we construct the
Hamiltonian matrix for $H_{N+1}$:
\begin{equation}
   H_{N+1}(rs,r^\prime s^\prime) =\,
   _{N+1}\langle r;s \vert H_{N+1} \vert r^\prime; s^\prime
      \rangle_{N+1} \ .
  \label{eq:Hmatrix}
\end{equation}
For the calculation of these matrix elements it is useful
to decompose $H_{N+1}$ into three parts
\begin{equation}
   H_{N+1} = \sqrt{\Lambda} H_N + \hat{X}_{N,N+1} + \hat{Y}_{N+1} \ ,
\label{eq:NextHam}
\end{equation}
(see, for example, Eq.~(\ref{eq:HN-rec}))
where the operator $\hat{Y}_{N+1}$ only contains the degrees of
freedom of the added site, while $\hat{X}_{N,N+1}$ mixes these
with the ones contained in  $H_{N}$. Apparently, the structure of
the operators $\hat{X}$ and $\hat{Y}$, as well as the equations
for the calculation of their matrix elements, depend on the model
under consideration.

\begin{figure}[!t]
\centerline{
  \includegraphics*[width=3.4in]{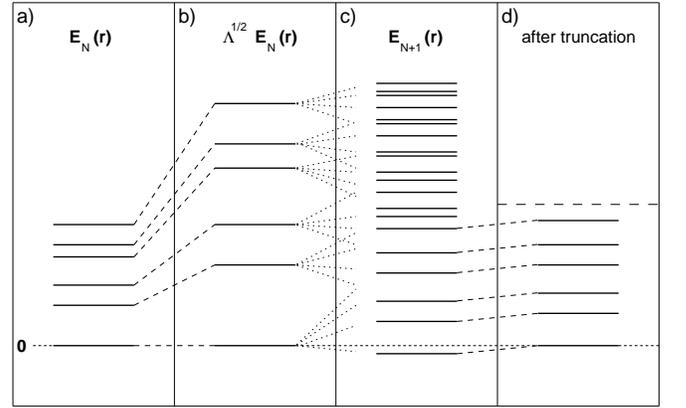}}
\caption{
(a): Many-particle spectrum $E_N(r)$ of the Hamiltonian $H_N$ with the
ground-state energy set to zero. (b): The relation between successive 
Hamiltonians, Eq.~(\ref{eq:HN-rec}), includes a scaling factor
$\sqrt{\Lambda}$. (c) Many-particle spectrum $E_{N+1}(r)$ of
$H_{N+1}$, calculated by diagonalizing the Hamiltonian matrix 
Eq.~(\ref{eq:Hmatrix}). (d) The same spectrum after truncation
where only the $N_{\rm s}$ lowest-lying states are retained; the
ground-state energy has again been set to zero.
}
\label{fig:it-diag2}
\end{figure}

The following steps are illustrated in Fig.~\ref{fig:it-diag2}: 
In Fig.~\ref{fig:it-diag2}a we
show the many-particle spectrum of $H_N$, that is the sequence
of many-particle energies $E_N(r)$. Note that, for convenience,
the ground-state energy has been set to zero. Figure \ref{fig:it-diag2}b
shows the overall scaling of the energies by the factor
$\sqrt{\Lambda}$, see the first term in Eq.~(\ref{eq:HN-rec}).

Diagonalization of the matrix Eq.~(\ref{eq:Hmatrix}) 
gives the new eigenenergies
$E_{N+1}(w)$ and eigenstates $\vert w \rangle_{N+1}$ which are
related to the basis $\vert r; s\rangle_{N+1}$ via the
unitary matrix $U$:
\begin{equation}
     \vert w \rangle_{N+1} =  \sum_{rs} U(w,rs) \vert r;
     s\rangle_{N+1} \ .
\label{eq:DiagH}
\end{equation}
The set of eigenenergies $E_{N+1}(w)$ of $H_{N+1}$ is displayed in
Fig.~\ref{fig:it-diag2}c (the label $w$ can now be replaced by $r$). 
Apparently, the number of states increases by adding the new degree of
freedom (when no symmetries are taken into account, the factor is
just the dimension of the basis $\vert s(N+1) \rangle$). The ground-state
energy is negative, but will be set to zero in the following step.

The increasing number of states is, of course, a problem for the numerical
diagonalization; the dimension of $H_{N+1}$ grows exponentially with
$N$, even when we consider symmetries of the model so that the full
matrix takes a block-diagonal form with smaller submatrices. This problem
can be solved by a very simple truncation scheme: after diagonalization
of the various submatrices of $H_{N+1}$ one only keeps the $N_{\rm s}$
eigenstates with the lowest many-particle energies. In this way,
the dimension of the Hilbert space is fixed to $N_{\rm s}$ and the
computation time increases linearly with the length of the chain. 
Suitable values for the parameter $N_{\rm s}$ depend on the model; for
the single-impurity Anderson model, $N_{\rm s}$ of the order of a few hundred
is sufficient to get converged results for the many-particle
spectra, but the accurate calculation of static and dynamic
quantities usually requires larger values of $N_{\rm s}$. The
truncation of the high energy states is illustrated in 
Fig.~\ref{fig:it-diag2}d.

Such an ad-hoc truncation scheme needs further explanations. 
First of all, there is no guarantee that this scheme will work in
practical
applications
and its quality should be checked for each individual application.
Important here is the observation that the neglect of the
high-energy states does not spoil the low-energy
spectrum in subsequent iterations -- this can be easily seen
numerically by varying $N_{\rm s}$. The influence of the high-energy
on the low-energy states is small since the addition of a new
site to the chain can be viewed as a perturbation of relative strength
$\Lambda^{-1/2}<1$. This perturbation is small for large values of
$\Lambda$ but for $\Lambda \to 1$ it is obvious that one has to
keep more and more states to get reliable results.
This also means that the accuracy of the NRG results is
getting worse when $N_{\rm s}$ is kept fixed and $\Lambda$
is reduced (vice versa, it is sometimes possible to improve the
accuracy by {\em increasing}  $\Lambda$ for fixed $N_{\rm s}$).

From this discussion we see that the success of the truncation scheme
is intimately connected to the special structure of the chain
Hamiltonian (that is $t_n\propto \Lambda^{-n/2}$) which in
turn is due to the logarithmic discretization of the original model.
A direct transformation of the single-impurity Anderson model to
a one-dimensional chain results in $t_n\to \rm const$ \cite{Hewson:1993},
and the above truncation scheme fails. A similar observation is made
when such a truncation is applied to the one-dimensional Hubbard model,
see the brief discussion in Sec.~\ref{sec:dmft}.

Let us now be a bit more specific on how to construct the basis
$\vert r; s\rangle_{N+1}$. For this we have to decide, first of all,
which of the symmetries of the Hamiltonian should be used in the
iterative diagonalization. In the original calculations of
\textcite{Wilson:1975} and 
\textcite{Krishnamurthy:1980a,Krishnamurthy:1980b} the
following quantum numbers were used: total charge $Q$ (particle
number with respect to half-filling), total spin $S$ and $z$-component
of the total spin $S_z$. It has certainly been essential in the
1970's to reduce the size of the matrices and hence the computation
time as
much as possible by invoking as many symmetries
as possible. This is no longer necessary to such an extent on the
modern computer systems, i.e.\ one can, at least for single-band
models, drop the total spin $S$ and classify the subspaces with the
quantum numbers $(Q,S_z)$ only. This simplifies the program considerably
as one no longer has to worry about reduced matrix elements and the
corresponding Clebsch-Gordan coefficients, see, for example
\textcite{Krishnamurthy:1980a}. As we use this representation in 
Sec.~\ref{subsec:thermodynamics}, let us here explicitly state the form of $\vert r; s\rangle_{N+1}$:
\begin{equation}\label{eq:basisQSz}
  \begin{array}{rcl}
   \left\vert Q,S_z,r;1 \right>_{N+1} &=& \left\vert Q+1,S_z,r \right>_N \ ,\\
   \left\vert Q,S_z,r;2 \right>_{N+1} &=& 
          c_{N+1\uparrow}^\dagger \left\vert Q,S_z-\frac{1}{2},r \right>_N \ , \\
   \left\vert Q,S_z,r;3 \right>_{N+1} &=& 
          c_{N+1\downarrow}^\dagger \left\vert Q,S_z+\frac{1}{2},r \right>_N \ , \\
   \left\vert Q,S_z,r;4 \right>_{N+1} &=& 
            c_{N+1\uparrow}^\dagger c_{N+1\downarrow}^\dagger
             \left\vert Q-1,S_z,r \right>_N \ .\\
   \end{array}
\end{equation}
Note that the quantum numbers $(Q,S_z)$ on both sides of these equations
refer to different systems, on the left-hand side they are for the
system including the added site, and on the right-hand side without the
added site.
We do not go into the details of how to set up the Hamiltonian
matrices Eq.~(\ref{eq:Hmatrix}), as this procedure is described in
great detail in Appendix B in \textcite{Krishnamurthy:1980a}).

For fermionic baths, the discretization parameter $\Lambda$ and the
number of states $N_{\rm s}$ kept in each iteration are the only
parameters which govern the quality of the results of the NRG
procedure. As discussed in more detail in Sec.~\ref{subsec:bosons},
for the case of a bosonic bath the infinite dimensional basis 
$\vert s(N+1)\rangle$ for the added bosonic site requires an
additional parameter $N_{\rm b}$, which determines the dimension of
$\vert s(N+1)\rangle$. 

\subsection{Renormalization group flow}

The result of the iterative diagonalization scheme are the
many-particle energies $E_N(r)$ with $r=1,\ldots,N_{\rm s}$
(apparently, the number of states is less than $N_{\rm s}$ for the
very first steps before the truncation sets in). The index $N$ goes
from $0$ to a maximum number of iterations, $N_{\rm max}$, which
usually has to be chosen such that the system has
approached its low-temperature fixed point.

As illustrated in Fig.~\ref{fig:it-diag2}, the set of many-particle
energies cover roughly the same energy range independent of $N$,
due to the scaling factor $\Lambda^{(N-1)/2}$ in 
Eq.~(\ref{eq:HN}). The energy of the first excited state of
$H_N$ is of the order of $\Lambda^{(N-1)/2}t_{N-1}$, a constant 
according to Eq.~(\ref{eq:tn-inf}). The energy of the highest
excited state kept after truncation depends on $N_{\rm s}$ --
for typical parameters this energy is approximately 5-10 times
larger as the lowest energy.


Multiplied with the scaling factor
$\Lambda^{-(N-1)/2}$, see Eq.~(\ref{eq:H-HN}), the energies $E_N(r)$ are 
an approximation to the many-particle spectrum of the chain Hamiltonian
Eq.~(\ref{eq:H_si}) within an energy window decreasing
exponentially with increasing $N$.  Note, that the energies for higher
lying excitations obtained for early iterations are not altered in
later iteration steps due to the truncation procedure. Nevertheless one
can view the resulting set of many-particle energies and states from all NRG
iterations $N$ as approximation to the spectrum of the full
Hamiltonian and  use them to calculate physical properties in the
whole energy range, see Sec.~\ref{sec:nrg-calc}.

Here we want to focus directly on the many-particle energies 
$E_N(r)$ and show how
one can extract information about the physics of a given model
by analyzing their flow, that is the dependence of $E_N(r)$
on $N$.

\begin{figure}[!t]
\centerline{
  \includegraphics*[width=3.1in]{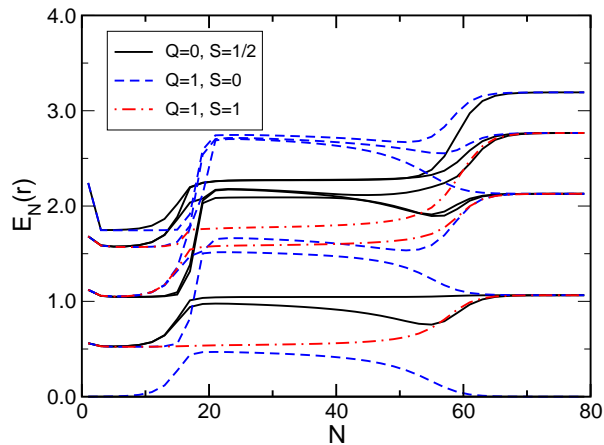}}
\caption{
Flow of the lowest-lying many-particle energies of the
single-impurity Anderson model for parameters 
$\varepsilon_f=-0.5\cdot 10^{-3}$, $U=10^{-3}$, $V=0.004$,
and $\Lambda=2.5$. The states are labeled by the quantum numbers
total charge $Q$ and total spin $S$. See the text for a discussion
of the fixed points visible in this plot.
}
\label{fig:flow1}
\end{figure}

As a typical example for such an analysis, we show in
Fig.~\ref{fig:flow1} the flow of the many-particle energies
for the symmetric single-impurity Anderson model, with parameters
$\varepsilon_f=-0.5\cdot 10^{-3}$, $U=10^{-3}$, $V=0.004$,
and $\Lambda=2.5$ (the
same parameters as used in Fig.~5 in \textcite{Krishnamurthy:1980a};
note that we show here a slightly different selection of the
lowest-lying states). The energies are plotted for 
odd $N$ only, that is an odd total number of sites (which is $N+2$). 
This is necessary, because the many-particle spectra
show the usual even-odd oscillations of a fermionic finite-size
system (the patterns for even $N$ look different but contain,
of course, the same physics). The data points are connected by lines to
visualize the flow. As in \textcite{Krishnamurthy:1980a}, the
many-particle energies are labeled by total charge $Q$ and
total spin $S$.

What is the information one can extract from such a flow diagram?
First of all we note the appearance of three different fixed
points of the RG transformation for early iteration numbers
$N<10$, for intermediate values of $N$ and for $N>60$
(strictly speaking, because we look at $N$ odd only, these are fixed
points of $R^2$, not of $R$). 
The physics of these fixed points cannot be extracted by just looking
at the pattern of the many-particle energies. This needs some further
analysis, in particular the direct diagonalization of fixed
point Hamiltonians (which usually have a simple structure) and
the comparison of their spectrum with the numerical data. An excellent
account of this procedure for the symmetric and asymmetric 
single-impurity Anderson model has been given by
\textcite{Krishnamurthy:1980a,Krishnamurthy:1980b}, and there is no
need to repeat this discussion here. The analysis shows that
for $N \approx 3-9$, the system is very close to the
free-orbital fixed point, with the fixed point Hamiltonian given
by Eq.~(\ref{eq:H_si}) for $U=0$ and $V=0$. This fixed point is unstable and
for $N\approx 11-17$, we observe a rapid crossover to the local-moment
fixed point. This fixed point is characterized by a free spin 
decoupled from the conduction band. The local-moment fixed
point is unstable as well and after a characteristic crossover (see
the discussion below) the system approaches the stable strong-coupling
fixed point of a screened spin. Note that the terminology
`strong-coupling' has been introduced originally because the 
fixed point Hamiltonian can be obtained from the limit $V\to\infty$,
so `coupling' here refers to the hybridization, not the Coulomb
parameter $U$.

The NRG does not only allow to match the structure of the numerically
calculated fixed points with those
of certain fixed point Hamiltonians. One can in addition identify the
deviations from the fixed points (and thereby part of the crossover) 
with appropriate perturbations of the fixed point
Hamiltonians. Again, we refer the reader to 
\textcite{Krishnamurthy:1980a,Krishnamurthy:1980b} for a detailed
description of this analysis. The first step 
is to identify the leading perturbations around the fixed
points. The leading operators can be determined by expressing them
in terms of the operators which diagonalize the fixed point
Hamiltonian; this tells us directly how these operators transform
under the RG mapping $R^2$. One then proceeds with the usual
classification into relevant, marginal, and 
irrelevant perturbations. The final results of this analysis 
perfectly agree with the flow
diagram of Fig.~\ref{fig:flow1}: There is a relevant perturbation
which drives the system away from the free-orbital fixed point,
but for the local-moment fixed point there is only a marginally
relevant perturbation, therefore the system only moves very slowly
away from this fixed point. Note that this marginal perturbation --
which is the exchange interaction between the local moment and the
spin of the first conduction
electron site -- gives rise to the logarithms observed in
various physical quantities. Finally, there are
only irrelevant operators which govern the flow to the strong-coupling
fixed point. These are responsible for the Fermi-liquid properties
at very low temperatures \cite{Hewson:1993}.

Having identified the leading operators for each fixed point,
it is possible to calculate physical properties close to the
fixed points perturbatively. We do not want to go into the
calculational details here, see \textcite{Krishnamurthy:1980a}
and also Sec.~4 in \textcite{Hewson:1993}. Recently, 
\textcite{Hewson:2004} and \textcite{Hewson:2005} developed
an alternative
approach based on the renormalized perturbation theory. 
This approach is much easier to implement, has been used to describe the physics
close to the strong-coupling fixed point and is, in principle, applicable also
on all energy scales and for non-equilibrium \cite{Hewson:2005b}. 

Flow diagrams as in Fig.~\ref{fig:flow1} also give information about the
relevant energy scales for the crossover between the fixed points.
For example, an estimate of the Kondo temperature $T_{\rm K}$ (the
temperature scale which characterizes the flow to the strong-coupling 
fixed point) is given by 
$T_{\rm K}\approx \omega_{\rm c} \Lambda^{-\bar{N}/2}$, with 
$\bar{N}\approx 55$ for the parameters in Fig.~\ref{fig:flow1}.

The discussion of flow-diagrams as in Fig.~\ref{fig:flow1} concludes
our introduction to the basics of the NRG approach. An important
part is still missing, of course, that is the calculation of
physical quantities from the flow of the many-particle
energies (and from certain additional matrix elements). This
is the topic of the following section.

In Sec.~\ref{sec:imp} we will come back to the discussion of flow
diagrams and the structure of fixed points when studying various other
quantum impurity systems, in particular the two-channel Kondo model
which displays a non-Fermi liquid fixed point, see 
Sec.~\ref{subsec:two-channel}, and the soft-gap Anderson model which
has a quantum critical point separating the strong-coupling and local-moment
phase, see Sec.~\ref{subsubsec:loc-crit}.

\section{Calculation of Physical Properties}
\label{sec:nrg-calc}

In the previous section~\ref{sec:nrg-intro} we discussed the
information that can be gained from the 
the low-lying energy levels during the RG flow. Apparently, a lot can
already be learned on this level about the physical properties of the
system. However, an obvious aim of any method is also to calculate
thermodynamic quantities like specific heat, susceptibilities or
even dynamical properties.

%
%
%
%

Let us start by reminding the reader that the coefficients $t_{n}$ appearing
in the transformed Hamiltonian Eq.~(\ref{eq:H_si}) decay like
$\Lambda^{-n/2}$ for large $n$. 
This aspect can be used in the
following way \cite{Wilson:1975,Krishnamurthy:1980a,Oliveira:1994}:
Diagonalizing the Hamiltonian (\ref{eq:HN}) for a given chain length $N$
yields a set of eigenvalues $\eta_l^{(N)}\propto \pm\Lambda^l$.
Obviously, eigenvalues $\Lambda^{-(N-1)/2}\eta_l^{(N)}\gg \kB T,\,\omega$ 
will not contribute significantly to the calculation of physical properties
anyway. On the other hand, for those $l$ where $\Lambda^{-(N-1)/2}\eta_l^{(N)}\ll \kB
T,\,\omega$ one can safely approximate $\eta_l^{(N)}\approx0$, which
means that {\em for the calculation of impurity properties} these
contributions will drop out.
With $\beta=(\kB T)^{-1}$,
\begin{equation}
  \label{eq:epsL}
  \beta\Lambda^{-(N-1)/2} =: \bar{\beta}
\end{equation}
and $\bar{\beta}$ chosen properly, it will thus be sufficient to use
$H_{N}$ instead of the full Hamiltonian
(\ref{eq:H_si}). 

Provided we can keep enough states in 
the truncation scheme introduced in Sec.~\ref{sec:nrg-intro}
to ensure convergence of the partition function on the scale $\kB T$,
it is thus
permissible to use the truncated Hamiltonian on the level $N$ obtained
from the iterative diagonalization 
to calculate physical properties {\em for the
impurity} on the temperature or energy scale $\Lambda^{-(N-1)/2}/\bar{\beta}$.

  \subsection{Thermodynamic and static properties}
  \label{subsec:thermodynamics}

\subsubsection{Entropy, specific heat and susceptibility}
The simplest physical quantities related to the impurity degrees of
freedom are the impurity contribution to the entropy, $S_{\rm imp}$,
specific heat, $C_{\rm imp}$, and magnetic
susceptibility, $\chi_{\rm imp}$. 

The entropy and specific heat are the first derivative of the free energy $F=-\kB T\ln
Z$ and internal energy $U=\langle H\rangle$ with respect to temperature, i.e.
$$
S = -\frac{\partial F}{\partial T} \ ,
$$
and
$$
C = \frac{\partial U}{\partial T}\;\;.
$$
From a numerical point of view, performing differentiations is
something to avoid if possible. For the numerical implementation of
the NRG another complication arises. To avoid an exponential increase
of energies, it is necessary to subtract the ground state energy at
each NRG-level $N$, i.e.\ one would
have to keep track of these
subtractions. Apparently, a much more convenient approach is to
evaluate the derivative analytically, yielding
$$
S/k_{\rm B} = \beta \langle H\rangle + \ln Z \ ,
$$
for the entropy and
$$
C/\kB = \beta^2\left[\langle H^2\rangle-\langle H\rangle^2\right] \ ,
$$
for the specific heat.

The prescription to calculate the impurity contribution to the
magnetic susceptibility requires some more thought. The standard
definition for the magnetic susceptibility is (we set $g\mu_{\rm B}=1$)
\begin{equation}
  \label{eq:SuszGeneral}
  \chi(T) = \int\limits_0^\beta\langle S_z[\tau]S_z\rangle d\tau -
  \beta\langle S_z\rangle^2\;\;,
\end{equation}
with
$$
\langle S_z[\tau]S_z\rangle = \frac{1}{Z}{\rm Tr}
\left[e^{-\beta H}e^{\tau
  H}S_ze^{-\tau H}S_z\right]\;\;.
$$
However, the evaluation of the latter expectation value is equivalent
to the calculation of a {\em dynamical} correlation function. This is
in general a much more complex task and will be discussed in detail in
the next section. Here, we employ a different approach, which in
turn is also closer related to the experimental definition of this
quantity. 

In general, experiments address the susceptibility of
the whole system. Since the total spin commutes with the Hamiltonian,
the expression (\ref{eq:SuszGeneral}) simplifies to
$$
\chi_{\rm tot}(T) = \beta\left[ \langle S_{{\rm tot},z}^2\rangle
- \langle S_{{\rm tot},z}\rangle^2\right] \ ,
$$
in this case. From this, one subtracts the susceptibility of a
reference system, i.e.\ without impurity, leading to Wilson's
definition \cite{Wilson:1975} of
the {\rm impurity contribution to the susceptibility}
\begin{equation}
  \label{eq:WilsonChi}
  \chi_{\rm imp}(T) = \chi_{\rm tot}(T) - \chi_{\rm tot}^{(0)}(T)\;\;.
\end{equation}
Since $S_{{\rm tot},z}$ is a quantum number used to classify the
states in the calculation, the expectation values in
(\ref{eq:WilsonChi}) can be evaluated straightforwardly.

Similarly, the impurity contributions to the entropy and specific heat
can be calculated as
\begin{equation}
  \label{eq:SImp}
  S_{\rm imp}(T) = S_{\rm tot}(T)-S_{\rm tot}^{(0)}(T) \ ,
\end{equation}
and 
\begin{equation}
  \label{eq:CImp}
  C_{\rm imp}(T) = C_{\rm tot}(T)-C_{\rm tot}^{(0)}(T)\;\;,
\end{equation}
where $S_{\rm tot}^{(0)}(T)$ and $C_{\rm tot}^{(0)}(T)$ are again entropy
and specific heat of a suitable reference system.

Let us discuss the details of the actual calculation for the entropy as specific example.
Following the introductory remarks, we can 
-- for a given temperature $\kB T$ --
restrict the Hilbert space to the NRG iteration $L$ fulfilling
(\ref{eq:epsL}). If we denote the corresponding Hamiltonian by
$H^{(N)}$, we can introduce the quantity
\begin{equation}\label{eq:Entropy}
S^{(N)}/k_{\rm B} := \beta \langle H^{(N)}\rangle^{(L)}+\ln Z^{(N)}\;\;,
\end{equation}
where, using the notation of Sec.~\ref{sec:nrg-intro} 
(see, for example, Eq.~(\ref{eq:basis})), 
\begin{equation}
  \label{eq:ThermAv}
  \langle\ldots\rangle^{(N)}:= \frac{1}{Z^{(N)}}
\sum\limits_{Q,S_z}\sum\limits_r\begin{array}[t]{l}\displaystyle
e^{-\beta E_L(Q,S_z,r)}\times\\[5mm]
\displaystyle\phantom{}_N\langle Q,S_z,r|\ldots|Q,S_z,r\rangle_N \ ,
\end{array}
\end{equation}
and
\begin{equation}
  Z^{(N)} := \sum\limits_{Q,S_z}\sum\limits_r
e^{-\beta E_N(Q,S_z,r)}\;\;.
\end{equation}
The impurity contribution to the entropy for a temperature
$\kB T_N := \Lambda^{-(N-1)/2}/\bar{\beta}$ can then be obtained as 
\begin{equation}\label{eq:SImpAppr}
S_{\rm imp}(T_N)/k_{\rm B} \approx  S^{(N)}/k_{\rm B} - S_{\rm cb}^{(N)}/k_{\rm B}\;\;.
\end{equation}
Here we introduced the ``free entropy'' 
\begin{equation}
  \label{eq:Entropy0}
S_{\rm cb}^{(N)}/k_{\rm B} := \beta \langle H_{\rm cb}^{(N)}\rangle^{(N)}+\ln
Z_{\rm cb}^{(N)} \ ,
\end{equation}
obtained from the bare conduction Hamiltonian 
\begin{equation}
  H_{\rm cb}^{(N)} = 
\sum_{\sigma n=0}^N
 \begin{array}[t]{l}\displaystyle
 \Bigl[ \epsilon_n c_{n\sigma}^\dagger c_{n\sigma}+\\[5mm]
\displaystyle t_n \left( c_{n\sigma}^\dagger c_{n+1\sigma}
                  + c_{n+1\sigma}^\dagger c_{n\sigma}\right)\Bigr]\;\;.
\end{array} 
\label{eq:H_cb_L}
\end{equation}
 
Similarly, for $\kB T_N = \Lambda^{-(N-1)/2}/\bar{\beta}$ the specific heat and magnetic susceptibility are obtained
as
\begin{equation}
  \label{eq:CImpAppr}
  C_{\rm imp}(T_N)/\kB \approx C^{(N)}_{\rm tot} - C^{(N)}_{\rm cb} \ ,
\end{equation}
and
\begin{equation}
  \label{eq:ChiImpAppr}
  \chi_{\rm imp}(T_N) \approx \chi^{(N)}_{tot} - \chi^{(N)}_{\rm cb}\;\;.
\end{equation}
Since the Hamiltonian (\ref{eq:H_cb_L}) is a non-interacting system, 
these quantities $S^{(N)}_{\rm cb}$ etc.\ can be
expressed via the eigenenergies $\eta_{l\sigma}$ of (\ref{eq:H_cb_L})
in standard fashion.

For $T\to0$ the behavior of $S_{\rm imp}(T)$ and $\chi_{\rm imp}(T)$
given by Eqs.~(\ref{eq:CImpAppr}) and (\ref{eq:ChiImpAppr}) can be obtained analytically from the fixed point spectra. We refer the reader interested in this derivation
to \textcite{Wilson:1975} and concentrate here on the actual
numerical calculations.

Another aspect is 
that the fixed points and
the flow to them are different for
$N$ even and odd. This in turn means, that one in principle has to
calculate thermodynamic properties either for $N$ even or odd only and
thus loose half of the temperature values. One can, however,
use all information by properly averaging odd and even steps: 
\begin{itemize}
\item For a given $N$, calculate the quantities $O^{(N-1)}$,
  $O^{(N)}$ and $O^{(N+1)}$. 
\item Approximate $O(T_N)$ as
$$
O(T_N) \approx \frac{1}{2}\Bigl[
\begin{array}[t]{l}\displaystyle
O^{(N)}+O^{(N-1)}+\\[5mm]
\displaystyle
\frac{O^{(N+1)}-O^{(N-1)}}{T_{N+1}-T_{N-1}}\left(T_N-T_{N-1}\right)\Bigr] \ .
\end{array}
$$
The first term in the square bracket is the observable calulated at
step $N$. The second and third term are a linear interpolation of the
values at $N-1$ and $N+1$ to iteration $N$.
\item Continue with $N+1$.
\end{itemize}
As a positive side effect, this averaging also improves the accuracy
of the thermodynamic quantities calculated.

At this point some remarks about potential numerical
problems one can encounter should be made. The arguments given in the introduction to
this section rely on the assumption, that one can keep states
with high enough energy to ensure (i) the accuracy of the states at
medium and low energies and (ii) the convergence of the partition
function and expectation values. Depending on the actual quantity to
be calculated, the latter point can in principle lead to problems. 
As an example, consider $\langle H\rangle$ and $\langle H^2\rangle$. 
While for a given energy cut-off $E_{\rm cut}$ the contribution $\bar{\beta}E_{\rm cut} e^{-\bar{\beta}E_{\rm cut}}$
to $\langle H\rangle$ can already be small enough to use the sum up
to $E_{\rm cut}$ as approximation to $\langle H\rangle$, this must not be
necessarily true for $\langle H^2\rangle$. Thus, the resulting values
for the specific heat, $\bar{\beta}\left[\langle H^2\rangle-\langle H\rangle^2\right]$, can be rather poor although entropy and
susceptibility come out much more accurate.

\subsubsection{Other local properties}
While entropy, specific heat and impurity susceptibility can be obtained directly
from the spectra of the Hamiltonian, other local quantities require
the calculation of the corresponding local matrix elements. As an
example, we want to discuss here the local occupancy $n_\sigma=\langle
f^\dagger_\sigma f^{\phantom{\dagger}}_\sigma\rangle $ and double
occupancy $D=\langle
f^\dagger_\uparrow f^{\phantom{\dagger}}_\uparrow f^\dagger_\downarrow
f^{\phantom{\dagger}}_\downarrow\rangle$ for the single-impurity
Anderson model
Eq.~(\ref{eq:siam}). Both quantities are of interest in actual
applications. Expectation values of
other local operators can be calculated in a similar manner.

As before, on a given temperature scale $\kB T_N =
\Lambda^{-(N-1)/2}/\bar{\beta}$, we approximate the expectation values
by
\begin{eqnarray}
  \label{eq:Occupancy}
  n_\sigma(T_L) &\displaystyle\approx
 \frac{1}{Z^{(N)}}
\sum\limits_{Q,S_z}\sum\limits_r&
e^{-\beta E_N(Q,S_z,r)}\times\\
&&
\phantom{}_N\langle Q,S_z,r|f^\dagger_\sigma f^{\phantom{\dagger}}_\sigma|Q,S_z,r\rangle_N\nonumber \ ,
\end{eqnarray}
for the occupancy and a corresponding expression for the double
occupancy. The matrix elements
\begin{equation}
  \label{eq:MEforN}
n_\sigma(Q,S_z,r,r';N):=  \phantom{}_N\langle Q,S_z,r|f^\dagger_\sigma f^{\phantom{\dagger}}_\sigma|Q,S_z,r'\rangle_N  \ ,
\end{equation}
at a given step $N$ can be calculated from those of the previous step
$N-1$ with the help
of the basis transformation (\ref{eq:DiagH})
for the step $N$. The same 
scheme works for the matrix elements of the
double occupancy $D(Q,S_z,w,w';-1)$ and the 
matrix elements of general local operators
 -- like $f_\sigma^\dagger$ needed in the calculation of the single-particle
Green function (see Sec.~\ref{subsec:dynamics}).

All we are left to specify are the initial values for
$n_\sigma(Q,S_z,w,w';-1)$ and $D(Q,S_z,w,w';-1)$ on the level of the impurity. For the
Anderson model Eq.~(\ref{eq:siam}) they are explicitly given as
\begin{eqnarray}
n_\sigma(0,0,0,0;-1) & = & 0 \ ,\nonumber\\
n_\sigma(1,\sigma,0,0;-1) & = & 1 \ ,\nonumber\\
n_\sigma(2,0,0,0;-1) & = & 2 \ ,\nonumber\\
D(0,0,0,0;-1) & = & 0 \ ,\nonumber\\
D(1,\sigma,0,0;-1) & = & 0 \ ,\nonumber\\
D(2,0,0,0;-1) & = & 1\;.
\end{eqnarray}
With these prerequisites we are now in the position to do actual calculations
for the thermodynamic properties of quantum impurity models using NRG.

\subsubsection{Example: The Kondo model}

As example for the method let us present results for the Kondo
model Eq.~(\ref{eq:TwoChannelKondo}). 
Depending on the number of bands coupling
to the local spin, one observes a conventional Kondo effect with
the formation of a local Fermi liquid  or a non-Fermi
liquid fixed point with anomalous temperature
dependencies of specific heat and susceptibility as well as a residual entropy
$S(0)=\frac{1}{2}\ln2$ at $T=0$ \cite{Nozieres:1980,Cragg:1980}
(this will be discussed in more detail in Secs.~\ref{subsec:kondo} and
\ref{subsec:two-channel}). 
\begin{figure}[htb]
\begin{center}
\includegraphics[width=0.43\textwidth,clip]{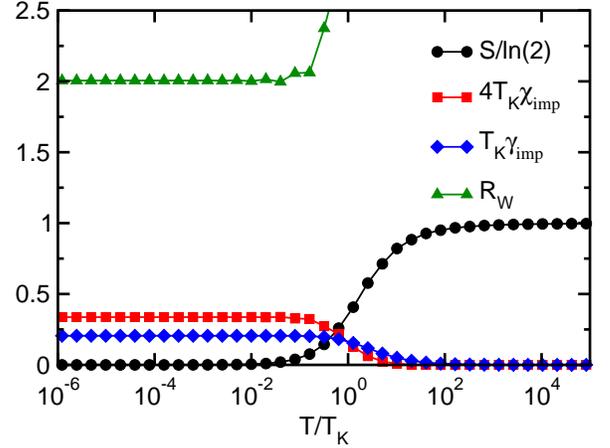}
\end{center}
\caption[]{Entropy $S_{\rm imp}(T)$, susceptibility $\chi_{\rm
    imp}(T)$, Sommerfeld coefficient $\gamma_{\rm imp}=C_{\rm imp}(T)/T$, and 
    Wilson ratio $R_{\rm W}$ for the
    single-channel Kondo model. The Kondo temperature is defined by
     the Wilson relation $\chi_{\rm imp}(0) = \frac{0.413}{4T_{\rm K}}$.\label{fig:III-1}}
\end{figure}
In Fig.~\ref{fig:III-1} we show the entropy $S_{\rm imp}(T)$,
susceptibility $\chi_{\rm imp}(T)$, Sommerfeld coefficient
$\gamma_{\rm imp}=C_{\rm imp}(T)/T$ and Wilson ratio $R_{\rm W} :=
4\pi^2\chi_{\rm imp}(T)/(3\gamma_{\rm imp}(T))$ as function of
$T/T_{\rm K}$ for the single-channel Kondo model. As Kondo coupling we
choose $J=0.05D$, where $D$ is the half-bandwidth of the conduction
band, for which we assume a density of states
$\rho_{\rm cb}(\epsilon)=N_F\!\Theta(D-|\epsilon|)$. The value of $T_{\rm K}$ is obtained from Wilson's
definition \cite{Wilson:1975} $4T_{\rm K}\chi_{\rm imp}(0)=0.413$.
The calculations are performed with a discretization parameter $\Lambda=4$, keeping
$400$ states at each NRG step. Although this value of $\Lambda$ seems
to be fairly large, 
experience tells that for static
properties
such large values of $\Lambda$ are still permissible, considerably
reducing
the number of states one has to keep in the truncation procedure.

One nicely sees in Fig.~\ref{fig:III-1} the quenching of the
local moment by the Kondo effect for temperatures of the order of
$T_{\rm K}$. Also the high-temperature values for the entropy $S_{\rm
  imp}(T\to\infty)=\ln2$ and the Wilson ratio $R_W=2$ \cite{Wilson:1975} are
obtained with high precision. 

\begin{figure}[htb]
\begin{center}
\includegraphics[width=0.43\textwidth,clip]{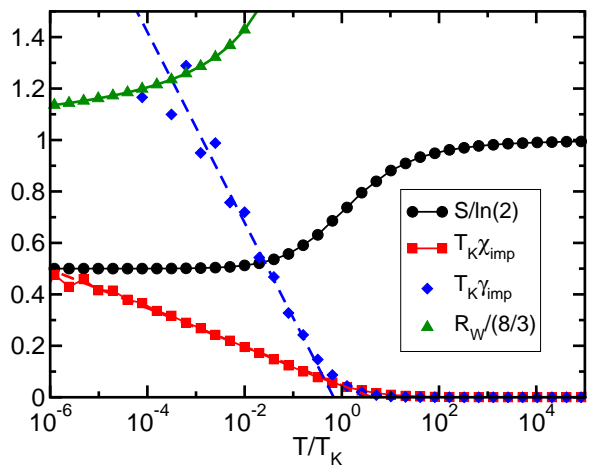}
\end{center}
\caption[]{Entropy $S_{\rm imp}(T)$, susceptibility $\chi_{\rm
    imp}(T)$, Sommerfeld coefficient $\gamma_{\rm imp}=C_{\rm imp}(T)/T$, and 
  Wilson ratio $R_{\rm W}$ for the
  two-channel Kondo model. The value for the Kondo temperature is the
  same as in Fig.~\ref{fig:III-1}.\label{fig:III-2}}
\end{figure}
If one adds a second screening channel to the Kondo model, one arrives
at the so-called two-channel Kondo model. The rather interesting
physics of this model will be discussed in detail in
Sec.~\ref{subsec:two-channel}. Here we merely want to demonstrate
that NRG calculations for this model are possible, too; however, the
additional bath degrees of freedom, which lead to Hilbert spaces
larger by a factor of four, make calculations more cumbersome and for
some quantities also less accurate. In Fig.~\ref{fig:III-2} we show as
before the impurity contributions to the entropy, susceptibility and
Sommerfeld coefficient as well as the Wilson ratio as function of
$T/T_{\rm K}$. The impurity
parameters are the same as in Fig.~\ref{fig:III-1}; for the NRG we again choose
$\Lambda=4$ but keep $8100$ states per iteration. 
The value of $T_{\rm K}$ is that of the corresponding 
single-channel model. As emphasized
before, entropy and susceptibility come out quite accurately, in
particular the residual entropy $S(0)=\frac{1}{2}\ln2$ is obtained as
well as the logarithmic increase of $\chi_{\rm
  imp}(T)\propto\ln\left(T/T_{\rm K}\right)$ for $T<T_{\rm
  K}$ \cite{Cragg:1980,Pang:1991,Affleck:1992}. The specific heat, however, is less accurate,
but also shows the logarithmic increase as expected, although with strong
oscillations superimposed. Fitting both
quantities with a logarithmic form, one can recover the
correct Wilson ration $R_W=8/3$ for $T\to0$. Note, however, that the
latter value is approached only logarithmically.
%
%
\subsubsection{Improving the accuracy: The $Z$-averaging\label{subsubsec:Zaverage}}

%
%
For more complex quantum impurity models, like the two-channel Kondo model
(discussed briefly in the previous section and in more detail in
Sec.~\ref{subsec:two-channel}), or multi-orbital models, the Hilbert 
space per NRG step increases more strongly than for single-channel models.
Consequently, the fraction of states kept in the truncation procedure has
to be reduced.
As has been pointed out by \textcite{Oliveira:1994}, this
leads to an exponential decrease of accuracy, which can however be compensated
by an increase of the discretization parameter $\Lambda$. However, the use
of a large $\Lambda$, (i), takes one further away from the continuum limit
$\Lambda\rightarrow 1$ of interest, 
and, (ii), introduces oscillations into the thermodynamic expectation
values. A way out of this dilemma, proposed by 
\textcite{Oliveira:1994}, is as follows:
\begin{itemize}
\item Instead of the discretization Eq.~(\ref{eq:discretization_points}) choose
\begin{equation}\label{eq:discretization_points_Z}
x_n=\Lambda^{-n+Z}\,,\;n\ge1\,,\;Z\in[0,1)\;.
\end{equation}
The mapping to a semi-infinite chain is done as before (see Sec.~\ref{subsec:semi-infinite}) with the replacement $\Lambda^{-n}\to\Lambda^{-n+Z}$ for $n\ge1$.
\item For fixed $Z\in[0,1)$ perform a NRG calculation for a {\em fixed set of
temperatures} $T_L=\Lambda^{-(L-1)/2}/\bar\beta$ as before.
\item Average over several calculations for different $Z$. This
  averaging is meant to reintroduce the continuum limit to some extent
\cite{Oliveira:1994} and also can be shown to remove oscillations
introduced by the use of a large $\Lambda\gg 1$.
\end{itemize}
Already for two different values of $Z$ this procedure removes
spurious oscillations in thermodynamic quantities and reproduces the exact result
with good accuracy for $\Lambda$ as large as $\Lambda=10$.
This technique can be incorporated straightforwardly 
into the NRG code (for applications, see
\textcite{Silva:1996,Costa:1997,Paula:1999,Ramos:2003,Campo:2003,Campo:2004}).
%
%
%
%
%
%
%
%
%
%

  \subsection{Dynamic properties}
  \label{subsec:dynamics}

\subsubsection{Equilibrium dynamics and transport}
\label{subsubsec:dyn+trans}
We consider now the application of the NRG to the calculation of
dynamic and transport properties of quantum impurity models 
\cite{Frota:1986,Sakai:1989,Costi:1992b,Costi:1994b}. For
definiteness we shall consider the Anderson impurity model and illustrate
the procedure for the impurity spectral density
$A_{\sigma}(\omega,T)=-\frac{1}{\pi}{\rm Im}
G_{\sigma}(\omega,T)$, with 
\begin{eqnarray}
G_{\sigma}(\omega,T) &=&\int_{-\infty}^{+\infty} d(t-t') 
e^{i\omega (t-t')} G_{\sigma}(t-t')\ , \label{eq:fourier-rep}\\
G_{\sigma}(t-t') & = & -i\theta(t-t')\langle 
[f_{\sigma}(t),f_{\sigma}^{\dagger}(t')]_{+}\rangle_{\varrho} \ ,
\label{eq:two-times}
\end{eqnarray}
with $\varrho$ the density matrix of the system.
Suppose, for the moment, that we have {\it all} the many-body eigenstates,
$|r\rangle$, and eigenvalues, $E_{r}$, of the Anderson
impurity Hamiltonian, $H$, exactly. Then the density matrix, $\varrho(T)$, and partition
function, $Z(T)$, of the full system at temperature $k_{\rm B}T=1/\beta$ can be written
\begin{eqnarray}
\varrho(T) &=& \frac{1}{Z(T)}\sum_{r} e^{-\beta E_{r}}|r\rangle \langle r|,
\label{eq:density-matrix}\\
Z(T) &=& \sum_{r} e^{-\beta E_{r}},
\end{eqnarray}
and the impurity spectral density, $A_{\sigma}$, can
be written in the Lehmann representation as
\begin{eqnarray}
A_{\sigma}(\omega,T) &=& \frac{1}{Z(T)}\sum_{r,r'}
|M_{r,r'}|^{2}
(e^{-E_{r}/k_{\rm B}T} + e^{-E_{r'}/k_{\rm B}T})\nonumber\\
&&\times \delta(\omega-(E_{r'}-E_{r})).
\label{eq:lehmann-finite}
\end{eqnarray}
with $M_{r,r'}=\langle r|f_{\sigma}|r'\rangle$ the relevant many-body
matrix elements.

Consider first the $T=0$ case ($T>0$ is described below), then 
\begin{eqnarray}
&&A_{\sigma}(\omega,T=0) =
\frac{1}{Z(0)}\sum_{r}
|M_{r,0}|^{2}\delta(\omega+(E_{r}-E_{0}))\nonumber\\
&&+\frac{1}{Z(0)}\sum_{r'}
|M_{0,r'}|^{2}\delta(\omega-(E_{r'}-E_{0}),
\label{eq:lehmann-zero}
\end{eqnarray}
with $E_{0}=0$ the ground state energy. In order to
evaluate this from the information which we actually obtain from
an iterative diagonalization of $H$, we consider the impurity
spectral densities corresponding to the sequence of Hamiltonians
$H_{N}$, $N=0,1,\dots$, whose characteristic scale is 
$\omega_{N}=\frac{1}{2}(1+\Lambda^{-1})\Lambda^{-(N-1)/2}$,
 \begin{eqnarray}
&&A_{\sigma}^{N}(\omega,T=0) =
\frac{1}{Z_{N}(0)}\sum_{r}
|M_{r,0}^{N}|^{2}\delta(\omega+E_{r}^{N})\nonumber\\
&&+\frac{1}{Z_{N}(0)}\sum_{r'}
|M_{0,r'}^{N}|^{2}\delta(\omega-E_{r'}^{N}). \label{eq:lehmann-zero-cluster}
\end{eqnarray} 
Here, $E_{r}^{N}$ and $|r\rangle_{N}$ are the eigenvalues and eigenstates of $H_{N}$, 
i.e.
\begin{equation}
H_{N}|r\rangle_{N}=E_{r}^{N}|r\rangle_{N},
\end{equation}
and,
\begin{equation}
M_{r,r'}^{N}={_N}\langle r|f_{\sigma}|r'\rangle_{N},
\end{equation}
are the relevant many-body matrix
elements, whose calculation will be outlined below.  
Since the spectrum of $H_{N}$ is truncated, the range of excitations
it describes is limited to $0\le \omega \le K(\Lambda)\omega_{N}$, 
where $K(\Lambda)$ depends on both $\Lambda$ and the actual number 
of states retained at each iteration and is typically $5-10$ 
for $\Lambda=1.5-2.0$ for $N_{\text s}=500-1000$ retained states. 
Moreover, excitations and eigenstates
below the characteristic scale $\omega_{N}$ of $H_{N}$ 
will only be approximations to the excitations and eigenstates
of the infinite system described by $H$. These excitations and eigenstates are 
refined in subsequent iterations. Hence, for each $N=1,2,\dots$ we can 
evaluate the spectral density from $A_{\sigma}^{N}$ at a frequency 
$\omega$ chosen to lie in the window
$\omega_{N}\le \omega \le K(\Lambda)\omega_{N}$,
\begin{equation}
A_{\sigma}(\omega,T=0) \approx A_{\sigma}^{N}(\omega,T=0).
\label{eq:approx-dynamics}
\end{equation} 
A typical choice, for $\Lambda=1.5-2.0$, is $\omega=2\omega_{N}$. 

The above procedure still only yields discrete spectra. 
For comparison with experiment, smooth spectra are required, so 
we replace the delta functions $\delta(\omega\pm E_{r}^{N})$ appearing in 
(\ref{eq:lehmann-zero-cluster})
by smooth distributions $P(\omega\pm E_{r}^{N})$. A natural choice
for the width $\eta_{N}$ of $P$ is $\omega_{N}$, the characteristic
scale for the energy level structure of $H_{N}$. Two commonly used
choices for $P$ are the Gaussian, $P_{G}$, and the Logarithmic Gaussian, $P_{LG}$,
distributions \cite{Sakai:1989,Costi:1994b,Bulla:2001}:
\begin{eqnarray}
P_{G}(\omega \pm E_{r}^{N}) &=& \frac{1}{\eta_{N}\sqrt{\pi}}
e^{-\left[\frac{(\omega \pm E_{r}^{N})}{\eta_{N}}\right]^{2}}\ , \\
P_{LG}(\omega \pm E_{r}^{N}) &=& \frac{e^{-b^{2}/4}}{bE_{r}^{N}\sqrt{\pi}}
e^{-\left[\frac{\ln(\omega/E_{r}^{N})}{b}\right]^{2}} \ , 
\end{eqnarray}
\begin{figure}[!t]
\centerline{
  \includegraphics*[width=3.0in]{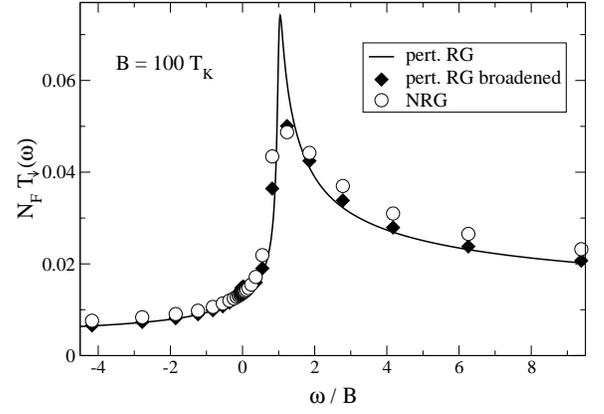}}
  \caption{The spin-resolved Kondo resonance at high magnetic fields calculated with
NRG and perturbative RG. The large Gaussian broadening used at 
$\omega=g\mu_{\rm B}B=100T_{\rm K}$ reduces the height of the sharp peak 
and overestimates its width, as is evident on applying the NRG broadening
procedure to the analytic perturbative RG result \protect{\cite{Rosch:2003}}.
}
\label{fig:Kondo-largeB}
\end{figure}
For the Gaussian, a width $\eta_{N}=0.3\omega_{N}-0.8\omega_{N}$ is typically used 
\cite{Costi:1994b}, whereas, for
the logarithmic Gaussian, a typical width parameter $b=0.3-0.7$ is used 
\cite{Sakai:1989,Bulla:2001}. Note that the logarithmic Gaussian gives little weight
to excitations below $\omega_{N}$ and more weight to the higher energy excitations. Due
to the logarithmic discretization, this might appear to be the better choice. 
In practice, the difference to using a Gaussian is small. 

In general, spectra
for even and odd $N$ differ by a few \% at most as a result of finite-size effects
(see also the discussion of the
reduced density matrix approach in Sec.~\ref{subsubsec:se+dm}), 
so generally either even $N$ or
odd $N$ spectra are calculated (as for thermodynamics). It is also possible to
combine information from shell $N$ and $N+2$ by an appropriate weighting 
\cite{Bulla:2001}. We note also, that since the broadening is proportional to energy,
a peak of intrinsic width $\Gamma$ at frequency $\Omega_{0}$ will 
be well resolved by the above procedure provided that 
$\Omega_{0}\ll \Gamma$, which is the
case for the Kondo resonance and other low energy resonances. In
the opposite case, the low (logarithmic) resolution at higher
frequencies may be insufficient to resolve the intrinsic widths
and heights of such peaks, although their weights are correctly 
captured. In cases where the width of such high energy peaks is
due to single-particle effects, e.g. the resonant level in the empty orbital
regime of the Anderson model, one can use the representation of the spectral
density in terms of the correlation self-energy, as described in the next section,
with the single-particle broadening being put in explicitly so that
essentially the correct peak widths and heights is obtained. 
In other cases, when the width
of such peaks is due to correlations, one inevitably has some over-broadening.
An extreme example is the spin-resolved Kondo resonance
at high magnetic fields, $B\gg T_{\rm K}$, which is sharply peaked 
at $\omega=B$ and is highly asymmetrical, as shown in Fig.~\ref{fig:Kondo-largeB}.
The extent of the problem is quantified here by comparison with analytic perturbative
results with and without the NRG broadening procedure. 

A procedure for obtaining smooth spectra, which resolves finite frequency 
peaks without broadening the discrete spectra, involves an averaging over 
many different discretizations of the band (the Z-averaging discussed in the
previous section on thermodynamics). We refer the reader to 
\textcite{Yoshida:1990} for details.

\begin{figure}[!t]
\centerline{
  \includegraphics*[width=3.0in]{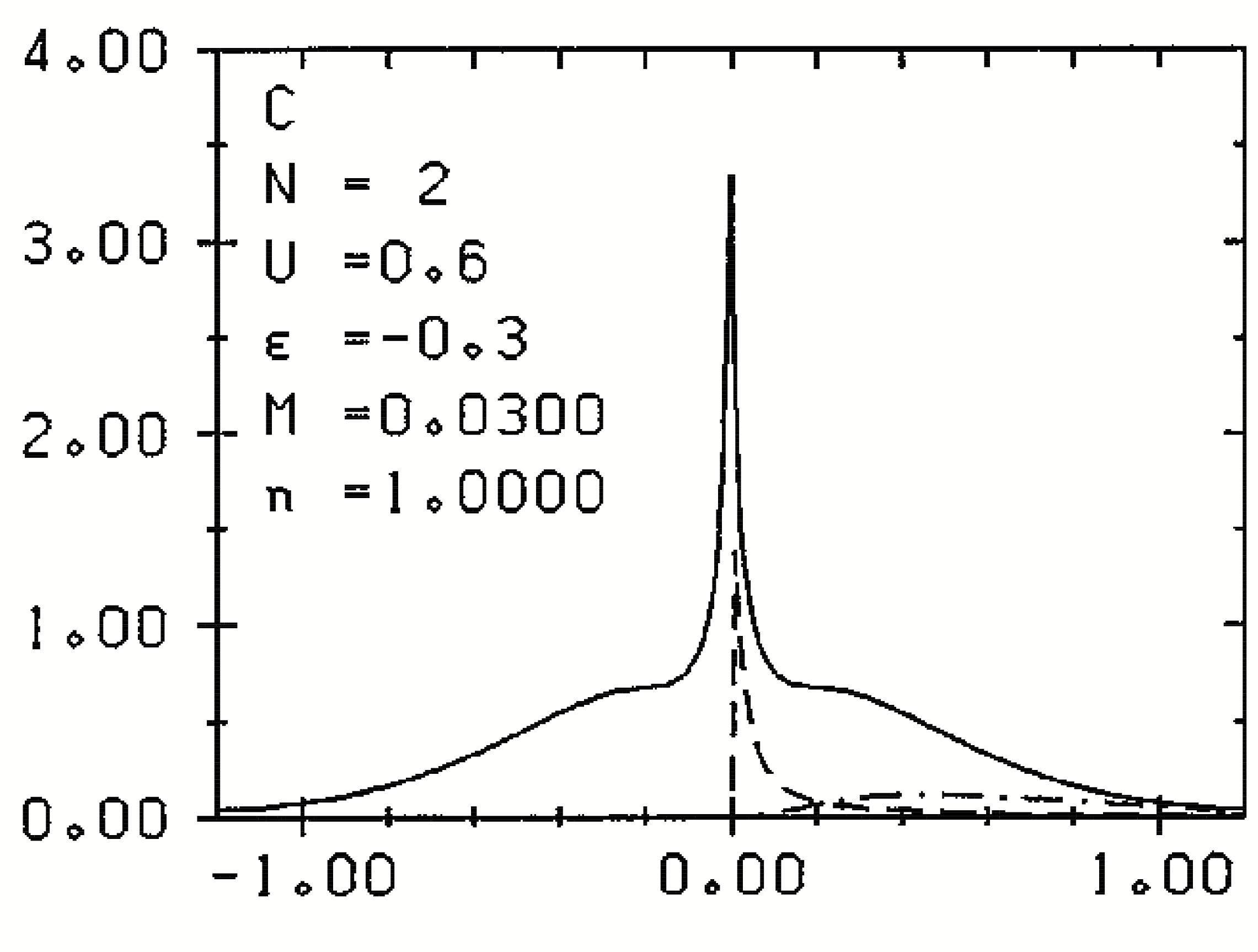}}
  \caption{$T=0$ spectral densities for single-particle (solid line), 
magnetic (dashed line) and charge (dot-dashed line) excitations of 
the spin degenerate symmetric Anderson model versus energy $\omega/D$ 
for $U=0.6D$, $D=1.0$ and $\Delta/\pi = M = 0.03D$ \protect{\cite{Sakai:1989}}.
}
\label{fig:SakaiSD}
\end{figure}

In calculating the impurity spectral density, one
requires also the matrix elements $M_{r,r'}^{N}$ at each iteration. These are
obtained recursively by using the unitary transformation 
Eq.~(\ref{eq:DiagH}) yielding
\begin{eqnarray}
M_{r,r'}^{N} &=&
\sum_{p,s_{N}}\sum_{p',s_{N}'}U_{N}(r,ps_{N})U_{N}(r',p's_{N}')\nonumber\\
&&\times \delta_{s_{N},s_{N}'}M_{p,p'}^{N-1}.
\end{eqnarray}
Hence, the matrix elements $M_{r,r'}^{N}$ can be evaluated recursively from a knowledge
of the eigenstates of all finite size Hamiltonians up to $H_{N}$ starting from 
the initial matrix elements 
$_{-1}\langle r|f_{\sigma}|r'\rangle_{-1}$ of  the isolated impurity 
$H_{\rm imp}=\varepsilon_{f}\sum_{\sigma} f_{\sigma}^{\dagger}f_{\sigma}+Uf_{\uparrow}f_{\uparrow}f_{\downarrow}^{\dagger}f_{\downarrow}$. Similar considerations apply to other local dynamical quantities 
such as dynamical spin and charge susceptibilities. 
Figure \ref{fig:SakaiSD} shows
$T=0$ spectral densities for single-particle, magnetic and charge excitations 
calculated using the above procedure. These NRG calculations have been shown to
satisfy exact Fermi liquid relations, such as the Friedel sum rule for the 
single-particle spectral density and the Shiba relation for the magnetic excitation
spectrum, to within a few percent irrespective of the interaction strength 
$U/\pi\Delta$ in the Anderson model or the value of the exchange $J$ in the Kondo
model, see \textcite{Costi:1994b} and \textcite{Costi:1998} for a discussion.

\begin{figure}[!t]
\centerline{
  \includegraphics*[width=3.0in]{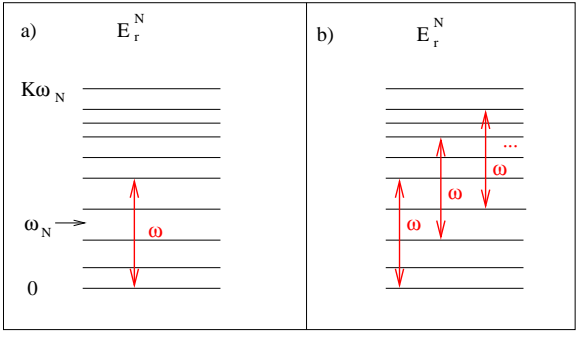}}
  \caption{Excitations of $H_{N}$ contributing to the spectral function at frequency 
$\omega$ for, (a), $T=0$, and, (b), $T>0$.
}
\label{fig:SDexcitations}
\end{figure}

The case of finite temperature dynamics is more complicated. Contributions
to the spectral density at frequency $\omega\sim \omega_N$ now arise from excitations
between arbitrary excited states, i.e. $\omega=E_{r}-E_{0}=E_{r'}-E_{p'}=E_{r''}-E_{p''}=\dots$ with $E_{0}=0<E_{p'}<E_{p''}<\dots$ see Fig.~\ref{fig:SDexcitations}. 
Consequently, the finite-T spectral 
density at $\omega \sim \omega_{N}$ will have contributions from all 
energy shells $n=1,\dots,N$. These need to be summed up, as in the calculation of
transient quantities described in Sec.~\ref{subsubsec:x-ray+transients}. 
It is clear, however, that in the
case of equilibrium spectral densities, the contributions from shells $n<N$ 
will be suppressed by Boltzmann factors. This motivates the 
following approximation: at
$\omega = 2\omega_{n} > k_{\rm B}T$ one can calculate $A_{\sigma}(\omega,T)$ 
as in the $T=0$ case
\begin{eqnarray}
&&A_{\sigma}(\omega_{n},T) \approx A_{\sigma}^{n}(\omega_n,T)\nonumber\\
&&=\frac{1}{Z_{n}(T)}\sum_{r,r'}
|M_{r,r'}^{n}|^{2}
(e^{-E_{r}^{n}/k_{\rm B}T} + e^{-E_{r'}^{n}/k_{\rm B}T})\nonumber\\
&&\times \delta(\omega_{n}-(E_{r'}^{n}-E_{r}^{n})).\label{eq:finite-T-approx1}
\end{eqnarray} 
In the other limit, $\omega=2\omega_{n}\leq T$, there is no completely satisfactory 
procedure. One approach assumes that the main contribution to the 
spectral density for $\omega=2\omega_{n}\leq T$
comes from the energy window containing thermal excitations 
${\cal O}(k_{\rm B}T)$ \cite{Costi:1992b,Costi:1994b}. In this case, 
the relevant shell, $M$, is determined by temperature
via $\omega_{M}\approx \bar{\beta}k_{\rm B}T$, as in the evaluation of 
thermodynamic properties in the previous section, 
so that, for $\omega = 2\omega_{n}\leq k_{\rm B}T$, we use 
\begin{eqnarray}
&&A_{\sigma}(\omega_{n},T) \approx A_{\sigma}^{M}(\omega_n,T)\nonumber\\
&&=\frac{1}{Z_{M}(T)}\sum_{r,r'}
|M_{r,r'}^{M}|^{2}
(e^{-E_{r}^{M}/k_{\rm B}T} + e^{-E_{r'}^{M}/k_{\rm B}T})\nonumber\\
&&\times \delta(\omega_{n}-(E_{r'}^{M}-E_{r}^{M})).\label{eq:finite-T-approx2}
\end{eqnarray} 
In practice, this procedure gives a smooth crossover as $\omega$ is lowered
below $k_{\rm B}T$ for temperatures comparable to the Kondo scale and higher, but
becomes less reliable at $\omega < k_{\rm B}T\ll T_{\rm K}$.

\begin{figure}[!t]
\centerline{
  \includegraphics*[width=3.0in]{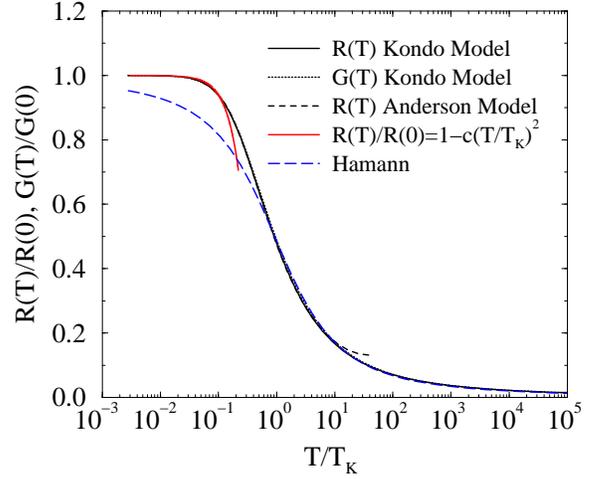}}
  \caption{Scaled resistivity and conductance of the $S=1/2$ Kondo model. Adapted from 
\textcite{Costi:2000}. For comparison
the resistivity of the symmetric Anderson model for $U/\pi\Delta=4$ 
is also shown \cite{Costi:1993} and is seen to be identical to that for
the Kondo model, up to non-universal corrections
arising from charge fluctuations at higher temperatures (for $U/\pi\Delta=4$ these
corrections occur for $T>10T_{\rm K}$). 
}
\label{fig:ResCond}
\end{figure}

Once the finite-T spectral density is known, one can also calculate transport
properties, since the transport time, $\tau_{\rm tr}$, for electrons scattering 
from a small concentration, $n_{i}$, of magnetic impurities is given in terms
of the spectral density by
\begin{equation}
\frac{1}{\tau_{\rm tr}(\omega,T)}=\frac{2n_{i}}{N_{F}} \Delta A_{\sigma}(\omega,T),
\end{equation}
where $N_{F}$ is the conduction electron density of states and $\Delta$ is 
the hybridization strength. For example, the resistivity $R(T)$ due to Kondo 
impurities in a clean metal is given by
\begin{equation}
R(T) = \frac{1}
{e^{2}\int d\omega \left(-\frac{\partial f}{\partial \omega}\right)
 \tau_{\rm tr}(\omega,T)} \ ,
\end{equation}
and the conductance through a quantum dot (or the resistivity of Kondo impurities
in a dirty metal) modeled by an Anderson impurity model is given by
\begin{equation}
G(T)/G(0) = \sum_{\sigma}\int d\omega 
\left(-\frac{\partial f}{\partial \omega}\right) A_{\sigma}(\omega,T) \ .
\label{eq:conductance-formula}
\end{equation}
Figure \ref{fig:ResCond} compares the scaled resistivity $R(T)/R(0)$ for the Kondo and
Anderson models with the scaled conductance $G(T)/G(0)$ for the Kondo model. 
The conductance and resistivity are seen to be almost
identical universal functions of $T/T_{\rm K}$. At finite magnetic field, the two quantities
deviate from each other in the region $T\approx B$ \cite{Costi:2000}. 
The NRG results can be compared to analytic results at low and 
high temperature. The resistivity of the Anderson model in the low-temperature Fermi 
liquid regime is given by the result of \textcite{Nozieres:1974}, 
\begin{equation}
R(T)/R(0) = G(T)/G(0) = 1 - c \left( \frac{T}{T_{\rm K}}\right)^{2}, T\ll T_{\rm K},
\label{eq:R-R0}
\end{equation}
where $c=\pi^{4}/16=6.088$ and $T_{\rm K}$ is the low-temperature Kondo scale 
defined from the static spin susceptibility via 
\begin{equation}
\chi(T=0)=(g\mu_{\rm B})^2/4k_{\rm B}T_{\rm K}.\label{eq:TK-chi}
\end{equation}
At high temperatures, $T\!>\!T_{\rm K}$, Hamann used the Nagaoka-Suhl approximation
\cite{Hewson:1993} to obtain for the resistivity of the Kondo model
\begin{equation}
R(T)/R(0) = \frac{1}{2}(1 - \frac{\ln(T/T_{\rm KH})}{(\ln(T/T_{\rm KH})^{2}+\pi^{2}S(S+1))^{1/2}}),
\end{equation}
where $S$ is the impurity spin and $T_{\rm KH}$ is a Kondo scale defined by 
\begin{equation}
R(T=T_{\rm KH}) = R(0)/2.
\end{equation}
\textcite{Micklitz:2006a} found
numerically that $T_{\rm KH}\approx 0.91T_{\rm K}$.  
We see from Fig.~\ref{fig:ResCond}
that the NRG result for the resistivity of the Kondo model agrees with the Hamann result
for $T \ge T_{\rm K}$. The $T^{2}$ Fermi liquid behavior at low-temperature
$T\ll T_{\rm K}$ is also recovered. In contrast, the Hamann result violates 
Fermi liquid behavior and cannot be trusted for $T < T_{\rm K}$. 
A numerical determination of 
the coefficient $c$ in Eq.~(\ref{eq:R-R0})
requires obtaining $\tau_{\rm tr}(\omega,T)$ accurately up 
to second order in both $\omega$ and $T$ \cite{Costi:1994b}.  
Typical errors for $c$ can be as large as 10-30\% so there is 
room for further improvement of the finite $T$ dynamics in the Fermi liquid regime 
$T\ll T_{\rm K}$. For a discussion of other transport properties of Kondo systems, 
such as thermopower and thermal conductivity, the reader is referred to 
\textcite{Costi:1993,Costi:1994b,Zlatic:1993}.

\subsubsection{Self-energy and reduced density matrix approach}
\label{subsubsec:se+dm}
We now describe two improvements to the calculation of dynamical quantities. 
The first of these, a direct calculation of the correlation
part of the self-energy of the Anderson impurity model \cite{Bulla:1998}, 
is particularly important for applications to DMFT, where the impurity 
self-energy plays a central role. The
second, the introduction of the reduced density matrix into the calculation
of dynamics,  is important, for example, in correcting large finite-size errors 
in spin-resolved spectra of the Anderson and Kondo models when a 
magnetic field perturbs the ground state \cite{Hofstetter:2000b}.

The correlation part of the self-energy for the Anderson impurity model, 
$\Sigma_{\sigma}$, is
defined via
\begin{equation}
G_{\sigma}(\omega,T) = \frac{1}{\omega-\varepsilon_{f}
+i\Delta - \Sigma_{\sigma}(\omega,T)},\label{eq:siam-gf}
\end{equation}
and can be expressed, via the equation of motion for 
$G_{\sigma}$ \cite{Bulla:1998}, as the ratio of a two-particle and a one-particle
Green function
\begin{equation}
\Sigma_{\sigma}(\omega,T) = U \langle\langle f_{\sigma}f_{-\sigma}^{\dagger}f_{-\sigma}
 ; f_{\sigma}^{\dagger}\rangle\rangle/\langle\langle
 f_{\sigma};f_{\sigma}^{\dagger}\rangle\rangle \ .
\label{eq:self-energy-expression}
\end{equation}
Evaluating the spectral densities of the two Green functions in 
(\ref{eq:self-energy-expression}) as in the previous section, 
and calculating from these, 
via a Kramers-Kronig transformation, the corresponding real parts of the Green functions
one obtains the self-energy. Using this in (\ref{eq:siam-gf}) one is able to
obtain the impurity spectral density with improved resolution of high-energy peaks, since
in this procedure, the single-particle broadening $\Delta$ is included exactly. 
In particular, this scheme recovers the limit $U\rightarrow 0$ exactly. 
It is also found to improve the spectral sum rule with typical errors as low
as 0.1\% or less. 
 
\begin{figure}[!t]
\centerline{
  \includegraphics*[width=3.0in]{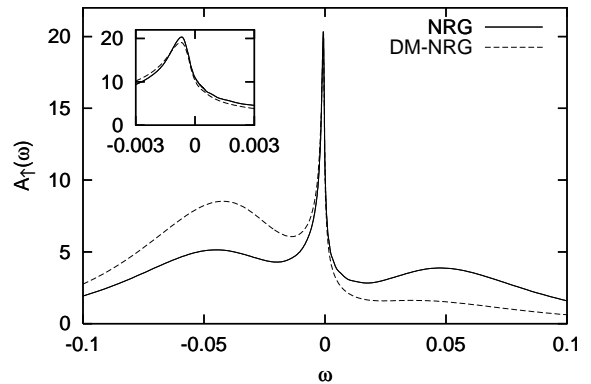}}
  \caption{Comparison of the spectral function for the Anderson impurity
model calculated in a magnetic field with and without the reduced
density matrix: $\Delta=0.01, U=0.1, \varepsilon_{f}=-0.05, B=0.001$ from
\textcite{Hofstetter:2000b}. 
}
\label{fig:SDHofstetter}
\end{figure}

The evaluation of spectral densities described in the previous section
is subject to systematic errors due to neglect of high-energy states 
in constructing $H_{N}$. These are the same as for thermodynamic 
properties, and they can be controlled by increasing $\Lambda$ and 
the number of retained states $N_{s}$.
Another source of error, specific to the method used to calculate dynamics, 
is that while we chose the frequency $\omega$ in evaluating spectra from
$H_{N}$ carefully so that $\omega>\omega_{N}$, nevertheless 
the eigenstates in the range $0\leq E_{r}^{N} \leq \omega_{N}$, which for
small $N$ are only crude approximations to the eigenstates of $H$, are
also used in the evaluation. They enter the calculation directly, as can be
seen from Fig.~\ref{fig:SDexcitations}, and, also, via the density matrices (e.g.
at $T=0$ via $\varrho_{N}=|0\rangle_{N}{_N}\langle 0|$) which are
used to arrive at (\ref{eq:lehmann-zero-cluster}) and
(\ref{eq:finite-T-approx1}-\ref{eq:finite-T-approx2}). 
As a result, the spectral density is subject to errors
for small $N$, i.e. for high energies, due to the use of low-lying 
states which are not converged \cite{Hofstetter:2000b}. With increasing $N$, i.e. 
lower energies, this error will decrease. An improvement, due to 
\textcite{Hofstetter:2000b}, is to use in place of $\varrho_{N}$ 
the reduced density matrix, $\varrho^{N,\rm red}$, 
of $H_{N}$, obtained from the density matrix of the largest finite size Hamiltonian 
diagonalized, $H_{N_{\rm max}}$, i.e.,
\begin{equation}
\varrho^{N,\rm red}=Tr^{}_{s^{}_{N+1},\dots,s^{}_{N_{\rm max}}}[ \varrho^{}_{N_{\rm max}}]
\end{equation}
where $s^{}_{N+1},\dots,s^{}_{N_{\rm max}}$ are the extra degrees of freedom contained
in $H_{N_{\rm max}}$ but absent in $H_{N}$. As
 $\varrho^{N,\rm red}$ is not diagonal in the eigenbasis of $H_{N}$, the resulting
spectral function takes on a more complicated form
\begin{eqnarray}
A_{\sigma}^{N}(\omega, 0) &=& \sum_{r',r}C_{r',r}
M_{r,r'}^{N}\delta(\omega-(E_{r'}^{N}-E_{r}^{N}))\\
C_{r',r} &=& \sum_{p}\rho_{p,r}^{N,\rm red}M_{r',p}^{N}
+\sum_{p}\rho_{r',p}^{N,\rm red}M_{p,r}^{N}.\label{eq:lehmann-hofstetter}
\end{eqnarray}
The reduced density matrices $\varrho^{n,\rm red}$ are calculated iteratively backwards
starting from the density matrix of $H_{N_{\rm max}}$. One situation where the reduced
density matrix is important is in obtaining correctly the spin-resolved spectral
density of the Anderson model in a magnetic field \cite{Hofstetter:2000b}. A magnetic
field comparable to $T_{\rm K}$ changes the magnetization and therefore the occupation
of up/down states by ${\cal O}(1)$ so large shifts in spectral weight occur at
high energies in the impurity spectral density which are only captured correctly 
by using the reduced density matrix, see Fig.~\ref{fig:SDHofstetter}. Results for
dynamical susceptibilities in a magnetic field using the reduced density matrix have
been discussed by \textcite{Hewson:2006a}. The reduced density matrix
also eliminates to a large extent the difference between the spectra calculated for
even/odd $N$ and allows a correct description of the asymptotics of the Kondo resonance
in high magnetic fields \cite{Rosch:2003}.

\subsubsection{The x-ray problem and transient dynamics}
\label{subsubsec:x-ray+transients}
We consider here the calculation of the response of a system to
a sudden local perturbation, such as the excitation of an electron from a 
core-level to the conduction band of a metal in  the x-ray problem, or, 
the time-dependent response of a spin in the spin-boson model 
following an initial-state preparation. The NRG approach to the
x-ray problem \cite{Oliveira:1981} was the first application of the method
to dynamical quantities. 
In common with previous treatments of the x-ray
problem, for a review see \cite{Mahan:1975}, the approach developed 
by \textcite{Oliveira:1981} calculates the absorption spectrum within 
linear response theory, and, therefore, belongs logically to 
Sec.~\ref{subsubsec:dyn+trans}.
We include it here for two reasons, (i), because the response of the 
electrons to the appearance of the core-hole potential is a problem 
of transient dynamics, and, (ii), because it uses the idea of 
formulating the calculation of the absorption spectrum in terms 
of initial and final state 
Hamiltonians \cite{Nozieres:1969}, which is also inherent to the 
recent NRG approach to transient dynamics beyond linear response 
theory \cite{Costi:1997a,Anders:2005c}.

A simple model for describing the x-ray absorption spectra in metals, is
given by the following spinless Hamiltonian
\begin{equation}
H = \sum_{k}\varepsilon_{k}c_{k}^{\dagger}c_{k} + E_{d}d^{\dagger}d +\sum_{k,k'}U_{dc}c_{k}^{\dagger}c_{k'}dd^{\dagger},\label{eq:x-ray-spinless}
\end{equation}
where $d^{\dagger}$ creates a core-electron with energy $E_{d}$ and the attractive 
screening interaction, $U_{dc}$, acts only when the core-level 
is empty ($dd^{\dagger}=1$). The core-level lifetime is assumed infinite, and
the interaction with the x-ray field is taken to be of the form 
\begin{equation}
H_{x}=w(c_{0}^{\dagger}de^{-i\omega t}+H.c.),
\end{equation}
where $c_{0}=\sum_{k}c_{k}$. 
The x-ray absorption spectrum, $\mu(\omega)$, is
obtained using linear response theory from the imaginary part
of the optical conductivity, 
$\chi_{cd}=\langle\langle c_{0}^{\dagger}d;d^{\dagger}c_{0} \rangle\rangle$.
At zero temperature, 
one finds for the absorption spectrum a power law singularity of the form 
\begin{equation}
\mu(\omega)\sim (\omega-E_{T})^{-\alpha},\label{eq:ND-absorption}
\end{equation}
where, $E_{T}$ is the absorption 
threshold, and, $\alpha$, is an exponent which 
depends on the strength of the core-hole potential.
The exponent $\alpha$ has two contributions $\alpha=\alpha'- \varepsilon$,
an excitonic part, $\alpha'$, due to \textcite{Mahan:1967} and an orthogonality part, 
$\varepsilon$, which follows from Anderson's orthogonality catastrophe theorem
\cite{Anderson:1967}. 
An exact solution of the 'x-ray problem' has been obtained 
by \textcite{Nozieres:1969} by using the decomposition of (\ref{eq:x-ray-spinless}) into
single-particle initial-state, $H_{I}$, and final-state, $H_{F}$, 
Hamiltonians, corresponding to the situations before 
($dd^{\dagger}=0$) and after ($dd^{\dagger}=1$) a 
core-electron is excited to the conduction band:
\begin{eqnarray}
H_{I} &=& \sum_{k}\varepsilon_{k}c_{k}^{\dagger}c_{k} + E_{d},
\label{eq:initial-state}\\
H_{F} &=&\sum_{k}\varepsilon_{k}c_{k}^{\dagger}c_{k} +
\sum_{k,k'}U_{dc}c_{k}^{\dagger}c_{k'}.\label{eq:final-state}
\end{eqnarray}
For the spinless model, (\ref{eq:x-ray-spinless}),
\textcite{Nozieres:1969} found the exponents
\begin{equation}
\alpha'=2\delta/\pi,\;\;  \varepsilon=(\delta/\pi)^{2}, 
\end{equation}
where the phase-shift $\delta=\arctan(-\pi N_{F}U_{dc})$ is that for
conduction electrons scattering from the additional potential created by the core-hole,
and, $N_F$ is the conduction electron density of states at the Fermi level.
In addition to the absorption spectrum, the core-level photoemission spectrum, 
$A_{d}(\omega)=-Im[\langle\langle d; d^{\dagger}\rangle\rangle]/\pi$, 
is also of interest. In the core-level photoemission 
spectrum, only the orthogonality effect is operative and the core-electron
spectral function, which has the quasiparticle form 
$A_{d}(\omega) = \delta(\omega-E_{d})$
in the absence of screening, is replaced by an incoherent spectrum of the
form 
\begin{equation}
A_{d}(\omega)\sim \theta(\tilde{E_{d}}-\omega)(\tilde{E_{d}}-\omega)^{-(1-\varepsilon)}\ ,
\end{equation} 
in the presence of screening \cite{Doniach:1970}.

\begin{figure}[!t]
\centerline{
  \includegraphics*[width=3.0in]{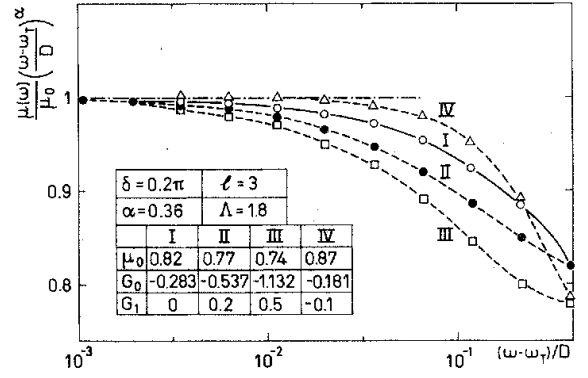}}
  \caption{Absorption spectrum normalized by the Nozi\'eres-De~Domenicis result 
(Eq.~(\ref{eq:ND-absorption})) for several screening
potentials of the form $U_{dc}(k,k')=G_{0}+G_{1}(k+k')$ \cite{Oliveira:1981}.
}
\label{fig:Oliveira81}
\end{figure}

\textcite{Oliveira:1981} applied the NRG to the initial-state and 
final-state Hamiltonians (\ref{eq:initial-state}-\ref{eq:final-state}) 
and calculated the zero temperature linear response absorption spectrum,
\begin{eqnarray}
&&\mu(\omega) = 2\pi w^{2}\sum_{m_{F}}|\langle m_{F}|c_{0}^{\dagger}d|m_{I,GS}\rangle|^{2}\nonumber\\
&&\times\delta(\omega - (E_{m_{F}}-E_{m_{I,GS}})),\label{eq:absorption-oliveira}
\end{eqnarray}
with $H_{I,F}|m_{I,F}\rangle = E_{m_{I,F}}|m_{I,F}\rangle$ and $|m_{I,GS}\rangle$ the 
ground state of $H_{I}$. 
In evaluating $\mu(\omega)$,  truncated Hamiltonians, $H_{I,F}^{N}$, were used 
and the spectrum was evaluated at an appropriate frequency $\omega=\omega_{N}$
as in Sec.~\ref{subsubsec:dyn+trans} 
but with a box broadening function on a logarithmic scale.
They were able to recover the exact threshold exponent 
of \textcite{Nozieres:1969} (see \textcite{Libero:1990a} for similar
calculations of the photoemission spectra). 
They also extended the calculation of absorption spectra to 
core-hole potentials of finite range $U_{dc}\rightarrow U_{dc}(k,k')$, 
finding (see Fig.~\ref{fig:Oliveira81}) that the threshold exponent remains universal, 
i.e. it depends only on the phase shift $\delta$ at the Fermi level, but 
that the asymptotic scale for the onset of power-law behavior 
depends on both the range \cite{Oliveira:1981} and the strength of the core-hole
potential \cite{Cox:1985}. It also depends on any energy dependence in the
density of states \cite{Chen:1995b}. These results reflect the fact that 
the crossover scale to the low energy fixed point, determining universal properties,
depends on details of the density of states and the core-hole potential. 

X-ray singularities also
play an important role in the dynamics of auxiliary particles in the slave-boson
approach to the infinite $U$ Anderson impurity model 
\cite{Coleman:1984,MH:1984}. A NRG calculation of the $T=0$ photoemission spectra
for slave-bosons, $A_{b}(\omega)$, and pseudo-fermions, $A_{f\sigma}(\omega)$, 
showed that these diverge with the exponents given
above for the photoemission and absorption spectra respectively, generalized to
include spin, \cite{Costi:1994a,Costi:1996a}:
\begin{eqnarray}
A_{f\sigma}(\omega) &\sim& (\omega-E_{T})^{-\alpha_{f}},\\
A_{b}(\omega) &\sim& (\omega-E_{T})^{-\alpha_{b}},\\
\alpha_{f} & = & 2\frac{\delta_{\sigma}}{\pi} -
\sum_{\sigma}(\frac{\delta_{\sigma}}{\pi})^{2},\label{eq:fermion-exp}\\
\alpha_{b} & = & 
1 - \sum_{\sigma}(\frac{\delta_{\sigma}}{\pi})^{2},\label{eq:boson-exp} 
\end{eqnarray}
with the phase shift $\delta_{\sigma}=\pi n_{f\sigma}$ and $n_{f\sigma}$ the
occupancy per spin of the local level. Here, also, the 
scale for the onset of power-law behavior is determined by the relevant
low energy crossover scale, e.g., the Kondo scale in the Kondo regime. 

We turn now to a problem which is formally similar to the x-ray problem, 
namely the dynamics of a spin subject to an initial state preparation,
as in the case, for example, of the dynamics of the spin in the 
spin-boson model Eq.~(\ref{eq:sbm}).
Further discussion of the effects of screening on the spectra of impurity models
is given in Sec.~\ref{subsubsec:screening}. 
In the spin-boson model one is interested in the dynamics
of a spin, $\sigma_{z}$, described by $P(t)=\langle \sigma_{z}(t)\rangle_{\varrho_{I}}$, 
following an initial-state preparation of the system described by an 
initial density matrix $\varrho_{I}$ \cite{Leggett:1987}. 
For example, the spin $\sigma_{z}$ in (\ref{eq:sbm}) could
be prepared in state $|\uparrow\rangle$ at $t<0$ by an 
infinite bias $\varepsilon=\infty$ with the environment fully relaxed
about this state, and the bias could subsequently be switched off at $t=0$
allowing the spin to evolve. The time evolution of the spin for $t>0$ is then described
by 
\begin{equation}
\langle \sigma_{z}(t)\rangle_{\varrho_{I}} = 
\frac{1}{Tr \varrho_{I}}Tr[\varrho_{I}e^{-iH_{F}t/\hbar}\sigma_{z}e^{iH_{F}t/\hbar}] \ ,
\end{equation} 
where $\varrho_{I}=Tr[e^{-\beta H_{I}}]$ and the initial and final state Hamiltonians
are given by
\begin{eqnarray}
H_{I} &=& H_{\rm SB}(\varepsilon=\infty) \ ,\\
H_{F} &=& H_{\rm SB}(\varepsilon=0)\ ,
\end{eqnarray}
where $H_{\rm SB}$ is the spin-boson Hamiltonian  (\ref{eq:sbm}).

This approach has been investigated within NRG by \textcite{Costi:1997a} using,
for the Ohmic case, in place of $H_{\rm SB}$, the equivalent anisotropic Kondo model.
Despite the formal similarity with the x-ray problem, the exact formulation indicates
the difficulties that have to be overcome in calculating transient dynamical quantities
beyond linear response. Consider the spectral function
\begin{eqnarray}
P(\omega) & = & \frac{1}{Z_{I}}\sum_{m_{I},m_{F},m_{F}'}e^{-\beta E_{m_{I}}}
\langle m_{I}|m_{F} \rangle \langle m_{F}'|m_{I} \rangle\nonumber\\
& \times &
\langle m_{F}|\sigma_{z}|m_{F}'\rangle\delta(\omega-(E_{m_{F}}-E_{m_{F}'})),\label{eq:lehmann-p}
\end{eqnarray}
with
\begin{equation}
P(t)=\int_{0}^{+\infty}P(\omega)\cos(\omega t)d\omega .
\end{equation}
We see that even at $T=0$, no ground state enters the delta functions in 
(\ref{eq:lehmann-p}), in contrast to the linear-response expression 
(\ref{eq:absorption-oliveira}) for $\mu(\omega)$ in the x-ray problem. 
This implies that in evaluating $P(\omega)$ at a frequency $\omega\sim \omega_{N}$, 
contributions will arise from all energy shells 
$n=1,2,\dots,N$ as discussed previously for finite-$T$ dynamics. 
In the present situation, however, the contributions from higher
energy shells (i.e. $n<N$) are not suppressed by Boltzmann factors, 
so it is not clear a priori that
using a single shell approximation will give meaningful results. Such an approximation
shows that the short-time dynamics of the spin-boson model can be recovered 
and that in order to obtain the long-time dynamics one has to sum up contributions from
all shells \cite{Costi:1997a}. Adding up such 
contributions using the retained states 
of successive Hamiltonians $H_{I,F}^{N},H_{I,F}^{N+1},\dots$ 
is problematical due to the overlap of the spectra at low energies. 
An elegant solution of this problem, allowing multiple-shell NRG calculations 
to be carried out, has
recently been found by \textcite{Anders:2005c}. Their idea, was to recognize that
the set of states discarded, $|r\rangle_{m,\rm disc}$, at each NRG iteration 
$m$, supplemented with the degrees of freedom 
$|e;m\rangle=
|s_{m+1}\rangle\otimes\dots \otimes|s_{N_{\rm max}}\rangle$, for $m=1,\dots, N_{\rm max}$, with
$N_{\rm max}$ the largest Hamiltonian diagonalized, forms a complete basis set
\begin{equation}
\sum_{m=1}^{N_{\rm max}}\sum_{r\in \{\rm disc\}}|r,e;m\rangle\langle r,e;m| = 1 \ .
\label{eq:complete-basis-set}
\end{equation}
Using this identity, their result for $P(t)$ in the basis of final states is
\begin{eqnarray} 
&&P(t) = \sum_{m=1}^{N}\sum_{m_{F},m_{F}'}^{\rm trun}e^{i(E_{m_{F}}^{m}-E_{m_{F}'})t/\hbar}\nonumber\\
&&\times\langle m_{F}|\sigma_{z}|m_{F}' \rangle\varrho_{m_{F},m_{F}'}^{m,\rm red} \ ,
\end{eqnarray}
where $\varrho_{r_{F},s_{F}}^{m,\rm red}$ are the matrix elements of the 
reduced density matrix $\varrho_{I}$ for $H_{I}^{m}$ introduced in 
the previous section and the
sum $\sum_{m_{F},m_{F}'}^{\rm trun}$ implies that at least one of the states $m_{F},m_{F}'$
is in the discarded sector for iteration $m$. 
Rotating $\varrho_{r_{F},s_{F}}^{m,\rm red}$ to the initial state 
basis gives overlap matrix elements 
$\langle m_{I}|m_{F} \rangle$, $\langle m_{I}|m_{F}' \rangle$ as 
in (\ref{eq:lehmann-p}) above. Within this approach the time-dependent transient
dynamics of a number of models has been investigated, including the Anderson and
resonant level models \cite{Anders:2005c}, the Kondo model \cite{Anders:2006b}
and the sub-ohmic spin-boson model \cite{Anders:2006c}. 

The use of a complete basis set (\ref{eq:complete-basis-set}) has potential
to improve also the finite-T calculation of spectral densities,
particularly in the problematical range $\omega<T$ \cite{Weichselbaum:2006}. 
A further improvement in using a complete basis set is that the sum rule 
for spectral densities $\int_{-\infty}^{\infty}d\omega A_{\sigma}(\omega,T)=1$ 
is, by construction, fulfilled exactly \cite{Peters:2006b,Weichselbaum:2006}.  
%
%
%
%
%
%
%
%

\section{Application to Impurity Models}
\label{sec:imp}

In this section we review the application of the NRG to 
a range of quantum impurity models. Section \ref{subsec:kondo} reviews work on models with
conduction electron screening (Sec.~\ref{subsubsec:screening}), underscreened
and fully screened Kondo models (Sec.~\ref{subsubsec:bulk}), 
and models exhibiting the Kondo effect in nanostructured devices 
(Sec.~\ref{subsubsec:nano}). 
Section \ref{subsec:two-channel} deals with the prototypical
overscreened Kondo model, the two-channel
Kondo model, which is often encountered as an effective model describing the quantum 
criticial point of more complex quantum impurity models, 
e.g.~certain double quantum dot systems \cite{Zarand:2006} or two-impurity systems 
\cite{Jones:1987}. Impurity quantum phase transitions are reviewed in 
Sec.~\ref{subsec:qpt} in the contexts of multi-impurity systems 
(\ref{subsubsec:mult-imp}), soft-gap systems (\ref{subsubsec:loc-crit}) 
and in the context of magnetic impurities in superconductors (\ref{subsubsec:sc}).
Sec.~\ref{subsec:orbital} reviews work on multi-orbital systems, including the
effects of crystal-field splittings in Anderson impurity models. 
Finally, models with bosonic degrees of freedom are reviewed in 
Sec.~\ref{subsec:bosons}. Note also, that a number of models of nanostructured
devices, for example, the single-electron box, quantum dots with phonons, 
or multi-orbital quantum dots, are to be found in  
Sec.~\ref{subsec:two-channel}-\ref{subsec:bosons}.

  \subsection{Kondo effect and related phenomena}
  \label{subsec:kondo}

\subsubsection{Screening and photoemission}
\label{subsubsec:screening}
Screening effects are important whenever an electron is excited 
from a localized core- or 
valence-state into the conduction band or removed completely, leaving 
behind a hole which attracts the conduction electrons. 
Such effects can have a drastic influence on the photoemission and absorption
spectra of impurity systems. In this section we consider a number of extensions to
the basic screening model (\ref{eq:x-ray-spinless}) introduced in 
Sec.~\ref{subsubsec:x-ray+transients},
\begin{equation}
H = \sum_{k}\varepsilon_{k}c_{k}^{\dagger}c_{k} + E_{d}d^{\dagger}d +\sum_{k,k'}U_{dc}c_{k}^{\dagger}c_{k'}dd^{\dagger}.\label{eq:x-ray-spinless2}
\end{equation}
We consider first the generalization of (\ref{eq:x-ray-spinless2})
to a simplified model of an atom adsorbed on a metallic surface 
\cite{Oliveira:1985}. 
In addition to a core-level, as in (\ref{eq:x-ray-spinless2}), the atom has a resonant
level (created by $b^{\dagger}$) whose position, $E_{bI}$ or $E_{bF}$, depends on the 
occupancy of the core-level according to the
term 
\begin{equation}
H_{db} = E_{bI}b^{\dagger}bd^{\dagger}d+E_{bF}b^{\dagger}b(1-d^{\dagger}d),
\end{equation} 
and the level hybridizes with the conduction band via a term 
\begin{equation}
H_{\rm mix}=V\sum_{k}(c_{k}^{\dagger}b+H.c.).
\end{equation}
Note that $H_{db}$ just represents a screening of the core-hole by electrons
in the resonant level.
Excitation of an electron from the core-level by an x-ray can proceed either
directly or via the resonant level, leading to a Fano anti-resonance in the
x-ray absorption at finite energy (in addition to the usual edge singularity at
$\omega=E_{T}$). This Fano anti-resonance, present also without the core-hole potential,
is found to be significantly narrowed and shifted in the presence of the 
core-hole potential \cite{Oliveira:1985}. 
It would be of interest to investigate also the core-level
photoemission spectrum of this model using the NRG, as both this, and the absorption
spectrum, are accessible in experiments. 
\textcite{Brito:1990} have carried out such a calculation for
an appropriately generalized spinfull version of the above model, 
i.e., the Anderson model (\ref{eq:siam}) in the presence of both an interaction, 
\begin{equation}
H_{dc}=U_{dc}dd^{\dagger}\sum_{kk'\sigma}c_{k\sigma}^{\dagger}c_{k'\sigma},
\end{equation} 
between the core-hole and the conduction electrons, and, an interaction, 
\begin{equation}
H_{df}=U_{df}dd^{\dagger}\sum_{\sigma}f_{\sigma}^{\dagger}f_{\sigma},
\end{equation} 
between the core-hole and the valence level. Signatures of the 
valence states could be identified
in the XPS spectra and their dependence on $U_{dc}$ was investigated in the 
mixed valence and empty orbital regimes. A reduction of the hybridization 
between the valence level and the conduction band, arising from 
orthogonality effects, was found with increasing $U_{dc}$.
The corresponding calculation in the Kondo regime is still lacking.
The x-ray absorption spectrum of the same model has been investigated in 
\cite{Helmes:2005} in the context of excitons in Kondo correlated quantum dots and
the expected absorption exponent from the Nozi\`eres -- De Domenicis theory was
recovered. 

In the models discussed so far, the core-level was assumed to have infinite lifetime.  
Consequently, the screening interaction gave rise to true singularities in 
the core-level absorption and photoemission spectra. These singularities are cut off 
as soon as the core-level lifetime is finite. Another situation where the
singularities due to screening are cut off, but where screening effects may nevertheless
be important, is in the valence band photoemission spectra of heavy fermions 
within a local impurity approach, which we now address. 

It is often assumed that the effects of conduction electron screening on the 
$f$ electron photoemission spectra of heavy fermions can be taken into account 
by renormalizing the bare parameters of an effective Anderson impurity model. 
However, this is not a priori clear as the screening interaction in these 
systems can be an appreciable fraction of the local Coulomb repulsion. One 
of the merits of the NRG, which allows
such questions to be investigated, is that it can deal with all
local Coulomb interactions on an equal footing and in a non-perturbative manner,
and some examples of this have already been given above. 
For the particle-hole symmetric Anderson model, (\ref{eq:siam}), it was shown in 
\textcite{Costi:1991,Costi:1992a} that the effect of a screening term of the form 
\begin{equation}
H_{fc}=U_{fc}\sum_{kk'\sigma\sigma'}f_{\sigma}^{\dagger}f_{\sigma}c_{k\sigma'}^{\dagger}c_{k'\sigma'}\ ,
\end{equation}
on the valence band photoemission spectrum could be well accounted for
by a renormalization of the bare parameters of the Anderson model, both the local
level position, $\varepsilon=-U/2$,  and the hybridization. An excitonic like
enhancement of the hybridization was found with increasing $U_{fc}$. Similar
effects are reflected in the STM conductance of a magnetic 
adatom modeled by the screened Anderson model \cite{Cornaglia:2003b}. Calculations 
for the infinite $U$ Anderson impurity model, for thermodynamics
\cite{Alascio:1986,Zhuravlev:2005}, and, dynamics \cite{Takayama:1993},
are consistent with the above findings.

The above model for screening in heavy fermions assumes that the largest 
contribution to screening of $f$ holes arises from conduction electrons in 
states that hybridize with the $f$ states. These hybridizing states are 
usually the $p$ levels from neighboring 
ligand ions, so the screening from these (denoted $U_{fc}$ above) should be expected
to be smaller than the onsite screening, $U_{df}$, from the $d$ electrons of 
the rare-earth ion. By symmetry, the latter do not hybridize with the $f$ states.
Neglecting $U_{fc}$ and representing the $d$ electrons by a spinless $s$ wave band,
we may represent the screening of $f$ holes by $d$ electrons by 
adding to the Anderson model (\ref{eq:siam}) the term 
\begin{equation}
H_{\text{scr}}=\sum_{k}\varepsilon_{k}d_{k}^{\dagger}d_{k} +
U_{df}\sum_{kk'}(n_{f}-1)
d^{\dagger}_{k}d_{k'},
\end{equation}
where $n_{f}=\sum_{\sigma}f_{\sigma}^{\dagger}f_{\sigma}$. 
The resulting model is a two-channel Anderson
model, in which one channel screens, but, does not hybridize, and, the other channel
hybridizes, but, does not screen. Assuming localized $d$ electrons gives the model
studied by \textcite{Brito:1990} and discussed above. For the full model,
\textcite{Takayama:1997,Takayama:1993} calculated the valence band 
photoemission spectrum and, surprisingly, found that the effect of 
$U_{dc}$ in the Kondo regime could be absorbed into a renormalization of 
the Anderson model parameters. This result was for infinite $U$, but it 
should remain valid in the Kondo regime for any finite $U$ provided 
$U_{dc}$ remains smaller or comparable to $U$. In contrast to the model described
above, where the screening occurs in the hybridizing channel, the effect 
of the screening interaction in the present model is to reduce
the effective hybridization of the valence electrons to the conduction
band, which can be understood as an orthogonality effect.  Nevertheless,
we conclude from these NRG calculations that in the Kondo regime, and 
with realistic values of $U_{df}$ ($U_{fc}$), the valence band 
photoemission spectra of both the above screening models can be 
well accounted for by an Anderson model with renormalized parameters.

The two-channel screening model above, has also been studied for finite $U$ 
\cite{Perakis:1993,Perakis:1994}. At particle-hole symmetry, increasing 
$U_{dc}$ reduces both $U$ and the effective hybridization, resulting, for 
sufficiently large $U_{dc}$, in an effective attractive local 
Coulomb interaction and a charge Kondo effect. 
For still larger $U_{dc}$, a Kosterlitz-Thouless transition to a non-Fermi 
liquid state occurs \cite{Perakis:1993,Perakis:1994} with a collapse of 
the Kondo resonance in the valence band photoemission spectrum \cite{Costi:1997b}.

\subsubsection{Kondo effect in the bulk and underscreened models}
\label{subsubsec:bulk}
Real magnetic impurities in metals have both orbital and spin degrees of
freedom and the resulting low-energy effective impurity models can be very 
complicated \cite{Hewson:1993}. The
NRG has been applied so far to models with at most three orbitals, see 
Sec.~\ref{subsec:orbital}.
In cases where the ground state is an orbital singlet, e.g. for 
dilute Mn ions in metals, \textcite{Nozieres:1980} have given a useful classification
of the resulting effective single-impurity Kondo models in terms 
of the size of the impurity spin $S$ and the number of conduction 
channels, $n$, which couple to the spin via the Kondo exchange. These 
multi-channel Kondo models are described by
\begin{equation}\label{eq:NChannelKondo}
   H = \sum_{k\sigma\alpha} \varepsilon_k c^\dagger_{k\sigma\alpha}
     c_{k\sigma\alpha} + J \sum_\alpha {\bf S} \cdot {\bf s}_\alpha \ ,
\end{equation}
where $\alpha=1,\dots,n$ is the channel index, and the exchange constant, 
$J$, is antiferromagnetic.  For $n=2S$, complete screening of the impurity spin
takes place leading to a local Fermi liquid at low temperatures. The overscreened
case, $n>2S$, exhibits non-Fermi liquid behavior and is reviewed in 
Sec.~\ref{subsec:two-channel}. 
In this section, we deal with some recent developments on the fully screened $S=1/2$
Kondo model, relevant to bulk Kondo impurities, and also describe work on the
the single-channel underscreened case, $n=1<2S$.

One of the signatures of the Kondo effect is the appearance in the impurity
spectral density of the Kondo resonance at the Fermi level. Point contact
spectroscopy on Cu wires containing magnetic impurities, using the mechanically 
controllable brake junction technique, show a zero bias anomaly, which is attributed 
to the Kondo resonance \cite{Yanson:1995}. In addition, these experiments show
that the Kondo resonance splits in a magnetic field. NRG calculations for the 
$S=1/2$ Kondo model in a magnetic field do indeed show that the Kondo resonance 
splits in a magnetic field, $B$, provided the Zeeman splitting $g_{i}\mu_{\rm B}B$ 
exceeds the Kondo scale $T_{\rm K}$ \cite{Costi:2000}.  Here
$g_{i},\mu_{\rm B}$ are the impurity $g$-factor and Bohr magneton and $T_{\rm K}$ is the Kondo
scale defined from the half-width at half-maximum of the $T=0$ Kondo resonance. 
The latter is obtained from the imaginary part of the many-body $T$-matrix, 
$T_{kk'\sigma}$, for spin $\sigma$, defined by
\begin{equation}
G_{kk'\sigma}(\omega) = \delta_{kk'}G_{kk'\sigma}^{0} + G_{kk\sigma}^{0}T_{kk'\sigma}
G_{k'k'\sigma}^{0} \ ,
\label{eq:T-matrix-equation}
\end{equation}
where $G_{kk'\sigma}(\omega)=\langle\langle 
c_{k\sigma};c_{k'\sigma}^{\dagger}\rangle\rangle$ is
the full conduction electron Green function and  $G_{kk'}^{0}$ is the corresponding
unperturbed Green function. From the equations of motion for $G_{kk'}$ one finds
for the orbitally isotropic Kondo model
\begin{eqnarray}
T_{kk'\sigma}(\omega) &=& \sigma\frac{J}{2}\langle S_{z}\rangle + (\frac{J}{2})^{2}\langle\langle O_{\sigma};O_{\sigma}^{\dagger} \rangle\rangle,\label{eq:T-matrix}\\
O_{\sigma} &=& \sum_{k}\vec{S} \cdot \vec{\tau}_{\sigma\sigma'}c_{k\sigma'},
\end{eqnarray}
with $\vec{\tau}$ the Pauli matrices. From the $T$-matrix one can also extract
the transport time and thereby the magnetoresistivity. The latter is found to 
agree well with experimental data on diluted Ce impurities in LaAl$_{2}$ 
\cite{Costi:2000}.

\begin{figure}[!t]
\centerline{
  \includegraphics*[width=3.0in]{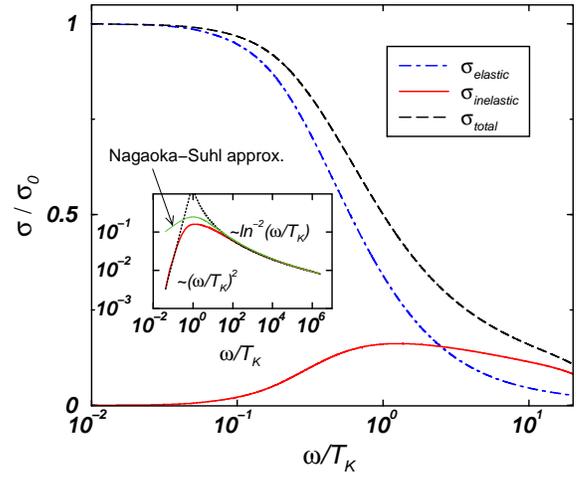}}
  \caption{Elastic, inelastic and total scattering rates for the $S=1/2$ 
fully screened Kondo model at $T=0$ (from \protect{\textcite{Zarand:2004}}). 
The Kondo scale $T_{\rm K}$ is that from the half-width at half-maximum of the Kondo
resonance and is approximately twice that from the $T=0$ susceptibility defined in 
(\ref{eq:TK-chi}).}
\label{fig:scattering-rates}
\end{figure}

A recent development has been the realization by \textcite{Zarand:2004} 
that one can use the NRG to extract from the many-body $T$-matrix both elastic and
inelastic scattering rates and cross sections. The total scattering
cross section, $\sigma_{\text{tot}}(\omega)$, is related to the imaginary 
part of the $T$-matrix by the optical theorem 
\begin{equation}
\sigma_{\text{tot}}(\omega=\varepsilon_{k}) = 
-\frac{2}{v_{k}}Im[T_{kk\sigma}(\omega)],
\end{equation}
with $v_{k}$ the velocity of electrons with wavevector $k$.
Consequently, by using the expression for the elastic scattering cross section
\begin{equation}
\sigma_{\text{el}}(\omega=\varepsilon_{k}) = \frac{2\pi}{v_{k}}\sum_{k'}
\delta(\varepsilon_{k'}-\varepsilon_{k})|T_{kk'\sigma}(\omega)|^{2} \ ,
\end{equation}
\textcite{Zarand:2004} were able to calculate the inelastic scattering 
cross section 
$\sigma_{\text{inel}} = \sigma_{\text{tot}} - \sigma_{\text{el}}$ and the
inelastic scattering time, $\tau_{\text{inel}}\sim \sigma_{\text{inel}}^{-1}$. 
In order to shed some light on the expression for $\sigma_{\text{inel}}$, consider the 
Anderson model for a flat band with density of states 
$N_{F}=\sum_{k'}\delta(\varepsilon_{k'}-\varepsilon_{k})$
and resonant level width $\Delta=\pi N_{F}V^{2}$. We have, $T_{kk'}=V^{2}G_{d}$, 
with $G_{d}=(\omega-\varepsilon_{d} +i\Delta -\Sigma(\omega))^{-1}$ and $\Sigma$
the correlation part of the self-energy. The inelastic scattering cross 
section for $\omega=\varepsilon_{k}$ reduces to \cite{Zarand:2004}
\begin{equation}
\sigma_{\text{inel}} = -\frac{2}{v_{k}}
\frac{V^{2}\Sigma''(\omega)}
{(\omega-\varepsilon_{d}-\Sigma'(\omega))^{2}+(\Delta-\Sigma''(\omega))^{2}},
\end{equation}
which shows that the inelastic scattering rate vanishes for electrons at the Fermi
level due to the Fermi liquid properties of the self-energy. \textcite{Zarand:2004}
evaluated $\sigma_{\text{inel}}$ for the $S=1/2$ Kondo model via the NRG using the 
T-matrix in (\ref{eq:T-matrix}) at $T=0$ and for both zero and finite magnetic 
fields. The maximum in the inelastic scattering rate occurs
close to the $\omega\approx T_{\rm K}$ (see Fig.~\ref{fig:scattering-rates}).

A quantity closely related to the inelastic scattering time, $\tau_{\text{inel}}\sim 
\sigma_{\text{inel}}^{-1}$, is the dephasing time,
$\tau_{\phi}$, for electrons scattering from magnetic impurities and measured
in weak-localization experiments on diffusive conductors 
\cite{Mohanty:1997,Schopfer:2003,Pierre:2003,Baeuerle:2005}. The two quantities
are, however, not identical. An exact expression for the dephasing rate of electrons
scattering from a dilute concentration of Kondo impurities in a 
weakly disordered metal has recently been derived \cite{Micklitz:2006a}. 
\begin{figure}[!t]
\centerline{
  \includegraphics*[width=3.0in]{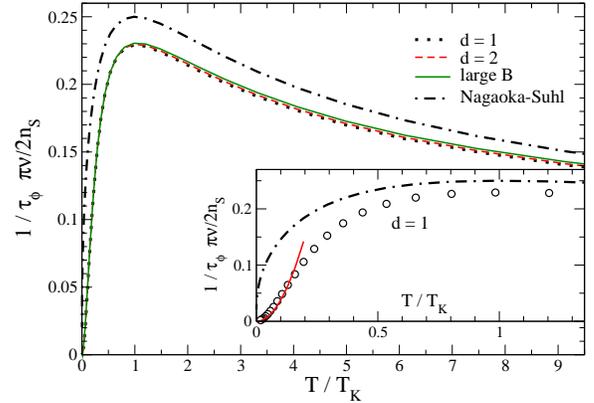}}
  \caption{Universal dephasing rate for the $S=1/2$ fully screened Kondo model
calculated via NRG for Kondo impurities in $d=1,2$ dimensional conductors 
(from \protect{\textcite{Micklitz:2006a}}). The solid line in the inset is
the analytic $T^{2}$ result from Fermi liquid theory valid for $T\ll T_{\rm K}$,
where $T_{\rm K}$ is the scale defined in (\ref{eq:TK-chi}).
}
\label{fig:dephasing}
\end{figure}
Fig.~\ref{fig:dephasing} 
shows the dephasing rate as a universal function of $T/T_{\rm K}$ for the
$S=1/2$ Kondo model, obtained by using the NRG for finite temperature
dynamics. The maximum dephasing rate occurs at 
$T\approx T_{\rm K}$ and decreases at first linearly with temperature below $T_{\rm K}$ and
eventually as $T^{2}$ in the Fermi liquid region $T\ll T_{\rm K}$. 
The magnetic field dependence of the dephasing time, $\tau_{\phi}(B,T)$, 
has also been calculated and the expression for the dephasing rate 
has been generalized to arbitrary dynamical scatterers \cite{Micklitz:2006b}. 
Recent experiments on Fe impurities in Ag wires show better than expected agreement
with the theoretical predictions for the dephasing rate of the $S=1/2$ Kondo model
\cite{Mallet:2006,Alzoubi:2006}. Fe impurities in Ag will have both an orbital moment and
a spin $S=2$ in the absence of crystal field and spin-orbit interactions. Inclusion of 
the latter, may, however, result in an effective $S=1/2$ single-channel Kondo model 
at the low temperatures $T\approx 1K$ of the experiments, thereby helping to explain
the good agreement with the $S=1/2$ theory. At the very lowest temperatures, 
$T<0.1T_{\rm K}$, a slower decay of the dephasing rate has been reported 
in these experiments, as compared to that expected from a fully screened model. 
One possible explanation for this is that a small fraction of Fe impurities 
is only partially screened. Underscreened Kondo models, to which we now 
turn, are known to give a much slower decay of the
dephasing rate below the Kondo scale, see below and  \textcite{Vavilov:2003}.

\textcite{Cragg:1979} investigated the single-channel $S=1$ underscreened Kondo model 
and showed that its low-energy fixed point corresponds to the spectrum of the
ferromagnetic $S'=1/2$ Kondo model. The deviations from the fixed point 
at iteration $N$ are of the form $\tilde{J}(N)\vec{S'}\cdot\vec{s}_{0}$ with 
$\tilde{J}(N)$ being ${\cal O}(-A/(N+C(J)))$ with $A$ being a constant and 
$C(J)$ depending on $J$, i.e. the deviations are marginally irrelevant. 
These calculations were extended by
\textcite{Koller:2005b} to models with $S=1,\dots,5/2$. They also determined
$C(J)$ explicitly for the different cases. 
Using the relation of $N$ to energy $\omega\sim\Lambda^{-N/2}$,
the effective coupling can be written as $\tilde{J}(\omega)\sim 1/\ln(\omega/T_{0})$ with
$T_{0}$ an appropriate Kondo scale \cite{Koller:2005b}. Consequently, there are 
logarithmic corrections to thermodynamic quantities at low temperature, instead of the
power-law corrections characteristic of fully screened Kondo models. Non-analytic
corrections are also found in dynamical quantities \cite{Mehta:2005,Koller:2005b}, so 
underscreened Kondo models have been termed ``singular Fermi-liquids'' 
\cite{Mehta:2005}. For, example, the spectral density, $\rho_{t}(\omega)$, obtained
from the T-matrix (\ref{eq:T-matrix}) takes a finite value at the Fermi level, 
but the approach to this value is non-analytic \cite{Koller:2005b}:
\begin{equation}
\rho_{t}(\omega) = \rho_{t}(0) - b/\ln(\omega/T_{0})^{2} \ ,
\end{equation}
Similarly, the $T=0$ inelastic cross section, also calculated by \textcite{Koller:2005b},
decays as $1/\ln(\omega/T_{0})^{2}$, 
at low energies, and consequently, the dephasing rate decays as 
$\tau_{\phi}^{-1}(T)\sim 1/\ln(T/T_{0})^{2}$. 
As mentioned above, a small fraction of underscreened Fe impurities may explain
the excess dephasing observed at the lowest temperatures in the experiments of 
\textcite{Mallet:2006} and \textcite{Alzoubi:2006}. Calculations for the 
temperature dependence of the resistivity and dephasing rates of the 
spin $S>1/2$ underscreened Kondo models and their relevance to Fe impurities
in Ag can be found in \cite{Mallet:2006}. 
It is also interesting to note, that calculations for ferromagnetic 
Kondo models \cite{Koller:2005b} show that all cross sections vanish at the
Fermi level with the inelastic part contributing nearly all the scattering
in this limit and with the elastic part being negligibly small.

Finally we mention recent work on calculating spatial correlations 
such as spin-density correlations, $C(x)=\langle\vec{S}\cdot\vec{s}_{x}\rangle$, 
around Kondo impurities, where $\vec{s_{x}}$ is the electron spin density
at distance $x$ from the impurity 
(for earlier work involving perturbative aspects combined
with NRG see \cite{Chen:1987,Chen:1992}). \textcite{Borda:2006b} 
works with Wannier states centered at both the impurity and at $x$, thereby
reducing the problem to a two-impurity type calculation 
(Sec.~\ref{subsubsec:mult-imp}). At $T=0$ and in one 
dimension, the decay of $C(x)$ is found to change from $1/x$ to $1/x^{2}$ 
around $x=\xi_{\rm K}=\hbar v_{F}/T_{\rm K}$, where the coherence length, $\xi_{\rm K}$,
describes the size of the Kondo screening cloud. At finite temperature,
the expected exponential decay of $C(x)$ for 
$x>\xi_{T}=\hbar v_{F}/k_{B}T$ is recovered.
\subsubsection{Kondo effect in nanostructures}
\label{subsubsec:nano}

\begin{figure}[!t]
\centerline{
  \includegraphics*[width=3.0in]{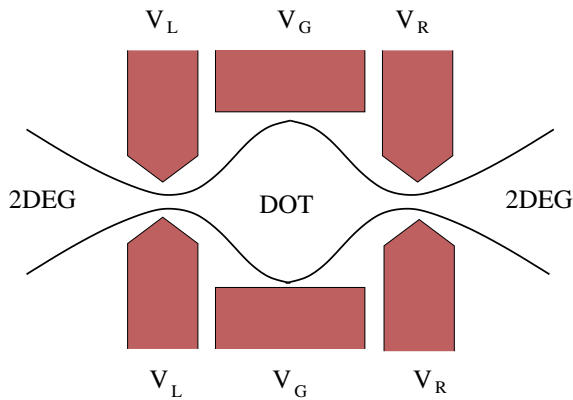}}
  \caption{Schematic top view of a lateral quantum dot, 
consisting of a confined region of typical size 10-100nm defined in the 
two-dimensional electron gas (2DEG) of a GaAs/AlGaAs heterostructure. The dot
is connected to left and right electron reservoirs. Gate voltages $V_{L,R}$
control the tunnel barriers into and out of the dot, while $V_{G}$ controls the
dot level positions.
}
\label{fig:settop}
\end{figure}

Recent experimental work has demonstrated the importance of the Kondo effect 
in determining the low-temperature transport properties of nanoscale size
devices such as quantum dots \cite{Goldhaber-Gordon:1998,Cronenwett:1998,Wiel:2000}. 
An example of such a device, a quantum dot, is shown in Fig.~\ref{fig:settop}.
More complicated devices, such as capacitively coupled double-dots or
dots contained in one or two arms of an Aharanov-Bohm interferometer can be
built up from this basic unit. A quantum dot consists of a confined region of electrons
coupled to leads via tunnel barriers. It may be viewed
as an artificial multi-electron atom, in which the different levels (filled, partially
filled or empty) couple to electron reservoirs via one or more 
channels.  A quantum dot can be described, in general, by the following 
multi-level Anderson impurity model
\begin{eqnarray}
  H &=& H_{\rm dot} + H_{\rm c} + H_{\rm tun} \ ,\label{eq:dot}\\
  H_{\rm dot} &=&   \sum_{i\sigma} \varepsilon_{\rm i\sigma} d^\dagger_{i\sigma}
                             d_{i\sigma}
                 + E_{C} \Big( \hat{N} - {\cal N}\Big)^{2} - J_{H}{\vec{S}}^{2} \ , \nonumber\\
H_{c}&=&\sum_{k\alpha\sigma}\varepsilon_{k\sigma}
c_{k\alpha\sigma}^{\dagger}c_{k\alpha\sigma} \ ,\nonumber\\
H_{\rm tun} &=&
       \sum_{k\alpha i\sigma} t_{\alpha i}
      \Big( d^\dagger_{i\sigma} c_{k\alpha \sigma}
                                    +   c^\dagger_{k\alpha\sigma} d_{i,\sigma} \Big) \ , \nonumber
\end{eqnarray}
where $\epsilon_{i\sigma},i=1,2,\dots$, are the dot level energies for spin $\sigma$
electrons,
${\cal N}=\langle \hat{N}\rangle=\sum_{i\sigma}\langle 
d^{\dagger}_{i\sigma}d_{i\sigma}\rangle$, is the dot occupancy, 
$E_{C}$ is the charging energy, $\vec{S}=\frac{1}{2}\sum_{i\mu\nu}d_{i\mu}^{\dagger}\vec{\sigma}_{\mu\nu}d_{i\nu}$ is the total spin of the dot and 
$J_{H}>0$ is the Hund's exchange coupling. In the above,
$\alpha=L,R$ labels left/right lead degrees of freedom and 
$k$ labels the wavevector of a single transverse channel propagating through 
the constriction between the 2DEG and the quantum dot. 
Electrons tunnel into and out of the dot with 
amplitudes $t_{\alpha i}$ and give rise to a single-particle broadening 
$\Gamma_{i}$ of the levels.

The model above is essentially the same model as the multi-orbital model of
Sec.~\ref{subsec:orbital}, used to describe bulk Kondo systems. 
The novel situation in quantum dots
is that parameters such as the tunnel couplings and level positions 
can be controlled by gate voltages. This allows such models to be 
experimentally investigated in all physically interesting regimes, such as
spin and charge fluctuation regimes, and in principle also to be tuned through
quantum phase transitions.
In addition, different realizations of quantum dots (nanotubes, vertical dots)
may have level degeneracies or near level degeneracies, allowing the effects of  
Hund's exchange to be investigated.
Finally, the devices described by (\ref{eq:dot}) can be driven 
out of equilibrium by a finite transport voltage, allowing the study
of non-equilibrium effects in relatively ``simple'' quantum many-body 
systems. This would be one motivation to further develop the NRG to 
steady-state non-equilibrium situations.

\paragraph{Single-level quantum dots}
In the low-temperature limit,
only one or two partially filled levels close to the Fermi level of the 
leads will be important for transport. The
remaining levels will be either filled or empty, and, at the low temperatures
of interest for quantum transport, they may be neglected. The simplest model, therefore, 
to describe low-temperature transport through a quantum dot, is 
the single-level Anderson impurity model (\ref{eq:siam}) 
with level position $\varepsilon_{1}=\varepsilon_{f}=-eV_{G}$ controlled by gate
voltage and Coulomb repulsion $U$ given by the charging energy $E_{C}=U/2$. 
Only one conduction channel, the even combination of left and right 
electron states, $a_{k\sigma}$, below, couples to the local level, 
as can be seen by using the canonical transformation 
\begin{eqnarray}
a_{k\sigma} &=& (t_{L}c_{kL\sigma}+t_{R}c_{kR\sigma})/\sqrt{t_{L}^2 + t_{R}^{2}},\\
b_{k\sigma} &=& (-t_{R}c_{kL\sigma}+t_{L}c_{kR\sigma})/\sqrt{t_{L}^2 + t_{R}^{2}},
\end{eqnarray}  
with $t_{\alpha}=t_{\alpha i}\delta_{i,1}$.
We note that treating the Coulomb interaction classically implies that, 
for an integer number of electrons on the dot, transport is blocked for 
large U, since transferring electrons through the dot requires 
overcoming the large Coulomb repulsion. \textcite{Glazman:1988} and 
\textcite{Ng:1988} pointed out, however, that in the situation
where the total spin on the dot is finite, as happens for an odd
number of electrons (i.e. for ${\cal N}=1$ in the effective single-level 
model), one should expect, on the basis of (\ref{eq:siam}),
 an enhancement of the conductance to its maximum possible 
value of $G=2e^2/h$ via the Kondo effect in the limit of zero temperature. 
A device, representing a tunable Anderson impurity model,
was realized \cite{Goldhaber-Gordon:1998,Cronenwett:1998} and the predicted
enhancement of the low-temperature conductance for dots
with an odd number of electrons was measured and compared \cite{Goldhaber-Gordon:1998} 
to quantitative NRG calculations \cite{Costi:1994a} such as those 
shown in Fig.~\ref{fig:ResCond} for the conductance in the Kondo regime 
(see also \textcite{Izumida:2001b}). Tuning the quantum dot to
the mixed valence and empty orbital regimes, has enabled also comparisons with
theory in those regimes \cite{Schoeller:2000,Costi:2003b}.

The frequency dependence
of the linear conductance, $G'(\omega)$, of a single-level quantum dot described by 
(\ref{eq:siam}) has been considered by several authors 
\cite{Izumida:1997,Campo:2003,Sindel:2005}. 
\textcite{Sindel:2005} calculated $G'(\omega)$, in the Kondo regime at $T=0$, and 
extracted also the current noise 
\begin{equation}
C(\omega) = \int_{-\infty}^{+\infty}dt e^{i\omega t} [\langle I(0)I(t)\rangle-
\langle I\rangle^{2}] \ ,
\end{equation}
by making use of the fluctuation-dissipation theorem
\begin{equation}
C(\omega) = \frac{2\hbar\omega}{e^{\hbar\omega/k_{B}T}-1}G'(\omega).
\end{equation}

The conductance and spin-resolved conductances of single-level quantum dots
in a magnetic field have also been calculated 
and a strong spin-filtering effect has been 
observed in the mixed valence regime \cite{Costi:2001}.
For spin-filtering effects in quantum dots with
ferromagnetic leads see  \textcite{Martinek:2003,Simon:2006}.


One of the hallmarks of the $S=1/2$ single-channel Kondo effect is the 
flow of the exchange coupling to strong coupling \cite{Wilson:1975}. This
can be interpreted as resulting in a phase shift of the conduction electrons 
at the Fermi level, at $T=0$, of $\delta_{\sigma}=\pi/2$ \cite{Nozieres:1974}.
A direct measurement of this phase shift is possible if one embeds a quantum dot 
in one arm of an Aharanov-Bohm interferometer. Assuming a single-level 
Anderson model for the quantum dot and a multi-terminal open geometry, 
\textcite{Gerland:2000} carried out NRG calculations for the interference 
term, $G_{\rm AB}$, whose measurement can be used
to extract $\delta_{\sigma}$. A similar set-up has been investigated by 
\textcite{Hofstetter:2001} for the flux dependence $G(\phi)$ of the conductance 
at $T=0$ and by \textcite{Kang:2005} for the complex transmission. 
\textcite{Izumida:1997} calculated $G(\phi)$ for
two single-level quantum dots embedded in the arms of an Aharanov-Bohm 
interferometer. This model, reduces, in general, 
to a two-channel two-orbital Anderson model, which we discuss next.

\paragraph{Two-level quantum dots}

\begin{figure}[!t]
\centerline{
  \includegraphics*[width=3.0in]{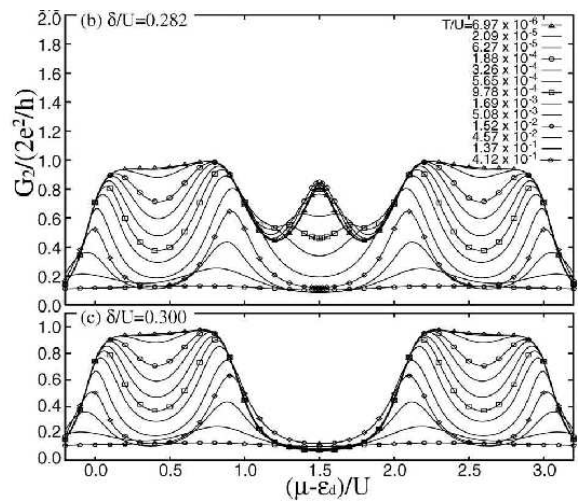}}
  \caption{The singlet-triplet crossover in the linear conductance of the two-channel 
two-orbital Anderson model including a Hund's exchange $J_{\rm H}$, intra- and inter-orbital
Coulomb energy, $U$, levels $\varepsilon_{d}\pm \delta/2$ and temperature $T$. Top/bottom
panels shows $G(T)$ on the triplet/singlet side of the crossover and $(\mu-\varepsilon_{d})/U=0.5,1.5,2.5$ correspond to ${\cal N}=1.0,2.0,3.0$ electrons on the dot.
Adapted from \textcite{Sakai:2003}.
}
\label{fig:IzumidaSTCrossover}
\end{figure}

A quantum dot with two active levels for transport
introduces some new physics due to the competition between the level 
spacing $\delta=\varepsilon_{2}-\varepsilon_{1}$, the charging energy 
$E_{C}$ and the Hund's exchange $J_{\rm H}$. In particular, a Kondo effect 
with an even number of electrons on the dot can be realized.
This can occur when the dot is occupied with two electrons and
$\delta<2J_{\rm H}$ so that the ground state of the dot has $S=1$. 
Such a two-level dot will, in general, couple to two channels so a 
$S=1$ Kondo effect will result, leading to a singlet ground state and 
an enhanced conductance $G(T)$ at low temperatures. In the opposite case
$\delta>2J_{\rm H}$ the dot will have $S=0$, the Kondo effect is absent and
the conductance will be low. This behavior is believed to have been measured
in the experiments of \textcite{Sasaki:2000} on vertical quantum dots, where 
a magnetic field was used to decrease the energy splitting, $\Delta_{TS}=\delta-2J_{\rm H}$, 
between the triplet and singlet states, thereby leading to the above mentioned 
crossover behavior in the conductance at ${\cal N}=2$. Theoretical calculations
by \textcite{Izumida:2001a}, shown in Fig.~\ref{fig:IzumidaSTCrossover}, are
consistent with the experimental results. 

The singlet-triplet crossover behavior in a two-level quantum dot can become a
quantum phase transition for the special case where only one conduction
channel couples to the leads, e.g. when all lead couplings are equal 
\cite{Pustilnik:2001,Hofstetter:2002}.
In this case, for large Hund's exchange, an effective single-channel 
$S=1$ underscreened Kondo model results which has a doubly degenerate many-body 
ground state. For small Hund's exchange, a model with $S=0$ results having 
a non-degenerate many-body ground state. A sharp transition separates these 
two different ground states. 
As discussed above, however, two channels will, in general, couple 
to the dot and this will result in perfect screening of the $S=1$ 
so that the ground state is always a singlet. Nevertheless, proximity
to the singlet-triplet transition can still be seen as signatures in
various quantities, such as a non-monotonic dependence of the 
conductance as a function of magnetic field on the triplet 
side of the crossover \cite{Hofstetter:2004}. 
Experiments on lateral quantum dots at the singlet-triplet crossover point
\cite{Wiel:2002} show behavior in the differential conductance similar 
to predictions for the spectral density \cite{Hofstetter:2002}.

The above is only a brief account of the simplest nanostructured devices
studied using the NRG. Further applications include numerous studies of
double-dot systems, including realizing an SU(4) Kondo state \cite{Borda:2003} and
quantum critical points of two-impurity Kondo models \cite{Zarand:2006,Garst:2004,Zhu:2006} 
(see Sec.~\ref{subsubsec:mult-imp}), static and dynamics of double-dots 
\cite{Galpin:2006a,Galpin:2006b}, 
double dots with only one dot coupled to the leads \cite{Cornaglia:2005c}, 
applications to quantum tunneling in molecular magnets 
\cite{Romeike:2006a,Romeike:2006b}, a novel Kondo effect in a $\nu=1$ integer
quantum Hall system \cite{Choi:2003a} and the conductance of ultrasmall tunnel
junctions \cite{Frota:1992,Frota:2004}.

  \subsection{Two-channel Kondo physics}
  \label{subsec:two-channel}

\textcite{Nozieres:1980} have proposed a variation of
the Kondo model in which the localized spin couples to {\em two} conduction
bands. The Hamiltonian of this two-channel Kondo model is
given by
\begin{equation}\label{eq:TwoChannelKondo}
   H = \sum_{k\sigma\alpha} \varepsilon_k c^\dagger_{k\sigma\alpha}
     c_{k\sigma\alpha} + J \sum_\alpha {\bf S} \cdot {\bf s}_\alpha \ ,
\end{equation}
with $\alpha=1,2$ the channel index and  ${\bf S}$ (${\bf s}_\alpha$)
the spin operators of the impurity (the conduction band electrons
at the impurity site with channel index $\alpha$).

An important feature of this model is the overscreening of
the impurity spin: in the strong-coupling limit, the spins of both
conduction bands try to screen the impurity spin, so that again
a net spin $1/2$ object is formed. In other words, the
strong-coupling fixed point at $J=\infty$ (which gives rise to the
Fermi-liquid fixed point in the single-channel case) is unstable and an
intermediate-coupling fixed point is realized. This new fixed
point shows a variety of non-Fermi liquid properties such as
\begin{itemize}
  \item a divergence of the specific heat ratio
        $C/T = \gamma \propto \ln T $ and of the spin susceptibility
        $\chi \propto \ln T$ for $T\to 0$;
  \item an anomalous Wilson ratio $R=\chi/\gamma=8/3$, in contrast
        to the result for the standard Kondo model $R=8/4=2$;
  \item a zero-point entropy of $\frac{1}{2} \ln 2$, indicating that
        `half'-fermionic excitations (Majorana fermions)
        play a crucial role for the
        structure of the fixed point.
\end{itemize}
We have discussed these features already in the section
on the calculation of thermodynamic and static quantities,
see Fig.~\ref{fig:III-2}.
An extensive review of the two-channel Kondo model, its physical
properties and its
relevance for non-Fermi liquid behavior in real materials has 
been given in \textcite{Cox:1998}. This paper also reviews
the earlier NRG-calculations for this model.

Historically, the two-channel Kondo model has been the first application
of the NRG to a quantum impurity model in which the physics is not
governed by the Fermi liquid fixed point of the standard Kondo model.
In this sense, the early work of \textcite{Cragg:1980} on the 
two-channel Kondo model opened the way for a variety of investigations
of more complex impurity models, displaying both Fermi liquid
and non-Fermi liquid fixed points. Due to the importance of this 
and following work, we want to focus this section purely on two-channel Kondo 
physics and shift the discussion of other multi-band models
to the section on orbital effects (Sec.~\ref{subsec:orbital}).

As discussed already in Sec.~\ref{sec:nrg-intro}, 
the truncation of states within
the iterative diagonalization scheme severely limits the applicability
of the NRG to multi-band models. In the calculations of \textcite{Cragg:1980},
the iterations were observed to break down only after a few (approximately
twelve) steps. The source of this problem is mainly the small number of states
($N_{\rm s} \approx 400$) used in this work, which would correspond
to keeping $N_{\rm s} \approx \sqrt{400}=20$ states in a one-channel
calculation. Specific symmetries of the two-channel Kondo model, such as
the total axial charge, have been used to reduce the matrix sizes
in the diagonalization \cite{Pang:1991}, but later calculations showed
that by simply increasing the number of states, 
the iterations can be stabilized sufficiently. 
Independent
of the value of $N_{\rm s}$, it is important to avoid any
symmetry breaking due to the truncation of states.

\begin{figure}[b]
\centerline{
  \includegraphics*[width=3.0in]{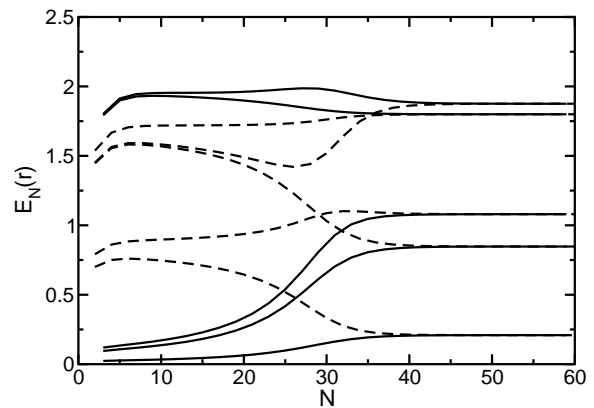}}
\caption{Flow diagram of the lowest lying many-particle levels
for the isotropic two-channel Kondo model (even iterations dashed curves,
odd iterations full curves). 
}
\label{fig:tckm}
\end{figure} 
In order to approach the non-Fermi liquid fixed point within only a
few iterations, \textcite{Cragg:1980} and \textcite{Pang:1991}
used large values of the exchange coupling $J$ or 
large values of the discretization parameter
(up to $\Lambda=9.0$). Nevertheless, these calculations give 
the correct fixed point spectrum of the (isotropic) two-channel Kondo model
with the characteristic structure of excitations at energies 
$1/8$, $1/2$, $5/8$, $1$, etc., at least for the lowest-lying excited 
states. 
Figure \ref{fig:tckm} shows a typical
flow diagram for parameters $J=-0.05D$, where  $2D$ is the bandwidth 
of the featureless conduction band density of states, $\Lambda=4$, 
and $N_{\rm s}=4900$, for both even (dashed curves) and odd NRG iterations (full curves)
(for similar
plots, see Fig.~1 in \textcite{Cragg:1980} and Figs.~1 and 2
in \textcite{Pang:1991}). 
After some initial
even-odd oscillations, the flow reaches the non-Fermi liquid 
fixed point which does not  show any even-odd effect. Note that this
feature is by no means related to the non-Fermi liquid properties of
the model; it just reflects the fact that in each iteration, two
sites are added to the chain so that (for particle-hole symmetry)
the number of electrons in the ground state is always even.

Comparison with conformal field theory calculations
\cite{Affleck:1992} gave an excellent agreement with the NRG
for both the excitation spectrum and the corresponding degeneracies.
Such a comparison, however, requires the extrapolation of the
NRG fixed point spectra for $\Lambda\to 1$ (see Fig.~9 in
\textcite{Affleck:1992}; the analysis is not quite satisfactory
for $\Lambda<2$ and it would be interesting to repeat these calculations
using larger values of $N_{\rm s}$). This work, and the previous
paper by \textcite{Pang:1991}, also focussed on the stability of the
non-Fermi liquid fixed point against various perturbations. As it
turns out, the non-Fermi liquid fixed point is stable against
anisotropy in the exchange interaction ($J_z\ne J_\perp$) but
unstable against both the presence of a magnetic field and the lifting of the 
exchange symmetry between the two channels ($J_a\ne J_b$). In the
latter case, a temperature scale $T^\ast \propto (J_a - J_b)^2$
for the crossover between the non-Fermi liquid fixed point
at intermediate temperatures and the stable Fermi liquid fixed point
at $T\to0$ has been found.
These instabilities have been later investigated in more detail
in \textcite{Yotsuhashi:2002}, via the calculation of the impurity
entropy and the crossover temperature.

Further investigations concerning the stability of the non-Fermi liquid fixed 
point have been performed in \textcite{Pang:1994} (flavor exchange
coupling) and \textcite{Kusunose:1996,Kusunose:1999} (effect of repulsion among conductions
electrons and potential scattering).

A `pedestrian' approach for the understanding of the two-channel
Kondo model was introduced by \textcite{Coleman:1995}.
The authors of this work
 argued that the two conduction bands in the two-channel
Kondo model can be replaced by a {\it single} conduction band, with a coupling
between impurity spin ${\bf S}$ 
to both spin ${\bf \sigma}$ and isospin ${\bf \tau}$ of the
conduction band. The isospin ${\bf\tau}$ 
takes into account the charge degrees
of freedom of the conduction band and the compactified 
`$\sigma$-$\tau$' model takes the form
\begin{equation}
   H = \sum_{k\sigma} \varepsilon_k c^\dagger_{k\sigma}
        c_{k\sigma} + 
    J \left( {\bf S} \cdot {\bf \sigma} + {\bf S} \cdot {\bf \tau}\right)
     \ .  \label{eq:tckm}
\end{equation}
This model can be related to an Anderson-type model (the
`O(3)-symmetric' Anderson model) via a
Schrieffer-Wolff trans\-for\-mation \cite{Coleman:1995b}.

It has been later verified with the NRG approach 
\cite{Bulla:1997c} that these
compactified models indeed show many of the anomalous non-Fermi
liquid properties of the two-channel Kondo model, although these
models do not allow for an overscreening of the impurity spin.
Furthermore, the structure of the non-Fermi liquid fixed
point has been studied in detail. It turns out that the 
many-particle spectrum of this fixed point is composed of single 
Majorana fermion excitations \cite{Bulla:1997b}. 
This information can then be extended to the fixed point structure
of the two-channel Kondo model which can be described by two
towers of excitations which are both composed of Majorana fermions,
see Sec.~VIII in \textcite{Bulla:1997b}.

Naturally, we expect that the non-Fermi liquid properties of the
two-channel Kondo model are also visible in its dynamic properties,
but, unfortunately, detailed and comprehensive NRG calculations
for the dynamics have not been performed so far. Apart from a 
brief sketch of the results for the $T$-matrix and the magnetic
susceptibility in \textcite{Sakai:1993a}, the published data
are only for models equivalent to the two-channel Kondo model in
certain limits.

It has been argued in \textcite{Bradley:1999} that the dynamical
spin-susceptibility $\chi(\omega)$ of the compactified models
introduced above is exactly equivalent to that of the 
two-channel Kondo model, and that this equivalence holds for the
{\em full} frequency range. The NRG-results show, for example, a
$\ln(\omega)$-divergence of $\chi^\prime(\omega)$ for $\omega\to 0$,
in agreement with the results of \textcite{Sakai:1993a}. On the
other hand, there is no counterpart of the single-particle dynamics
calculated by \textcite{Bradley:1999} in the two-channel Kondo model.

The two-channel Anderson model investigated in \textcite{Anders:2005a}
is connected to the two-channel Kondo model via a Schrieffer-Wolff
transformation (note that this only holds when the impurity
degrees of freedom in the Anderson model are written in terms
of Hubbard operators which include the channel index). This
connection is clearly visible in thermodynamic properties, such
as the zero-point entropy of $\frac{1}{2} \ln 2$. Again, the
single-particle dynamics (spectral function and self-energy)
do not have a counterpart in the two-channel Kondo model. Concerning
the results for the dynamic susceptibility  $\chi(\omega)$ presented
in \textcite{Anders:2005a}, a comparison to the corresponding
results of the two-channel Kondo model has not yet been done.

There is an ongoing discussion about the observability of
two-channel Kondo physics in experiments for a variety of systems.
Let us stress here that the instability of the non-Fermi liquid
fixed point itself does not exclude its observation. As for
any system with a quantum critical point, the corresponding
anomalous properties dominate a significant fraction
of the finite-temperature phase diagram 
(determined by the critical exponent)
so that a precise
tuning of the Hamiltonian parameters is not required.
Nevertheless, two-channel Kondo physics is now mainly discussed
within systems in which alternative degrees of freedom
(such as orbital quantum numbers) take the role of spin or
channel in the Hamiltonian Eq.~(\ref{eq:TwoChannelKondo});
one example is the quadrupolar Kondo model which is
discussed in detail in \textcite{Cox:1998}.

Here we want to briefly discuss NRG calculations for 
two-channel Kondo physics in quantum dot systems
\cite{Lebanon:2003a,Lebanon:2003b,Anders:2004,Anders:2005e}.
Within a model of a quantum box coupled to the leads via 
a single-mode point contact (see Fig.~1 in \textcite{Lebanon:2003b}),
the physics at the degeneracy points of the Coulomb blockade
staircase can be directly connected to that of the 
two-channel Kondo model. Here the two charge configurations
in the box play the role of the impurity spin and the physical
spin of the conduction electrons corresponds to the channel index.
For such a system, the NRG allows the non-perturbative calculation
of the charge inside the box and the capacitance in the whole
parameter regime. The results show, for example, that the shape
of the charge steps is governed by the non-Fermi liquid
fixed point of the two-channel Kondo model.

To conclude this section let us mention that
there are models involving a more complicated orbital
structure of the impurity -- including, for example,
excited crystalline electric field levels --
which reduce to the two-channel Kondo model in certain
limits or which display  non-Fermi liquid fixed points of the
two-channel Kondo type.
NRG-studies of such models can be found in 
\textcite{Koga:1999,Hattori:2005d,Sakai:1997,Koga:1996,Koga:1995}.
Overscreening can also be realized in single-channel
models when the conduction electron spin exeeds the impurity spin,
for a discussion of this issue see for example \textcite{Kim:1997}.
We note also a recent study \cite{Kolf:2006} showing an exponential
dependence of the Kondo scale on $-1/JN_{F}$ and $-JN_{F}$, for small and
large coupling cases respectively, which may explain the absence of a broad
distribution of Kondo scales in nanoconstrictions with 
two-channel Kondo impurities.

  \subsection{Impurity quantum phase transitions}
  \label{subsec:qpt}

In this section, we focus on models which, as a function of
one or more couplings in the model, give rise to a phase transition in
the ground state. Typically, this is due to a competition between the 
Kondo effect on the one hand, which
tends to favor a strong-coupling ground state with a screened or partially
screened local moment, and some competing mechanism, which leads to a ground state
with a free or almost free local moment. 
In general, such phase transitions
are termed `impurity quantum phase transitions' (for recent
reviews, see \textcite{Vojta:2004,Bulla:2003a}), 
as they are only observable in the
{\em impurity} contribution to physical properties and not
connected to possible phase transitions in the bulk to which 
the impurity couples.

As impurity quantum phase transitions are usually associated with a
vanishing low-energy scale, the NRG method is ideally suited to their
investigation, allowing  their detection and characterization with 
very high accuracy. 
This is most evident for continuous transitions where the critical exponents
connected to the quantum critical point can only be calculated
when a large range of energy or temperature scales is accessible.
In this section we give an overview on NRG-results for multi-impurity
models (Sec.~\ref{subsubsec:mult-imp}), models with locally critical behavior 
(Sec.~\ref{subsubsec:loc-crit}), and
models with magnetic impurities in superconductors 
(Sec.~\ref{subsubsec:sc}). Note
that impurity quantum phase transitions are also observed in models
which are discussed in other sections of this review: the non-Fermi
liquid fixed point of the two-channel Kondo model 
Sec.~\ref{subsec:two-channel} can be viewed
as a quantum critical point when the control parameter `channel anisotropy'
is tuned through zero; locally critical behavior is also connected to
models with coupling to a bosonic bath as discussed in 
Sec.~\ref{subsec:bosons}.

\subsubsection{Multi-impurity physics}
\label{subsubsec:mult-imp}
An early extension of the NRG to more complex systems was the study of
the two-impurity $S=1/2$ Kondo model \cite{Jones:1987,Jones:1988}, whose 
Hamiltonian is given by
\begin{equation}
  \label{eq:two_imp_km}
     H = \begin{array}[t]{l}\displaystyle
\sum_{k\sigma} \varepsilon_k c^\dagger_{k\sigma}
        c^{\phantom{\dagger}}_{k\sigma} + J_{\rm K} \sum_{l=1}^2 {\bf S}({\bf R}_l)\cdot
        {\bf s}({\bf R}_l)\\[5mm] 
\displaystyle
+I_{\rm D}{\bf S}({\bf R}_1)\cdot{\bf S}({\bf R}_2).
\end{array}
\end{equation}
Here, ${\bf s}({\bf R}_l)$
is the conduction electron spin density at the impurity 
site ${\bf R}_l$ and $J_{\rm K}>0$  is the antiferromagnetic Kondo exchange.
The first two terms in  Eq.~(\ref{eq:two_imp_km}) are sufficient to generate
an indirect RKKY interaction $I_{\rm RKKY}$ between the impurity spins. In some contexts
a direct exchange interaction among the impurity spins of strength $I_{\rm D}$ can
arise \cite{Jones:1987}), so the last term has been added. The net effective 
exchange interaction between the spins is given by 
$I_{\rm eff}=I_{\rm D}+I_{\rm RKKY}$ and can be either 
ferromagnetic $I_{\rm eff}<0$ or antiferromagnetic $I_{\rm eff}>0$.
The properties of the model then depend solely on the ratio $I_{\rm eff}/T_{\rm K}$,
where $T_{\rm K}$ is the single-ion Kondo scale, and the details of the 
dispersion relation $\varepsilon_{k}$.
The model in Eq.~(\ref{eq:two_imp_km}) also arises in the  Schrieffer-Wolff limit 
\cite{Schrieffer:1966} of the two-impurity Anderson model, which in the
notation introduced in Eq.~(\ref{eq:siam}) reads
\begin{eqnarray}
  H &=& \sum_{k\sigma} \varepsilon_k c^\dagger_{k\sigma}
        c^{\phantom{\dagger}}_{k\sigma}
+\sum\limits_{l=1}^2\sum\limits_\sigma \varepsilon_{l\sigma}
f^\dagger_{l\sigma}f^{\phantom{\dagger}}_{l\sigma}\nonumber\\
&&
+U\sum\limits_{l=1}^2
f^\dagger_{l\uparrow} f^{\phantom{\dagger}}_{l\uparrow}
                       f^\dagger_{l\downarrow}
                       f^{\phantom{\dagger}}_{l\downarrow}\nonumber\\
&&
+\frac{1}{\sqrt{N}}\sum\limits_{l=1}^2\sum\limits_{k\sigma} V_k
      \Big( e^{i{\bf R}_l\cdot{\bf k}}f^\dagger_{l\sigma} c^{\phantom{\dagger}}_{k\sigma}
                                    +\mbox{h.c.}\Big)\nonumber\\
&&
  \label{eq:two_imp_am}
+I_{\rm D}{\bf S}({\bf R}_1)\cdot{\bf S}({\bf R}_2)\ .
\end{eqnarray}
The motivation to study such two-impurity models originally arose in 
the context of heavy fermions.  In these systems, the competition 
between the local Kondo exchange and the intersite RKKY interaction
is expected to lead to a  phase transition between non-magnetic and 
magnetically ordered ground states as a function of $I_{\rm eff}/T_{\rm K}$
\cite{Doniach:1977}.   
The nature of this quantum phase transition remains 
an open question in heavy fermion physics \cite{Lohneysen:2006}. It is 
therefore of some interest to investigate the possibility of a transition
in the two-impurity problem as this might shed light on the physics of
heavy fermions.

\textcite{Jones:1988} have established that such a phase transition can 
occur under certain conditions, see below, in the particle-hole symmetric 
two-impurity Kondo model. This can be seen by considering
the strong-coupling limits $I_{\rm eff}\rightarrow \pm\infty$ \cite{Affleck:1995}. 
For $I_{\rm eff}\rightarrow -\infty$ the two spins combine to form a 
spin $S=1$ interacting 
antiferromagnetically with two conduction channels 
(characterized by even/odd parity) with, in general, energy dependent
coupling strengths $J_{e}(k),J_{o}(k)$ replaced in \cite{Jones:1988} 
by constants (see below). 
The resulting two-stage Kondo effect progressively screens
the $S=1$ spin down to a singlet and leads to a Fermi-liquid ground 
state characterized by phase shifts $\delta_{e,o}$ for electrons in the even/odd
parity channels. The assumed particle-hole symmetry and the nature of 
the strong-coupling ground state ensures that
these phase shifts will be exactly $\pi/2$ \cite{Millis:1990}. 
%
%
%
In the other limit, 
$I_{\rm eff}\rightarrow \infty$, the spins form an intersite singlet $S=0$ and the 
Kondo effect is absent so that the phase shifts are exactly zero. 
Since the fixed points at $I_{\rm eff}=\pm\infty$ are both stable and characterized by
different (constant) phase-shifts, it follows that there can be an
unstable fixed point at some critical intermediate value of the intersite exchange, 
$I_{c}$, at which the phase shifts change discontinuously. 
This phase transition has also
been found in the particle-hole symmetric two-impurity Anderson model 
\cite{Sakai:1992, Paula:1999}. 
\textcite{Jones:1988} estimates $I_{c}/T_{\rm K}\approx 2.2$.
The associated critical point has been characterized using conformal
field theory \cite{Affleck:1992b,Affleck:1995} and bosonization
\cite{Gan:1995}, and the physics is found to be similar
to that of the two-channel Kondo model. In particular, the 
staggered susceptibility, $\chi_{s}(T)$, diverges logarithmically
at low temperature and the residual entropy has the same value as in
the two-channel Kondo model $S(T=0)=\ln\sqrt{2}$ \cite{Gan:1995}. In
contrast $C(T)/T=\gamma$ is predicted to remain
finite exactly at the critical point, in contrast to the behavior
in the two-channel Kondo model. Close to the  critical point,
conformal
field theory predicts
$\gamma \sim (I_{\rm eff}-I_{c})^{-2}$, 
in agreement with numerical results \cite{Jones:1990}.
The level structure of the fixed point at quantum criticality agrees well
with NRG calculations and is rather complex, exhibiting
a hidden $SO(7)$ symmetry \cite{Affleck:1995}. 

For generic situations, the natural energy dependence of $J_{e/o}$ 
obtained from transforming the Kondo model (\ref{eq:two_imp_km}) or 
Anderson model (\ref{eq:two_imp_am}) as described below in 
Eq.~(\ref{eq:two_imp_km_eo}), as well as a charge transfer term of the form 
$t\sum_{\sigma}(f^\dagger_{1\sigma} f^{\phantom{\dagger}}_{2\sigma}+h.c.)$
in the two-impurity Anderson model, breaks particle-hole symmetry and 
destroys the critical point \cite{Sakai:1990,Affleck:1995}. A similar charge transfer
term involving conduction electrons has the same effect 
in the two-impurity Kondo model \cite{Zarand:2006, Zhu:2006}. Potential
scattering, if it does not induce charge transfer, breaks particle-hole symmetry
but may not affect the critical point, for a discussion see 
\textcite{Affleck:1995, Zarand:2006, Zhu:2006}.
Thus, in general the quantum phase transition discussed above will be
absent in the two-impurity models (\ref{eq:two_imp_km}) and
(\ref{eq:two_imp_am}), although signatures of it might
still be observable as crossover behavior in various properties.
We note briefly here the case of Ising coupled spins 
$I S_{z}({\bf R}_1)S_{z}({\bf R}_2)$. In this case, the ground state for large 
$I$ will be doubly degenerate as compared to that for
small $I$ where the Kondo effect screens the individual spins to singlets, 
so a quantum phase transition separating these two different ground states
arises and is found to be of the Kosteritz-Thouless type \cite{Garst:2004}.

In order
to formulate Eq.~(\ref{eq:two_imp_km}) or (\ref{eq:two_imp_am}) as a linear chain problem for treatment
with the NRG, one needs an orthonormal basis set. The local conduction electron states
on the impurity sites in Eq.~(\ref{eq:two_imp_km}) are not orthogonal. Following
\textcite{Jones:1987}, the Kondo exchange part of 
Eq.~(\ref{eq:two_imp_km}) is rewritten in terms of 
orthonormal even ($e$) and odd ($o$) parity states for the 
conduction electrons. This results in more complicated interaction terms; 
in particular, one will obtain two exchange couplings
$J_{e/o}(k)$, with $J_e(k)\ne J_o(k)$ in general, which will depend on momentum
or energy \cite{Jones:1987}. The precise form of $J_{e/o}(k)$ will
depend on the details of the band-structure of the conduction
electrons. For free electrons in $D=3$ it can be approximated as
\cite{Jones:1987,Sakai:1990} 
$$
J_{e/o}(k)\approx 
J_{\rm K}\left[1\pm\frac{\sin kR}{kR}\right]\;,
$$
where $R=|{\bf R}_1-{\bf R}_2|$. \textcite{Jones:1987} used constant
couplings $J_{e/o}(k)\approx J_{e/o}(k_{F})$ to obtain for the interaction part of the
Hamiltonian (\ref{eq:two_imp_km})
\begin{eqnarray}
  H_{\rm int} &=& {\bf S}^{(e)}\cdot
\sum\limits_{\alpha\beta}\left[
J_ec^{\dagger}_{e\alpha}{\bf\sigma}_{\alpha\beta}
c^{\phantom{\dagger}}_{e\beta}+
J_oc^{\dagger}_{o\alpha}{\bf\sigma}_{\alpha\beta}
c^{\phantom{\dagger}}_{o\beta}
\right]+\nonumber\\
&&{\bf S}^{(o)}\cdot
\sum\limits_{\alpha\beta}\left[
i\sqrt{J_eJ_o}c^{\dagger}_{e\alpha}{\bf\sigma}_{\alpha\beta}
c^{\phantom{\dagger}}_{o\beta}+\mbox{h.c}
\right]+\nonumber\\
  \label{eq:two_imp_km_eo}
&&I_{\rm D}{\bf S}({\bf R}_1)\cdot{\bf S}({\bf R})_2\;\;,
\end{eqnarray}
where ${\bf S}^{(e/o)}:={\bf S}({\bf R}_1)\pm{\bf S}({\bf R}_2)$. 
The conduction electron Hamiltonian now consists of two decoupled
linear chains with even and odd parity symmetry. By neglecting the
energy dependence of the couplings an particle-hole 
symmetric model results. This is
the form used by \textcite{Jones:1988} to investigate the phase transition
discussed above. The results of retaining the 
full energy dependence of the couplings, using for example the formulation 
of Sec.~\ref{sec:nrg-intro}, will be described below. We note here that
from the NRG point of view the two-impurity models
(\ref{eq:two_imp_am}) and (\ref{eq:two_imp_km}) present a challenging
task because, as in the case of the two-channel Kondo model (\ref{eq:TwoChannelKondo}),
the ``impurity'' now couples to  {\em two semi-infinite chains}. Consequently,
the Hilbert space grows by a factor $16$ in each NRG
step. While this is still manageable with modern computer resources, it
is apparent that larger clusters or more complex situations quickly
become too expensive to be treated with NRG with sufficient accuracy,
although the flow of the many-body eigenstates can still be used to
identify fixed points and thus qualitatively describe the physics of
more complicated systems, like the two-channel two-impurity Kondo
model \cite{Ingersent:1992} and the three-impurity Kondo model \cite{Ingersent:2005}. 
In high-symmetry situations even a 
calculation of thermodynamical quantities has been performed recently
for three-impurity models
\cite{Zitko:2006b}. However, for a reliable calculation of {\em
  dynamics} or in situations with less symmetries in the system --
e.g.\ in an external magnetic field -- additional tools
like the ones described in section~\ref{subsubsec:Zaverage} allowing one to
work with large $\Lambda\gg 1$ and so maintain low truncation errors should be
useful.

The generic two-impurity Anderson model (\ref{eq:two_imp_am}), including a charge
transfer term, has been studied by Sakai 
and coworkers \cite{Sakai:1990,Sakai:1992,Sakai:1992b} using the NRG. 
Single particle and magnetic excitation spectra were calculated and in the
case of particle-hole symmetry, \textcite{Sakai:1993a} showed that
on passing through the transition, a peak in the impurity single-particle spectra
sharpened at $I_{\rm eff}=I_{c}$ into a cusp and turned into a dip for 
$I_{\rm eff}>I_{c}$. 
In the generic case, the regime with Kondo screening, $|I_{\rm eff}|\ll T_{\rm K}$,
and the non-local singlet regime, $I_{\rm eff}\gg T_{\rm K}$, 
are connected via a smooth crossover 
\cite{Sakai:1990,Sakai:1992,Sakai:1992b,Silva:1996,Campo:2004}.

\begin{figure}[htb]
\begin{center}
\includegraphics[width=0.48\textwidth,clip]{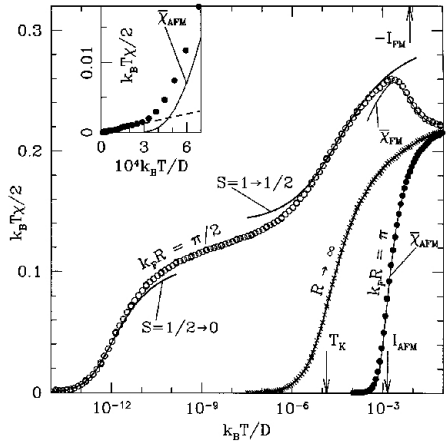}
\end{center}
\caption[]{Effective local moment $\mu^2(T):=T\cdot\chi_{\rm imp}(T)$
for the two-impurity Kondo model (\ref{eq:two_imp_am}) for the three regimes
described in the text.
 Figure taken from \textcite{Silva:1996}.\label{fig:IV-C-1} 
The arrows indicate the Kondo scale $T_{\rm K}=1.4\times 10^{-4}$, and the 
RKKY interactions for ferromagnetic ($I_{\rm FM}=-8\times 10^{-3})$) 
and atiferromagnetic cases ($I_{\rm AFM}=3\times 10^{-3}$).
}
\end{figure}

Results from thermodynamic calculations are shown in Fig.~\ref{fig:IV-C-1} 
for the squared effective magnetic moment taken from \textcite{Silva:1996}
for the two-impurity Kondo model with $I_{\rm D}=0$. In these calculations
$I_{\rm eff}=I_{\rm RKKY}$ and the energy dependence 
of $J_{e/o}(k)$ is crucial to generate the intrinsic RKKY exchange 
interaction $I_{\rm RKKY}$. Using the result for free electrons in
three dimensions \cite{Sakai:1990}, an approximate formula for the
energy dependence of the coupling constants is \cite{Silva:1996}
\begin{equation}\label{eq:Jeo}
J_{e/o}(\epsilon)= J_K\left(
  1\pm\frac{\sin\left[k_FR\left(1+\epsilon\right)\right]}{k_FR\left(1+\epsilon\right)}\right) 
\end{equation}
with $\epsilon\in[-1,1]$, $k_F$ the Fermi momentum of the
conduction states. For the derivation of Eq.~(\ref{eq:Jeo}) a
linearized dispersion relation $\epsilon_k\approx\frac{D}{k_F}(k-k_F)$
was assumed and $D=1$ used as energy scale.

Dependent on the value of $k_FR$,
different regimes can then be identified 
(see, for example, Fig.~\ref{fig:IV-C-1}): For
$k_FR\to\infty$ we have $I_{\rm RKKY}=0$, 
single-impurity physics dominates and no non-local
magnetic exchange is generated, as expected \cite{Jones:1987}. For $k_FR=\pi/2$, the
RKKY exchange $I_{\rm RKKY}=I_{\rm FM}$ is ferromagnetic, with 
$|I_{\rm RKKY}|/T_{\rm K}\gg 1$ and a two-stage screening scenario
arises: First, the system locks into an $S=1$ state at high
temperatures due to the RKKY interaction. In the intermediate
temperature regime this triplet is screened to a doublet via the even
channel, which then is further screened to a singlet by the odd 
channel. Finally, for $k_FR=\pi$, the RKKY exchange 
$I_{\rm RKKY}=I_{\rm AFM}$ is
antiferromagnetic with $I_{\rm RKKY}/T_{\rm K}\gg 1$ 
and a non-local singlet is formed eventually.
Similar results for entropy and specific heat of the two-impurity Kondo
model, exhibiting a smooth change of physical properties with changing $I_{\rm RKKY}$, 
can be found in \textcite{Campo:2004}.

We note, here, that while two-impurity models with energy independent coupling constants
are crude approximations in the context of bulk Kondo impurities and heavy
fermions, these can, however, be realized in quantum dots. 
Correspondingly, they have been
proposed to describe various extensions of single quantum dots and
studied in this context with NRG by several groups 
\cite{Vojta:2002d,Boese:2002,Hofstetter:2002,Borda:2003,Zitko:2006,Zarand:2006}
over the past years. Since modern nanostructure technology permits a
rather broad tailoring of such mesoscopic objects, the models
discussed typically introduce additional interactions as compared to the
conventional two-impurity Anderson model (\ref{eq:two_imp_am}) like
capacitive couplings \cite{Hofstetter:2002,Boese:2002,Borda:2003} or
direct hopping \cite{Zitko:2006,daSilva:2006}. Consequently, these extended models
show a much larger variety  in intermediate- and low-temperature
fixed points than the bare model (\ref{eq:two_imp_am}), ranging from
conventional Kondo effect over a two-stage Kondo effect
\cite{Jayaprakash:1981,Vojta:2002d}, two-channel physics as intermediate fixed-point
\cite{Zitko:2006,Zitko:2006b} to quantum-phase transitions 
\cite{Vojta:2002d,Zitko:2006,Zarand:2006,Zhu:2006}.








\subsubsection{Local criticality}
\label{subsubsec:loc-crit}

The term `local criticality' has been first used in the context
of phase transitions in certain heavy fermion systems, such as
CeCu$_{6-x}$Au$_x$ \cite{Si:2001,Si:1999,Lohneysen:2006}. 
It has been argued that the quantum critical point
separating the magnetically ordered and the paramagnetic phases
at $T=0$ is characterized by critical excitations which
are local. This observation raised a considerable interest
in models which show such locally critical behavior: these are either lattice
models studied within certain extensions of DMFT 
(see also Sec.~\ref{subsec:pam})
or impurity models as discussed in this section. Such impurity models
might not be directly connected to the locally critical behavior in
heavy fermion systems, nevertheless, the insights gained in studying
impurity models might be helpful in constructing theories for
lattice systems (for a general discussion of the relation between
quantum impurity physics and the physics of lattice systems, see
\textcite{Bulla:2004}).

Let us focus here on the soft-gap Anderson model, originally
proposed by \textcite{Withoff:1990}. The Hamiltonian is the
same as the one for the standard single-impurity Anderson model
Eq.~(\ref{eq:siam}), but the hybridization function is assumed to
have a power-law form
\begin{equation}
   \Delta(\omega) = \Delta_0 |\omega|^r  \ \ \ {\rm with} \ \ r>-1 \ ,
\end{equation}
either valid over the whole frequency range or restricted to
some low-frequency region. The competing mechanisms leading to
a quantum phase transition in this model are local moment formation
(favored by increasing $U$) and screening of the local moments.
For values of the exponent $r>0$, corresponding to a soft-gap in 
$\Delta(\omega)$, there are less degrees of freedom available to screen
the moment and a quantum phase transition occurs at some finite value
of $\Delta_0$.

This quantum phase transition and the physical properties in the whole
parameter regime have been studied in detail with a variety of techniques
(for an overview, see \textcite{Vojta:2004,Bulla:2003a}
and the introductory parts in 
\textcite{Lee:2005}). The NRG method has been particularly helpful
to clarify the physics of the soft-gap Anderson model (and the related
Kondo version of the model) as we shall briefly discuss in the following.
The technical details necessary to apply the NRG to the soft-gap Anderson
model have already been introduced in 
Sec.~\ref{sec:nrg-intro}, see also \textcite{Bulla:1997a}.

Thermodynamic and static properties of the various phases of the soft-gap 
Anderson and Kondo models have been presented in 
\textcite{Chen:1995a}, \textcite{Ingersent:1996}, \textcite{Bulla:1997a},
and \textcite{Gonzalez:1998}. The most comprehensive review of these
results is given in \textcite{Gonzalez:1998}. This paper covers
the discussion of thermodynamic properties and the analysis of the
various fixed points also for the underscreened spin-1 Kondo
model and the (overscreened) two-channel Kondo model (both with
a soft-gap in the conduction electron density of states).

The key role of particle-hole symmetry has been 
identified in \textcite{Ingersent:1996}
and investigated in more detail in \textcite{Gonzalez:1998}.
As shown, for example, in Fig.~5 in this work, the line of quantum critical
points separating the local moment (LM) and strong-coupling (SC)
phases is restricted to $0<r<1/2$ in the particle-hole
symmetric case (for $r>1/2$,
only the LM phase exists). This is different in the asymmetric case where
transition line extends up to $r\to \infty$.
Particle-hole symmetry also influences the physical properties of the
various fixed points. The symmetric SC fixed point, for example,
shows a residual magnetic moment of $\chi_{\rm imp}=r/(8k_{\rm B}T)$
 and a residual
entropy of $S_{\rm imp}=2rk_{\rm B}\ln 2$, whereas both values are zero in the
asymmetric SC fixed point. The appearance of unstable fixed points is
particularly complex in the asymmetric case, see, for example, the
schematic flow diagrams of Fig.~16 in \textcite{Gonzalez:1998}.

The impurity spectral function of the symmetric soft-gap Anderson model
was first investigated in \textcite{Bulla:1997a}: the spectral function
shows a divergence $A(\omega)\propto|\omega|^{-r}$ for both the
SC and quantum critical phases whereas it goes as 
$A(\omega)\propto|\omega|^{r}$ in the LM phase (for the behavior in
the asymmetric case, see the discussion in Sec.~\ref{subsubsec:sc}).

In the symmetric SC phase, the product $F(\omega)=c|\omega|^{r}A(\omega)$
(where the prefactor cancels the divergence in the spectral function)
contains a generalized Kondo resonance at the Fermi level with a
pinning of $F(\omega=0)$ (for a properly chosen constant $c$)
and a width that goes to zero upon approaching the quantum
critical point. This feature, together with the scaling properties
and the low-energy asymptotics has been discussed in detail in
\textcite{Bulla:2000a}, based on both the results from NRG and 
from the local moment approach (also described earlier in 
\textcite{Logan:2000}).

Dynamical properties at the quantum critical point are particularly
interesting: \textcite{Ingersent:2002} have shown that the dynamical
susceptibility at the critical point exhibits $\omega/T$-scaling with
a fractional exponent, similar to the locally critical behavior
in the heavy fermion systems mentioned above. This result also
implies that the critical fixed point is interacting, in contrast
to the stable fixed points (SC and LM) which both can
be composed of non-interacting
single-particle excitations.

\begin{figure}[htb]
\centerline{
  \includegraphics*[width=3.0in]{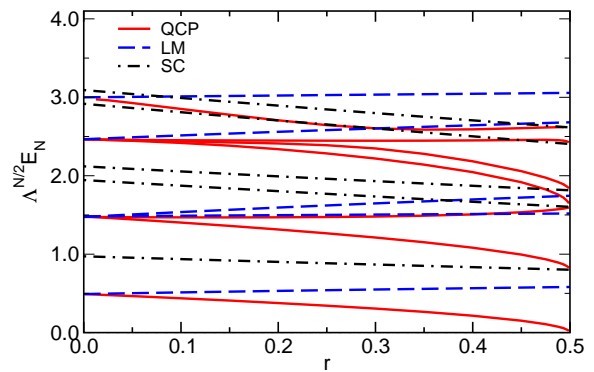}}
\caption{Dependence of the many-particle spectra for the three
            fixed points of the particle-hole 
            symmetric soft-gap Anderson model on
            the exponent $r$: SC (dot-dashed lines),
            LM (dashed lines),
            and the (symmetric) quantum critical point (solid lines).
(Figure adapted from \textcite{Lee:2005}).}
\label{fig:soft-gap}
\end{figure}

The interacting fixed point of the symmetric soft-gap model has been 
further analyzed in \textcite{Lee:2005}. The general idea of this
work can be best explained with Fig.~\ref{fig:soft-gap} 
which shows the dependence
of the many-particle spectra for the various fixed points on
the exponent $r$.

For the limits $r\to 0$ and $r\to 1/2$, the many-particle
spectra of the quantum critical point approach those of the LM
and SC fixed points, respectively. The deviations and splittings of the
spectra at the quantum critical point
close to these limits can then be understood from a
proper perturbational analysis using suitable marginal operators.
Information on these operators can be extracted from epsilon-expansion
techniques, as shown in \textcite{Lee:2005}.

The case of negative exponents in the hybridization function,
$\Delta(\omega)\propto|\omega|^r$ with $-1<r<0$, where the
soft-gap turns into a divergence at the Fermi level, has been
analyzed in \textcite{Vojta:2002b} in the context of the Kondo
model with both ferromagnetic and antiferromagnetic values of $J$.
The behavior of this class of models turns out to be rather
complex, see the schematic flow diagrams of Fig.~1 in this work.
A remarkable feature here is the appearance of a {\em stable} intermediate
coupling fixed point with universal properties corresponding to a 
fractional ground-state spin.

The case of a hard gap in the hybridization function, that is
$\Delta(\omega)=0$ within a certain gap region around the
Fermi level, can be viewed as the $r\to\infty$-limit of the
soft-gap case, provided the powerlaw is restricted to 
the gap region $|\omega|\le E_{\rm g}/2$, with $E_{\rm g}$
the width of the gap. 

From a technical point of view, two
different strategies have been developed to apply the NRG to the 
hard-gap case. \textcite{Takegahara:1992} and
\textcite{Takegahara:1993} considered the case of a small
but finite value of $\Delta(\omega)$ in the gap region,
$\Delta(\omega)=\bar{\Delta}$ for $|\omega|\le E_{\rm g}/2$, and based 
their conclusions on the extrapolation $\bar{\Delta}\to 0$.
In this approach, the standard NRG for non-constant hybridization
functions as described in Sec.~\ref{sec:nrg-intro} can be applied.

If, on the other hand, the value of $\bar{\Delta}$ is set to zero from
the outset, the NRG approach has to be modified. As discussed in
\textcite{Chen:1998}, the logarithmic discretization of a
$\Delta(\omega)$ with a hard gap results in a discretized model
which maps onto a chain with a {\em finite} number of sites, $M$,
with $E_{\rm g}$ of the order of $\Lambda^{-M}$. The iterative
diagonalization then has to be terminated at site $M$. Thermodynamic
properties at temperatures $T<E_{\rm g}$ can nevertheless be
computed using the Hamiltonian of the final iteration (see
\textcite{Chen:1998} where also a variety of correlations functions 
have been calculated for both the Kondo and the Anderson model
with a hard gap).

Certain features of the soft-gap case with finite $r$ are also visible
in the fully gapped case. As expected from the discussion above,
there is no transition in the particle-hole symmetric case, but a
transition exists as soon as one is  moving away from  
particle-hole symmetry 
\cite{Takegahara:1992,Ingersent:1996,Chen:1998}.
This transition turns out to be of first order.

\subsubsection{Kondo effect in superconductors}
\label{subsubsec:sc}

Let us now consider magnetic impurities in superconducting hosts.
In this case,
the screening of the magnetic moments competes with
Cooper pair formation of the conduction electrons. We therefore
expect a quantum phase transition from a screened phase to
a local moment phase upon increasing the value of the superconducting
gap, $\Delta$, similar to the phase transitions in the soft-gap
(and hard-gap) impurity models discussed in 
Sec.~\ref{subsubsec:loc-crit}. In fact, a relation between 
impurity models in superconductors and those in metallic hosts
with a soft or hard gap can be established as discussed below.

\begin{figure}[htb]
\centerline{
  \includegraphics*[width=3.2in]{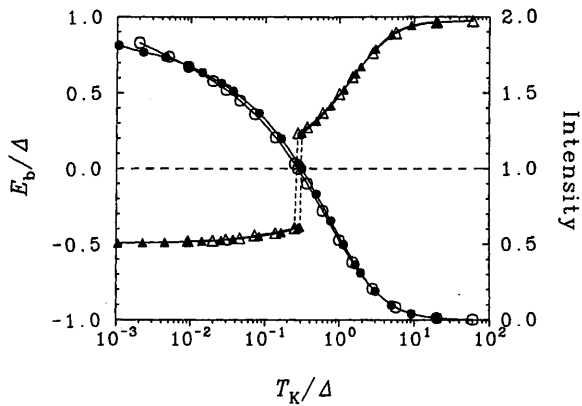}}
\caption{Position $E_{\rm b}$ and weight (intensity) 
of the localized excited state as a 
function of $T_{\rm K}/\Delta$ for the Kondo model
in an $s$-wave superconductor. 
At $T_{\rm K}/\Delta\approx 0.2$,
the position changes its sign and the weight jumps by a factor 
of 2, see also Fig.~2A in \textcite{Sakai:1993b}.}
\label{fig:sc1}
\end{figure}

The first applications of the NRG to magnetic impurities in
superconductors focused on the $s$-wave case 
\cite{Satori:1992,Sakai:1993b}, with the standard Kondo Hamiltonian
Eq.~(\ref{eq:NChannelKondo}) supplemented by the BCS pairing interaction
\begin{displaymath}
   -\Delta \sum_k 
   \left(
          c^\dagger_{k\uparrow}c^\dagger_{-k\downarrow} +
          c_{-k\downarrow}c_{k\uparrow}
   \right) \ .
\end{displaymath}
Several strategies have been developed to transform the Hamiltonian
including the BCS-term onto a semi-infinite chain which can then
be diagonalized iteratively in the usual way.
\textcite{Satori:1992} performed a sequence of transformations,
including a Bogoliubov and a particular particle-hole transformation,
to map the original model onto a Hamiltonian which conserves
particle number (this is somewhat easier for the numerical
implementation though not absolutely necessary). The
approach in \textcite{Sakai:1993b} leads to the same Hamiltonian,
the difference here is that the Bogoliubov transformation is performed
{\em before} the logarithmic discretization. In both cases, the
semi-infinite chain contains a staggered potential of the form
\begin{displaymath}
   -\Delta \sum_{n=0}^\infty (-1)^n
     \left(
            c^\dagger_{n\uparrow}c_{n\uparrow} +
            c^\dagger_{n\downarrow}c_{n\downarrow}  \right) \ .
\end{displaymath}
This term does not fall off exponentially as the other terms in the
chain-Hamiltonian so that the NRG-iterations should be terminated
a few steps after the characteristic scale $\omega_N$ of the
chain Hamiltonian $H_N$ has reached the superconducting gap $\Delta$.
This procedure still allows to access the properties of the localized
excited state within the energy gap whose
position and weight can now be determined in the full parameter space
(in contrast to previous investigations, see the references in
\textcite{Satori:1992}). Figure \ref{fig:sc1}
 shows position and weight of the localized
excited state as a function of $T_{\rm K}/\Delta$ ($T_{\rm K}$
is determined from the corresponding Kondo model with
$\Delta=0$).  The position changes its sign when 
$T_{\rm K}$ is of the order of $\Delta$ (the precise value depends on 
the model), corresponding to a change of the 
ground state from doublet for small $T_{\rm K}/\Delta$ to singlet
for large $T_{\rm K}/\Delta$. This quantum phase transition can be
characterized as a level crossing transition (see Fig.~5 in
\textcite{Satori:1992}) and is not connected to quantum critical
behavior.

These studies of impurities in $s$-wave superconductors have been later
extended to more complex impurity models.
\textcite{Yoshioka:1998} investigated the case of an anisotropic
interaction between impurity and conduction electron spin,
with basically the same NRG approach as in \textcite{Sakai:1993b}.
The phase diagram of this model turns out to be much more
complex than the one for the isotropic case. For example, {\em two}
localized excited states with different energies appear in certain regions
of the parameter space.

\textcite{Yoshioka:2000} considered the Anderson version of the
impurity model
with coupling to an $s$-wave superconductor. From a technical point of view,
this case is different to the corresponding Kondo model since
the sequence of transformations used in, for example,  
\textcite{Sakai:1993b} now produce an extra impurity term of the
form $\delta ( f_\uparrow^\dagger f_\downarrow^\dagger + h.c. )$,
so that the whole Hamiltonian does no longer conserve charge
(note that the parameter $\delta$ is zero for the
particle-hole symmetric case). The results for the Kondo regime
of this model are, as expected, the same as those obtained
previously, but the approach of \textcite{Yoshioka:2000}
also allows to study other parameter ranges of the model, 
such as the mixed valent regime.

It is important to note here that the final Hamiltonian used
in \textcite{Satori:1992} and \textcite{Sakai:1993b} for the
NRG iteration is the same as the one for an impurity in a
non-superconducting host with a gapped density of states
(which corresponds to the quasiparticle density of states
of the superconductor). In addition, the sequence of transformations
also generates a potential scattering term. In the light of the
results for the hard-gap impurity models 
(Sec.~\ref{subsubsec:loc-crit}) this potential scattering term
is essential to observe the quantum phase transition from
a screened to an unscreened phase.

The question now arises, whether the quasiparticle density of states
can be used as the sole bath characteristic (possibly supplemented
by a potential scattering term) in more general situations, such
as impurities in unconventional superconductors. Before we address
this issue, let us have a look at what happens when a similar sequence
of transformations as in the $s$-wave case is applied to impurity
models in $p$- or $d$-wave superconductors.

\textcite{Matsumoto:2001,Matsumoto:2002}
considered the Kondo model with a coupling of the impurity spin
to superconductors with $p_x + ip_y$ and $d_{x^2-y^2} + id_{xy}$
symmetry (with extensions to spin-polarized superconducting
states investigated in \textcite{Koga:2002a} and to 
$S\!=\!1$ impurities in \textcite{Koga:2002b}). The quasiparticle
density of states in these cases also shows a full gap, as for
$s$-wave superconductors, but the sequence of transformations now
results in a model with a coupling of the impurity to
{\em two} angular momenta of the conduction electrons. NRG calculations
for this two-channel model give a ground state which is always a
spin doublet for arbitrary values of $T_{\rm K}/\Delta$, in
contrast to the $s$-wave case, and no level crossing is observed.
This is supported by calculations of the impurity susceptibility
which show that the effective magnetic moment is always finite,
although strongly reduced with increasing $T_{\rm K}/\Delta$
\cite{Matsumoto:2002}. The authors of this work argued that the
orbital dynamics of the Cooper pairs is responsible for the
ground state spin.

This interpretation has been questioned in \textcite{Fritz:2005},
where is was shown that, indeed, the local quasiparticle density
of states of the superconductor is the only necessary ingredient
in a number of cases, in particular for unconventional superconductors.
Applied to the model studied in \textcite{Matsumoto:2001}, this means
that the results of the NRG calculations for the effective two-channel
model can also be understood from a single-band calculation where
screening is absent for a hard-gap density of states and particle-hole
symmetry.

The results of \textcite{Fritz:2005} also have important consequences
for the study of impurities in unconventional superconductors with 
$d_{x^2-y^2}$ symmetry. In this case, the mappings which have been 
used for the models discussed above result in an impurity model with
coupling to infinitely many bands to which the NRG clearly cannot be 
applied. For certain geometries, however, it is sufficient to
consider only the quasiparticle density of states which, for a point-like
impurity, shows a soft-gap with exponent $r=1$.

This simplification has already been used earlier in
\textcite{Vojta:2002a} (at that time it has been argued to be a
reasonable approximation). The results of this work are therefore
both valid for the soft-gap Kondo model and for impurities in
$d$-wave superconductors. \textcite{Vojta:2002a} have motivated
their investigations with experimental results for non-magnetic
impurities in cuprate superconductors which have been seen to
generate magnetic moments. As discussed in this work, an effective
model for this problem then takes the form of a Kondo model
in a $d$-wave superconductor. Connections to experimental
results can indeed be made within this framework.
For example, the $T$-matrix $T(\omega)$ displays a very narrow peak
at finite frequencies with the peak energy corresponding to the
energy scale which vanishes at the quantum phase transition from
a screened to an unscreened moment. A very similar peak has been
observed in STM-experiments.

This work has been later extended in \textcite{Vojta:2002c}, where
the effects of local and global magnetic fields have been investigated.
For the case of a local field $h_{\rm loc}$, the quantum phase transition for
zero field persists for $h_{\rm loc}\ne 0$, but for a global field,
the quantum phase transition turns into a sharp crossover since the
global field induces a finite spectral weight at the Fermi level.

The investigations described so far are mainly applicable to impurities
in the bulk or on the surface of a superconducting host. A different
geometry is realized in quantum dot systems 
(see Sec.~\ref{subsubsec:nano}). For superconducting
leads, such a setup introduces a new control parameter to the problem,
that is the phase difference, $\Phi = \Phi_{\rm L} - \Phi_{\rm R}$,
between the phases of the two superconducting leads. The resulting
Josephson current, in particular the transition from $0$- to
$\pi$-junction behavior, has been studied in detail in
\textcite{Choi:2004} and \textcite{Oguri:2004}.

\textcite{Choi:2004} investigated various static and dynamic properties
for this geometry with identical $s$-wave superconductors as the
two leads. For zero phase difference, $\Phi = 0$, the local pairing 
correlation shows a sign change at $T_{\rm K}/\Delta \approx 0.42$.
Physically, this is connected to the same quantum phase transition
as described above since for $\Phi = 0$ and identical leads the model
can be mapped onto the same model as discussed in
\textcite{Satori:1992}.
For finite phase difference (or for non-identical leads) the system
remains a two-channel problem and the NRG analysis is more complicated.
Nevertheless, detailed information on ground state properties such as
the single-particle excitation spectrum have been obtained in
\textcite{Choi:2004} and interpreted as a phase-dependent formation of 
Andreev bound states. 

In \textcite{Oguri:2004}, the Hamiltonian  of an Anderson impurity coupling
to two superconducting leads has been considerably simplified by studying
the limit $|\Delta_{\rm L}| \gg |\Delta_{\rm R}|$ in which the
model can be mapped exactly onto a single-channel one with an extra
superconducting gap on the impurity. Results for this limit show
that the phase difference changes both the energy and the wave function
of the bound state. In particular, the phase difference appears to
work against the screening of the local moment.

  \subsection{Orbital effects}
  \label{subsec:orbital}

\subsubsection{Multi-orbital Anderson model}
\newcommand{\rem}[1]{}
The physics of the Kondo effect requires the existence of local 
magnetic moments, as realized, for example, in systems with
open $d$ or $f$ shells, such as transition metal or rare-earth
impurities in non-magnetic host metals. For such systems, the local
Coulomb correlations and Hund's exchange determine the electronic 
structure of the impurity and they usually give rise to finite spin
and orbital magnetic moments. Thus, a realistic description 
of such impurities in solids, requires taking both spin and orbital magnetic
moments into account. The same is true for the compounds of 
transition metal, rare-earth and actinide elements, where the
interplay of orbital and spin degrees of freedom gives rise to
very rich phase diagrams \cite{Imada:1998}. Among the methods to
theoretically study the properties of these materials, the dynamical 
mean-field theory (see Sec.~\ref{sec:dmft}) has become a standard 
approach. Since in this approach one ends up with an
effective quantum impurity problem which retains the full local orbital and
spin structure of the original lattice system, the development 
of a reliable method to solve quantum impurity models with orbital and
spin degrees of freedom is of crucial importance.

In this section we, therefore, discuss the application of the NRG to 
situations where orbital and spin degrees of freedom are both present. 
Some orbital effects in quantum dots have been discussed in 
Sec.~\ref{subsubsec:nano} and will be discussed further below.
The appropriate model is again a suitable extension of the
single-impurity Anderson model (\ref{eq:siam}) and is given by
\begin{eqnarray}
  \label{eq:mosiam}
  H 
  &=&
  \sum_{k}\sum_{mm'\sigma}\epsilon_{k\sigma}^{mm'}
  c^{\dagger}_{km\sigma}c^{\phantom{\dagger}}_{km'\sigma}\nonumber\\
  &&
   + \sum_{m\sigma}\varepsilon_{m\sigma}n_{m\sigma}^f
  +\frac{U}{2}\sum_{m\sigma}n_{m\sigma}^fn_{m\bar{\sigma}}^f\nonumber\\
  &&
  +\frac{2U'-J}{4}\sum_{m\ne m'}\sum_{\sigma\sigma'}
  n_{im\sigma}^fn_{im'\sigma'}^f\nonumber\\
  &&
  -J\sum_{m\ne m'}\vec S_m^f\cdot\vec S_{m'}^f\nonumber\\
  &&
  -\frac{J}{2}
  \sum_{m\ne m'}\sum_\sigma 
  f_{m\sigma}^\dagger
  f_{m\bar\sigma}^\dagger 
  f_{m'\bar\sigma}^{\phantom{\dagger}}
  f_{m'\sigma}^{\phantom{\dagger}}\nonumber\\
  &&
  +\frac{1}{\sqrt{N}}\sum_{k}\sum_{mm'\sigma}
  \left(
    V_{k\sigma}^{mm'}c_{km\sigma}^\dagger 
    f_{m'\sigma}^{\phantom{\dagger}}
  +\mbox{h.c.}
  \right)\ , \nonumber\\
\end{eqnarray}
where $m$ labels the orbital degrees of freedom, and
$n_{m\sigma}^f=f^\dagger_{m\sigma}f^{\phantom{\dagger}}_{m\sigma}$, with
$\vec S_m^f =
\frac{1}{2}\sum\limits_{\alpha\beta}f^\dagger_{m\alpha}\vec{\sigma}_{\alpha\beta}f^{\phantom{\dagger}}_{m\beta}$.
In addition to the intra-orbital Coulomb term $U$ also occurring in
(\ref{eq:siam}), the following interaction terms are present now: An
inter-orbital Coulomb repulsion $U'$ and an exchange term $J$. The
exchange term we have split in accordance with standard notation \cite{Imada:1998}
into a Heisenberg-like spin-exchange term (Hund's coupling) and an
orbital exchange term. To account for the proper combination
of operators in the general exchange contribution, an additional part
proportional to $J$ appears in the inter-orbital Coulomb term. For
free atoms, rotational invariance usually imposes $U'=U-2J$ as
constraint for the different Coulomb parameters \cite{Imada:1998}.
Further modifications to the model (\ref{eq:mosiam}) can be made, to take
into account, for example, spin-orbit and crystal-field effects.

The mapping of (\ref{eq:mosiam}) onto a linear-chain model, see
Eq.~(\ref{eq:H_si}) of Sec.~\ref{sec:nrg-intro}, clearly leads 
to $m$ semi-infinite conduction chains coupled to
the local Hamiltonian, which in turn means that at each step 
of the iterative diagonalization, the Hilbert space increases
by a factor $4^m$. 
For large $m$, this exponential increase in the number of
states makes the NRG truncation scheme useless,
because the number of states one can keep is way too small to allow for a
reasonable accuracy. Thus, calculations for the model
(\ref{eq:mosiam}) involving a full $d$ shell ($m=5$),
or even a full $f$ shell ($m=7$), and taking into account all interactions
seem to be impossible.  

In practice, however, one is typically not interested in rotationally invariant
situations, as described by (\ref{eq:mosiam}), but in situations where 
the impurity is embedded in the crystalline environment of a solid.
The reduced point-group symmetry due to the crystalline electric field, 
then leads to a splitting of the orbital degeneracy. The energy associated with this
crystal-field splitting can be much larger than the temperatures one
is interested in experiments, for example in $3d$ transition
metals. Furthermore, the local Coulomb interaction can lead to a
localization of electrons in the lower crystal-field multiplets, as
happens, for example, in the case of manganese in a cubic environment
\cite{Imada:1998}. In this case, these states form a localized
spin according to Hund's rules. For manganese, for example, this
results in a high-spin state ($S=3/2$) of the threefold degenerate
$t_{2g}$ orbitals, which couples ferromagnetically to the twofold
degenerate $e_g$ electrons. Thus, the actual number of relevant
orbitals, and thus the number of semi-infinite chains coupling to the
local Hamiltonian, may be considerably reduced. Similar effects can be observed
in the higher rare-earth elements, for example in gadolinium.

In case the local point-group symmetry is reduced sufficiently, one may, in
fact, be left with a localized spin $S$ coupled to a single spin-degenerate,
correlated orbital hybridizing with conduction states. This Kondo-Anderson
model is given by (\ref{eq:siam}) supplemented with the ferromagnetic exchange
term $-J_{H}\vec{S}\cdot \vec{s}_{d}$, with $\vec{s}_{d}$ the spin-density of 
the correlated level in the Anderson model. Such a Kondo-Anderson
model was studied by \textcite{Sakai:1996} and
\textcite{Peters:2006} and shown to exhibit different types of
screening, ranging from conventional Kondo screening to two-stage
screening and local singlet formation or two-channel Kondo effect.

\subsubsection{NRG calculations -- an overview}
A first serious attempt to study effects of true orbital degeneracy with NRG can be found in
\textcite{Sakai:1989}. However, the authors did not study the full
Hamiltonian (\ref{eq:mosiam}), but a $SU(N)$ version of it, 
using values of $N$ ranging from two (i.e.\ the
standard single-impurity Anderson model 
(\ref{eq:siam})) to five, representative
for rare-earth ions like samarium or thulium in solids under the influence
of a crystalline field \cite{Shimizu:1990,Allub:1995}. Since the $SU(N)$
model has a large degeneracy of the individual levels, it allows for a
considerable reduction of the sizes of the individual Hilbert spaces
in the diagonalization. This enabled the authors to use the NRG to
calculate physical properties including dynamical quantities and 
to study, for example, the development of the Kondo temperature or the
behavior of the Abrikosov-Suhl resonance as function of degeneracy
\cite{Sakai:1989,Shimizu:1990}. Similar investigations for the model
(\ref{eq:mosiam}) with fixed orbital degeneracy $m=2$ in the presence
of a magnetic field were presented in the work by
\textcite{Zhuravlev:2004}, a comparison with STM experiments for
Cr(001) surface states by \cite{Kolesnychenko:2005}, and a
detailed study of the dependence of the low-energy properties of the
multi-orbital Anderson model for $m=2$ with $J>0$ by
\textcite{Pruschke:2005}.
Furthermore, from the fixed-point level structure, interesting
information about quantities like ``residual interactions'' in the
heavy Fermi liquid state can be extracted \cite{Hattori:2005}. In
particular for an $f^2$ ground state -- as possibly realized in
uranium compounds -- a subtle enhancement of interorbital interactions
can be observed \cite{Hattori:2005}, which can lead to superconducting
correlations in a triplet channel when used as effective interaction
in a model for heavy-fermion superconductivity.

While these studies deal with the conventional Kondo effect in
multi-orbital models,
it was noted \cite{Cox:1998} that for higher rare-earth and actinide
elements the orbital structure in connection with spin-orbit coupling
and crystal-field effects can
result in an orbital multiplet structure that leads to the
two-channel Kondo effect (see Sec.~\ref{subsec:two-channel}). 
Multi-orbital models of that type were
studied by \textcite{Sakai:1996}, \textcite{Shimizu:1998}, \textcite{Shimizu:1999b,Shimizu:1999},
\textcite{Koga:2000} and \textcite{Hattori:2005b}, covering a wide
range of aspects possibly realized in actinide heavy-fermion
systems. The authors could identify parameter regimes where non-Fermi
liquid properties related to the two-channel Kondo effect can be
observed and could, in addition, identify relevant symmetry breakings like
crystal-field splittings or external fields that eventually lead to
conventional Kondo physics below a temperature scale connected to the
energy scale of the symmetry breaking. 

As mentioned in the introductory remarks of this section, 
the crystal-field splitting
is usually much larger than the relevant low-energy scales. 
However, this does not need to be true in general. Besides
uranium-based compounds \cite{Kusunose:2005}, a possible example where the
crystal-field splitting can actually be of the order of the Kondo
scale is Ce$_{1-x}$La$_{x}$Ni$_{9}$Ge$_4$ \cite{Scheidt:2005}. For
higher temperatures, the ground state seems to be a quadruplet, i.e.\
it can be described by a multi-orbital Anderson model with $m=2$. The
states building this quadruplet are obtained from spin-orbit coupled
$f$ states, which results in different $g$-factors for its
members. Interestingly, these different $g$-factors in connection with
the small crystal-field splitting can lead to a behavior, where
specific heat and susceptibility seem to have different low-energy
scales \cite{Scheidt:2005}. However, this discrepancy can be resolved
by observing that the difference in $g$-factors leads to a
``protracted'' screening behavior for the specific heat in NRG
calculations, while the system as a whole has only one, but strongly
reduced Kondo temperature. In addition, NRG results for the low-energy
spin dynamics show an anomalous energy dependence, which again is
related to the difference in the $g$-factors entering the local
susceptibility \cite{Anders:2006a}. 

Conventionally, the Hund's exchange $J$ appearing in (\ref{eq:mosiam})
is positive, i.e.\ mediating a ferromagnetic interaction leading to
Hund's first rule. However, there may be circumstances, for example
a coupling to vibrational modes, Jahn-Teller distortions or crystal-field induced anisotropies
\cite{DeLeo:2004}, which can lead to an effective $J<0$, i.e.\ {\em
  antiferromagnetic} exchange. In this case we encounter a situation
similar to the multi-impurity problem (see section
\ref{subsubsec:mult-imp}), where the exchange was generated by the
RKKY effect. Under special conditions, an antiferromagnetic exchange
then could lead to a quantum phase transition between Kondo screening and
non-local singlet formation. Consequently, one may expect a similar
transition for the multi-orbital model, too, when one varies $J$ from
the ferromagnetic to the antiferromagnetic regime. Such a model was
studied by \textcite{Fabrizio:2003} and, with an additional 
single-ion anisotropy, by \textcite{DeLeo:2004} for two orbitals,
i.e.\ $m=2$. The model indeed shows the anticipated quantum phase
transition, which in this case is driven by the competition between
the local antiferromagnetic exchange coupling and the hybridization to
the band states. Furthermore, one can study the development of 
the spectral function across this transition \cite{DeLeo:2004}. One
finds that the impurity spectral function on
the Kondo screened side of this transition shows a narrow
Kondo peak on top of a broader resonance. As has also been observed by
\textcite{Pruschke:2005} and \textcite{Peters:2006} this broad resonance
is related to the exchange splitting $J$. The narrow peak
transforms into a pseudogap on the unscreened side of
the transition.

\textcite{DeLeo:2005} have demonstrated that 
NRG calculations are possible even for $m=3$, using the
symmetries of the model to reduce the size of the Hilbert space blocks. 
They studied
a realistic model for a doped C$_{60}$ molecule, taking into account
the orbitally threefold degenerate $t_{2u}$ lowest unoccupied
molecular orbitals. Again, coupling to vibrational modes can lead to
Hund's coupling with negative sign. In this regime, one observes
non-Fermi liquid behavior for half-filling $n=3$, associated with a
three-channel, $S=1$ overscreened Kondo model. Interestingly, the
critical susceptibilities associated with this non-Fermi liquid appear
to be a pairing in the spin and orbital singlet channel
\cite{DeLeo:2005}. Using conformal field theory, the authors could
deduce the residual entropy as 
$$
S(T=0)=\frac{1}{2}\ln\left[\frac{\sqrt{5}+1}{\sqrt{5}-1}\right] \ ,
$$
and also corresponding fractional values for the local 
spectral function $\rho_f(0)$
at the Fermi energy, which leads to non-unitary values in the
conductance, and a non-integer power law for $\rho_f(\omega)-\rho_f(0)$.
Away from half filling a quantum phase
transition between Kondo screening and local singlet formation as
function of filling occurs \cite{DeLeo:2005}. 
Multi-orbital models also arise in the context of quantum dots as described
in Sec.~\ref{subsubsec:nano} and Sec.~\ref{subsubsec:mult-imp}.

\subsubsection{Selected results on low-energy properties}
In the following we shall present some selected results for the properties
of the multi-orbital Anderson model (\ref{eq:mosiam}) with orbital
degeneracy $m=2$.

\begin{figure}[htb]
\centerline{
  \includegraphics*[width=3.0in,clip]{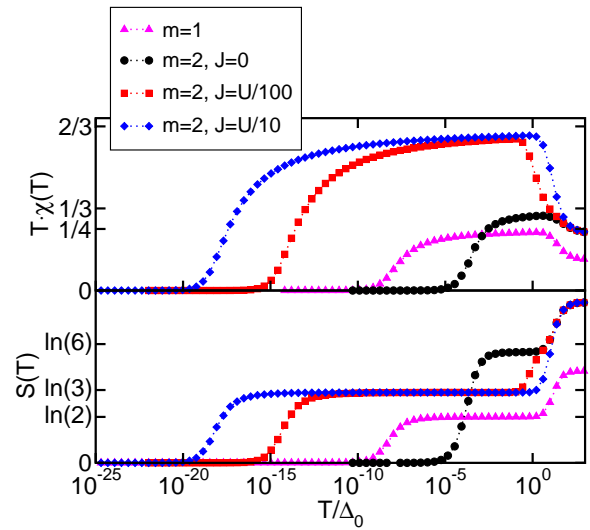}}
\caption{Effective local moment $T\cdot\chi(T)$ (upper panel) and
  entropy (lower panel) for a two-orbital impurity Anderson
  model. Model parameters are $U/\Delta_0=16\pi$ at particle-hole
  symmetry. For comparison the results for one orbital are included (triangles).}
\label{fig:mos_th}
\end{figure}

\paragraph{Effect of Hund's coupling}
As a first example we want to discuss the influence of Hund's exchange
$J$ on the low-energy properties of the model Eq.~(\ref{eq:mosiam}).
We consider only
$J\ge0$, i.e.\ the usual atomic ferromagnetic exchange, and in
addition use the constraint $U'=U-2J$. For the conduction electrons we
assume a band with a flat density of states
$\rho_c^{(0)}(\epsilon)=N_F\Theta(D-|\epsilon|)$ and use
$\Delta_0=\pi V^2N_F$ as energy scale. Results for thermodynamic
quantities are shown in Fig.~\ref{fig:mos_th}.
The calculations are for $U/\Delta_0=16\pi$ and particle-hole
symmetry. NRG parameters are $\Lambda=5$ and $2500$ retained states per
iteration. The triangles are results from a calculation with $m=1$ for the
same parameters. For $J=0$ (circles) we find the behavior expected for
a $SU(4)$ symmetry, i.e.\ an energy scale, $T_{\rm K}^m$, that is related 
to the Kondo scale at $m=1$,  $T_{\rm K}^{m=1}$, by
$T_{\rm K}^{m}=(T_{\rm K}^{1})^{1/m}$. However, even a small
Hund coupling $J=U/100$ leads to a dramatic reduction of $T_{\rm K}$,
which increases with increasing $J$. Let us note here, that for $J>0$
the effect of the orbital exchange term in 
(\ref{eq:mosiam}) is negligible, i.e.\
the results are indistinguishable if one does the calculation with and
without this contribution.
Very often, the orbital exchange term is neglected in 
theoretical studies of
transition metal compounds \cite{Imada:1998}, an approximation
which is supported by the above result.

\paragraph{Crystal-field effects} 
Recently, experiments showing unusual specific heat, magnetic
susceptibility and resistivity data for  Ce$_{1-x}$La$_{x}$Ni$_{9}$Ge$_4$
have drawn a lot of attention,
because this material has the ``largest ever recorded
value of the electronic specific heat at low temperature'' \cite{Killer:2004} 
of $\gamma(T)= \Delta C/T \approx 5 {\rm JK}^{-2} {\rm
 mol}^{-1}$. While the $\gamma$ coefficient continues to rise at the
lowest  experimentally accessible temperature, the magnetic
susceptibility tends to saturate at low temperatures.

One possible scenario \cite{Scheidt:2005,Anders:2006a} to account for the 
behavior of  Ce$_{1-x}$La$_{x}$Ni$_{9}$Ge$_4$ is a competition of
Kondo and crystal-field effects which leads to a crossover regime connecting
incoherent spin scattering at high temperatures and a conventional
strong-coupling Fermi liquid regime at temperatures much lower than
the experimentally accessible 30mK.
The Hund's rule ground state of Ce$^{3+}$ with
$j=5/2$ is split in a tetragonal symmetry \cite{Killer:2004} in three
Kramers doublets. If the  crystalline electric field (CEF) parameters are
close to those of cubic symmetry, the two low-lying doublets
$\Gamma_7^{(1)}$ and $\Gamma_7^{(2)}$, originating from the splitting
of the lowest $\Gamma_8$ quartet, are well separated from the higher lying
$\Gamma_6$ doublet. Ignoring this $\Gamma_6$ doublet, we can discuss
two extreme limits. In a cubic environment, the  CEF splitting vanishes
and the low-temperature physics is
determined by an $SU(4)$ Anderson model. In a strongly tetragonally
distorted crystal, on the other hand, the crystal-field splitting of
the quartets is expected to be large. In this case, the 
low-temperature properties are determined by an $SU(2)$ Anderson model. 
If, however, the material parameters lie in the crossover regime where the
effective low-temperature scale $T^*$  is of the order of the crystal-field
splitting $\delta_{\rm CEF}=E_{\Gamma_7^{(2)}} -E_{\Gamma_7^{(1)}}$, then 
the excited doublet will have significant
weight in the ground state so that the total magnetic response
differs  from a simple $SU(4)$ Anderson model. 

Such a situation can be captured by a $SU(4)$ Anderson model with 
infinite $U$ 
whose Hamiltonian is given by \cite{Scheidt:2005}
\begin{eqnarray}
  \label{eq:siam-cef}
H&=&  \sum_{k\alpha} \epsilon_{k\alpha\sigma} 
c^\dagger_{k\alpha\sigma} c_{k\alpha\sigma} 
+ \sum_{\alpha\sigma} E_{\alpha\sigma}
|\alpha\sigma\rangle\langle\alpha\sigma| 
\\
&& +
\sum_{k\alpha\sigma} V_{\alpha\sigma}\left( 
|\alpha\sigma\rangle\langle0|  c_{k\alpha\sigma} 
+c_{k\alpha\sigma}^\dagger |0\rangle\langle\alpha\sigma|
\right)
\nonumber
\; ,
\end{eqnarray}
where $|\alpha\sigma\rangle$ represents the state
$\Gamma_7^{(\alpha)}$ with spin $\sigma$ and energy $E_{\alpha\sigma}$
on the Ce $4f$ shell, and $ c_{k\alpha\sigma}$ annihilates a corresponding
conduction electron state with energy $\epsilon_{k\alpha\sigma}$.
Note that locally only fluctuations between an
empty and a singly occupied Ce $4f$ shell are allowed.

While the entropy and specific heat for the model (\ref{eq:siam-cef})
can be calculated in the usual way, the Ce contribution to the
susceptibility \rem{specific heat} requires some more thought, because 
the spin-orbit coupled states $\Gamma_7^{(\alpha)}$ have different $g$-factors, 
which we label by $g_\alpha$. Thus, the total susceptibility is given by
\cite{Scheidt:2005}
\begin{eqnarray}
  \chi_{\rm imp} &=& \mu_B^2 \sum_\alpha g_\alpha^2 
\chi^{(\alpha)}_{\rm imp}\;\;.
\end{eqnarray}
While the $g$-factors are, in principle, determined by the CEF states of the multiplets, we
view them as adjustable parameters and fix them together with
$E_{\Gamma_7^{(\alpha)}}$ by comparing with experiment
\cite{Scheidt:2005}. 

\begin{figure}[tb]
 \centering
 \includegraphics*[width=3.0in,clip=true]{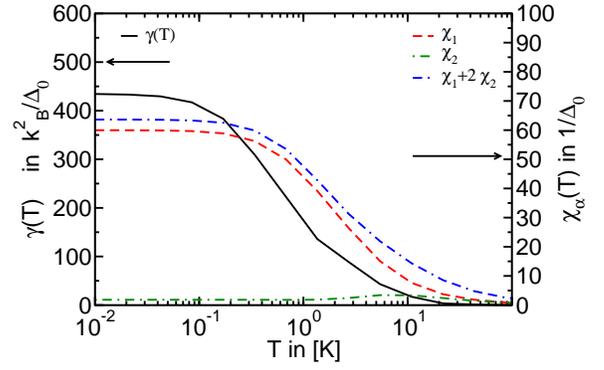}
 \caption{Comparison between $\gamma(T)=C(T)/T$  versus $T$ and the
   susceptibility contributions of the two doublets versus $T$ for
   $E_{\Gamma_7^{(1)}}/\Delta_0=-8.5$, $\delta_{\rm CEF}/\Delta_0=0.015$ and $g_2^2/g_1^2=2$. The 
   contribution of the lower doublet, $\chi_1$ is much larger than the
   one of the upper doublet, $\chi_2$. NRG parameters are 
   $\Lambda=4$ and 1500 states kept in each step.}
   \label{fig:dc-sus}
\end{figure}

The comparison between the temperature dependence of $\gamma(T)$ and
$\chi(T)$ is shown  in Fig.\ \ref{fig:dc-sus} assuming a ratio
of $g_2^2/g_1^2=2$ for a good fit to the experimental
data \cite{Scheidt:2005}. The ground state doublet
dominates the magnetic response at low temperature and tends to
saturate at temperatures higher than the $\gamma$-coefficient,
consistent with the experiments \cite{Killer:2004}. We 
find this behavior only for CEF-splittings 
$\delta_{\rm CEF}\approx T^*(\delta_{\rm CEF})$
while for much larger or much smaller values $\chi(T)$ and $\gamma(T)$
saturate simultaneously.


  \subsection{Bosonic degrees of freedom and dissipation}
  \label{subsec:bosons}

In the models discussed so far, the bath consists of non-interacting
fermionic degrees of freedom while the impurity is either represented
by a fermion or a spin. This section deals with quantum impurity
systems involving {\em bosonic} degrees of freedom. We distinguish 
between models in which only a small number of bosonic
degrees of freedom couples to the impurity and models where 
the impurity couples to a bosonic bath (corresponding to an infinite
number of bosonic degrees of freedom).

As we will see below, the first case can be dealt with in the usual
scheme, provided the subsystem consisting of impurity and bosons
can be treated as a large impurity which is then coupled to the
fermions. The second class, however, requires a different set-up
for the NRG-procedure.

Let us first consider the so-called Anderson-Holstein model 
\cite{Hewson:2002} in which the impurity is linearly coupled
to a single bosonic degree of freedom (typically a phonon mode):
 
\begin{equation}
  H =  H_{\rm SIAM}
       +  \lambda (b^\dagger + b )
          \sum_\sigma  f^\dagger_{\sigma}
                             f_{\sigma}   + 
         \omega_0 b^\dagger  b \ ,
\label{eq:anderson-holstein}
\end{equation}
with $H_{\rm SIAM}$ the Hamiltonian of the single-impurity
Anderson model as in Eq.~(\ref{eq:siam}).
The coupling to the bosonic operators ($b^\dagger$ and $b$)
does not influence the mapping of the conduction electron
part of the Hamiltonian to a semi-infinite chain. This means that
the bosons enter the iterative diagonalization only in the very
first step in which the coupled impurity-boson subsystem 
has to be diagonalized. The fact that only a limited number of
bosonic states $n_{\rm b}$ can be taken into account in this diagonalization
imposes some restrictions on the parameters $\lambda$ (the electron-phonon
coupling strength) and $\omega_0$ (the frequency of the phonon mode).
As discussed in \textcite{Hewson:2002}, it should be sufficient
to include a number of 
$n_{\rm b}\approx 4\lambda^2/\omega_0^2$ bosonic states for the
initial diagonalization. With an
upper limit of $n_{\rm b}\approx 1000$ this means that the limit
$\omega_0 \to 0$ (with fixed $\lambda$) cannot be treated within this
setup.

Apart from this minor restriction, \textcite{Hewson:2002} showed that the
NRG (which is non-perturbative in both $\lambda$ and $U$) works
very well for such type of impurity models. In particular, both electron
and phonon spectral functions as well as dynamic charge and spin 
susceptibilities can be calculated with a high accuracy.

As discussed in detail in \textcite{Jeon:2003} and
\textcite{Choi:2003b}, the calculation of the phonon spectral function
needs some extra care and the authors introduced an improved 
method (as compared to \textcite{Hewson:2002}). The proper
calculation of the phonon spectral function is important to
discuss the softening of the phonon mode, see also the discussion
in Sec.~\ref{subsec:phonons} in the context of lattice models
with coupling to phonons.

The low-energy features of the model Eq.~(\ref{eq:anderson-holstein})
can be partly explained by an effective single-impurity Anderson
model in which the coupling to the phonons is
included in an effective interaction $U_{\rm eff}$. An explicit form
of this interaction can only be given in the limit $\omega_0\to \infty$:
$U_{\rm eff}=U-2\lambda^2/\omega_0$. Interestingly, \textcite{Hewson:2004}
have shown that an effective quasiparticle interaction can be defined
for {\em any} value of $\omega_0$. This can be accomplished 
with the renormalized perturbation theory by fitting the lowest
lying energy levels obtained in the NRG calculations to those
from a renormalized Anderson model.

These investigations represent a starting point for various applications
of the NRG to coupled electron-phonon system.
For the investigation of transport properties of single molecule
devices, for which the coupling to local phonons is a natural ingredient,
similar models as  Eq.~(\ref{eq:anderson-holstein}) have been investigated in
\textcite{Cornaglia:2004} and \textcite{Cornaglia:2005b}. 
Not only the coupling to 
the electron density as in Eq.~(\ref{eq:anderson-holstein}),
but also the change of the hybridization
between molecule and leads due to the phonons has been shown to be important
for the conductance properties \cite{Cornaglia:2005a}.

Different physical phenomena can be expected in multi-orbital systems
when the impurity degrees of freedom couple to Jahn-Teller phonons.
Such a model has been investigated in \textcite{Hotta:2005}
and it was argued that within this model a new mechanism of Kondo phenomena
with non-magnetic origin can be established.

Two different strategies have been developed to study impurity
models with a coupling to a bosonic bath, i.e.~a bosonic
environment with a continuous spectral density $J(\omega)$. 
Let us discuss these strategies in the context of the
spin-boson model
\begin{equation}
H=-\frac{\Delta}{2}\sigma_{x}+\frac{\epsilon}{2}\sigma_{z}+
\sum_{i} \omega_{i}
     a_{i}^{\dagger} a_{i}
+\frac{\sigma_{z}}{2} \sum_{i}
    \lambda_{i}( a_{i} + a_{i}^{\dagger} ) \ .
\label{eq:sbm}
\end{equation}
This model naturally arises in the description of quantum
dissipative systems \cite{Leggett:1987}.
The dynamics of the two-state system, represented by the Pauli matrices $\sigma_{x,z}$,
is governed by the competition between the tunneling term $\Delta$ and the friction
term $\lambda_{i}(a_{i}+a_{i}^{\dagger})$.
The $a_i$ constitute a bath of harmonic oscillators responsible for the damping,
characterized by the bath spectral function
\begin{equation}
    J\left( \omega \right)=\pi \sum_{i}
\lambda_{i}^{2} \delta\left( \omega -\omega_{i} \right) \,.
\label{eq:sbm-J}
\end{equation}
A standard parametrization of this spectral density is
\begin{equation}
  J(\omega) = 2\pi\, \alpha\, \omega_c^{1-s} \, \omega^s\,,~ 0<\omega<\omega_c\,,\ \ \ s>-1 \,.
\label{eq:J-power}
\end{equation}
The case $s=1$, known as Ohmic dissipation, allows a mapping of the spin-boson
model onto the anisotropic Kondo model (for the definition of the
Hamiltonian and the relation of its parameters to those of the 
spin-boson model,
see \textcite{Costi:1996b}). 
The first strategy is then to apply
the NRG to the anisotropic Kondo model and to treat the fermionic conduction
band in the usual way. Though restricted to the Ohmic case, such calculations
have been shown to give very accurate results for dynamic and thermodynamic
quantities of the corresponding spin-boson and related models, as briefly
discussed in the following.

The focus of \textcite{Costi:1996b} has been the equilibrium dynamics of the
spin-boson model, in particular the spin-spin correlation function, 
$\chi(\omega)=\langle\langle \sigma_{z};\sigma_{z}\rangle\rangle$, at
temperature $T=0$. An interesting finding of this study, is that the 
spin relaxation function, $\chi''(\omega)/\omega$,
exhibits a crossover from inelastic to quasielastic behavior as $\alpha$ exceeds
the value $1/3$ signalling the onset to incoherent dynamics which occurs at 
$\alpha\ge 1/2$ \cite{Leggett:1987}. 
The accuracy of this approach has been shown via the
comparison to exactly solvable limiting cases, such as the Toulouse point
$\alpha=0.5$, and via the generalized Shiba relation. The issue of
scaling and universality, concepts which are quite naturally connected to
renormalization group treatments of the Kondo problem, have been discussed
in the context of the spin-boson model in \textcite{Costi:1998}.
Universal scaling functions have been calculated for thermodynamic
quantities (the specific heat), the static susceptibility and the
spin-relaxation function. Scaling as a function of temperature
or frequency has been 
observed in the limit $\Delta\to 0$ for {\em fixed} coupling
strength $\alpha$, this means that the scaling functions turn out to depend
on the value of $\alpha$. This is illustrated, for example, in Fig.~2 in 
\textcite{Costi:1998} which shows the temperature dependence of the specific
heat for different $\alpha$. In particular, a signature distinguishing weakly 
dissipative from strongly dissipative systems is found in $\gamma(T)=C(T)/T$, which
is found to exhibit a peak for $\alpha<1/3$ but is monotonic in temperature 
for $\alpha\ge 1/3$. This, together with the above mentioned behavior in the
spin-relaxation function, is reminiscent of measurements on 
Ce$_{1-x}$La$_{x}$Al$_3$ \cite{Goremychkin:2002}, but the applicability of
an anisotropic Kondo model here is controversial \cite{Pietri:2001}. 
Specific heats have also been calculated for more complicated tunneling models,
such as the ionic tunneling model with a spinless fermionic bath 
\cite{Ferreira:2000}, which shows similar behavior to the Ohmic
spin boson model for $\alpha\le 1/4$, and to an extension  of this
including an assisted tunneling term \cite{Ramos:2006}. The spectral density
of the former has also been investigated \cite{Libero:1990b}.

The mapping of a bosonic bath to a fermionic one has been also exploited
for various other problems. \textcite{Costi:2003} have used the NRG
to calculate the entropy of entanglement for the spin-boson model, 
a quantity which measures the entanglement between the spin and the 
environment. Interestingly, the entanglement appears to be highest for
$\alpha\to 1^-$, where the system undergoes a quantum phase transition
from a delocalized to a localized phase.

The case of {\em two} bosonic baths which couple to different components
of the impurity spin operator, has been discussed in
\textcite{CastroNeto:2003} and \textcite{Novais:2005}. The bosonic
baths in these models can be mapped onto two independent fermionic
baths, and a generalization of the anisotropic Kondo model is obtained.
These models are of interest to study the effect of quantum frustration
of decoherence.

%
%
The second strategy to investigate impurity models with a coupling to
a bosonic bath does not rely on a mapping to a fermionic impurity
model. This approach -- which has been termed ``bosonic'' NRG --
was introduced in \textcite{Bulla:2003b} in the context of the 
spin-boson model (for full details see \cite{Bulla:2005}). 
Let us just mention briefly the main differences
to the standard (fermionic) NRG: the logarithmic discretization is now
directly performed for the bosonic bath (for the spin-boson model, the
bath spectral function $J(\omega)$ Eq.~(\ref{eq:sbm-J}) is discretized);
the subsequent mapping onto a semi-infinite chain is technically very 
similar to the fermionic case but the resulting tight-binding chain
is built up of bosonic sites. This gives rise to additional difficulties
in setting up the iterative diagonalization scheme because only a finite
number of bosonic states $N_{\rm b}$ can be taken into account when adding
one site to the chain. 
Furthermore, the set of $N_{\rm b}$ states should in general be 
optimized to give the best description of the lowest lying many-particle 
states, see the discussion in \textcite{Bulla:2005}.

\begin{figure}[htb]
\centerline{
  \includegraphics*[width=3.0in]{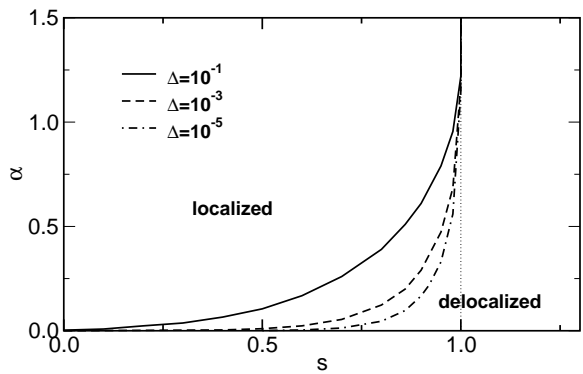}}
\caption{Phase diagram of the spin-boson model for $T=0$ calculated
with the bosonic NRG for various values of $\Delta$. A line of 
quantum critical points separates the delocalized from the localized
phase.
(Figure adapted from \textcite{Bulla:2003b}).}
\label{fig:sbm-pd}
\end{figure}

The first applications of the bosonic NRG focused on the
spin-boson model Eq.~(\ref{eq:sbm}), in particular the sub-Ohmic case
with exponents $0<s<1$ in the parametrization of the bath
spectral function $J(\omega)$ Eq.~(\ref{eq:J-power}). The sub-Ohmic case
does not allow for the mapping to a fermionic impurity model, in
contrast to the Ohmic case. Furthermore, the bosonic
NRG turns out to have certain advantages over other approaches
usually applied to the spin-boson model \cite{Leggett:1987}
as it is non-perturbative in both $\alpha$ and $\Delta$. As
an example of the success of the bosonic NRG we show in
Fig.~\ref{fig:sbm-pd} the $T=0$ phase diagram of the spin-boson
model with bias $\epsilon=0$ in the $\alpha$-$s$ plane for different
values of the tunneling amplitude $\Delta$.

The most remarkable feature of the phase diagram is the line of
quantum critical points for $0<s<1$ which terminates for $s\to 1^-$
in the Kosterlitz-Thouless transition of the Ohmic case. The 
critical exponents along this line have been discussed in detail in
\textcite{Vojta:2005} where it was shown that the exponents
fulfill certain hyperscaling relations.

There are still quite a number of open issues and possible applications
in the context of the sub-Ohmic spin-boson model which can be
investigated with the bosonic NRG, see for example \textcite{Tong:2005}.
In addition, the NRG can be generalized to more complex
impurity models with a coupling to a bosonic bath. Two recent
examples are the investigations of the Bose-Fermi Kondo model
\cite{Glossop:2005} and of models which are connected
to electron and exciton transfer phenomena \cite{Tornow:2006,Tornow:2006b}.

Technically the most challenging of these extensions is the study
of impurity models which couple to both fermionic and bosonic baths.
The first successful treatment of such a model, the Bose-Fermi Kondo model
with Ising-type coupling between spin and bosonic bath, has been
discussed in \textcite{Glossop:2005,Glossop:2006b}. 
Here the two baths are mapped on two
semi-infinite chains, one for the bosonic and one for the fermionic
degrees of freedom. Due to the competition between dissipation and
screening, quantum phase transitions occur which turn out to be
in the same universality classes as the transitions in the spin-boson
model. This can be understood through a mapping between the 
(Ising-)Bose-Fermi Kondo model and the spin-boson model where the
bath spectral function has both an Ohmic and a sub-Ohmic component
(the Ohmic part represents the coupling to the fermionic bath,
while the sub-Ohmic part is the same as the one in the original
model, see also \textcite{Li:2005}). On the other hand, the
(Ising-)Bose-Fermi Kondo model with an Ohmic bosonic bath
can also be mapped onto the anisotropic Kondo model, see
for example \textcite{Borda:2005}.

\section{Application to Lattice Models within DMFT}
\label{sec:dmft}

The application of the NRG is restricted to quantum impurity systems with the
impurity degree of freedom coupled to a {\em non-interacting} bath.
Therefore, the NRG cannot be directly applied to lattice models of
interacting particles, such as the Hubbard model (see Eq.\
(\ref{eq:hub}) in Sec.\
\ref{subsec:HubbardModel}). Early attempts to extend Wilson's concepts
to such models failed
\cite{Chui:1978,Bray:1979,Lee:1979}. The reason for this failure was
later found to be connected with boundary conditions between
``system'' and ``environment''
\cite{White:1992,Noack:2005},
and led to a novel scheme nowadays known as \emph{density-matrix
renormalization group} \cite{Schollwoeck:2005,Hallberg:2006}, which today is
a standard technique to study one-dimensional interacting quantum models.

There exists, however, an approximation for correlated lattice models,
where the interacting lattice problem is mapped onto an effective
quantum impurity model, for which the NRG can be applied. Underlying
this approach is the dynamical mean-field theory (DMFT). The DMFT has
its origin in the investigation of correlated lattice models in the
limit of infinite spatial 
dimensionality \cite{Metzner:1989}. A proper scaling of the hopping
matrix element $t$ in models such as the Hubbard model (\ref{eq:hub})
leads to a 
vanishing of all non-local self-energy sceleton diagrams. The resulting
purely local self-energy $\Sigma(z)$ can be identified with the
self-energy of an effective single-impurity
Anderson model.
In this sense we speak of a mapping of a lattice model onto an effective
quantum impurity model, typically the single-impurity Anderson model
as introduced in Eq.~(\ref{eq:siam}), 
supplemented by a self-consistency condition, which determines the
bath degrees of freedom of the effective quantum-impurity.
Since the technical details of the
DMFT are not the subject of this review, we refer the reader to
the review by \textcite{Georges:1996}.

To investigate lattice models in the DMFT we therefore need a technique
(analytical or numerical) to calculate the full frequency dependence
of the self-energy for a single-impurity Anderson model defined
by arbitrary input parameters ($\varepsilon_{\rm f}$, $U$, $T$, and
a manifestly energy-dependent hybridization function
$\Delta(\omega)$). There are many methods besides the NRG
available to calculate dynamic quantities for quantum impurity
models and we shall not give an overview here (for reviews see, 
for example, Sec.~VI in \textcite{Georges:1996}, Sec.~III in
\textcite{Maier:2005}, and \textcite{Bulla:2004}), but rather
concentrate on the application of the NRG method
to the Hubbard model (see Sec.~\ref{subsec:hubbard}), 
the periodic Anderson and Kondo lattice
models (see Sec.~\ref{subsec:pam}), 
and lattice models with coupling to phonons (see
Sec.~\ref{subsec:phonons}) within the DMFT approach.

Before we discuss the results obtained for those models, let us
comment on pecularities of the NRG when applied to
DMFT calculations. 
The DMFT self-consistency specifies at each iteration an input hybridization
function $\Delta(\omega)$, the form of which depends on the model under
investigation, its parameters and also on the history of the
previous DMFT-iterations. The frequency dependence of $\Delta(\omega)$
has to be taken into account within the logarithmic discretization scheme,
exactly as described in 
Sec.~\ref{sec:nrg-intro} and already employed in the NRG investigations
of the soft-gap Anderson model, see Sec.~\ref{subsubsec:loc-crit}.

Concerning the output, the quantity of interest is usually the self-energy
$\Sigma_{\rm AM}$ of the effective single-impurity Anderson
model, although in some cases, as for the Bethe lattice, 
the knowledge of the single-particle
Green function is sufficient for the DMFT iteration \cite{Georges:1996}.
It has proven advantageous to calculate,
within DMFT, the self-energy $\Sigma_{\rm AM}$ via
the ratio of a two-particle and a one-particle
Green function, see
Eq.~(\ref{eq:self-energy-expression}).
As discussed in 
\textcite{Bulla:1998} (see also Sec.~\ref{subsubsec:se+dm}) 
the calculation of the self-energy
via Eq.~(\ref{eq:self-energy-expression}) significantly improves the 
quality of the results.
This approach has been used in most NRG calculations within DMFT.

  \subsection{Hubbard model}
  \label{subsec:hubbard}

\label{subsec:HubbardModel}
The simplest model for correlated fermions on a lattice
is the single-band Hubbard model with the Hamiltonian
\begin{equation}
   H =  -t\sum_{<ij>\sigma} (c^\dagger_{i\sigma} c_{j\sigma} +
                   c^\dagger_{j\sigma} c_{i\sigma}) +
         U\sum_i c^\dagger_{i\uparrow} c_{i\uparrow}
            c^\dagger_{i\downarrow} c_{i\downarrow} \;.
\label{eq:hub}
\end{equation}
Consequently the first applications of the NRG within DMFT 
focused on this model; in particular on the Mott-transition,
which the Hubbard model displays in the half-filled paramagnetic
case. These investigations and further generalizations are
described in the following subsections.

\subsubsection{Mott metal-insulator transition}

Although the qualitative features of the Mott transition have been
correctly described very early in the development of the DMFT
(see the review by \textcite{Georges:1996}), the NRG helped to clarify
a number of conflicting statements in the literature (see
the discussion in \textcite{Bulla:1999} and \textcite{Bulla:2001}). 
The NRG method appears to be ideally suited to investigate
the Mott transition because (i) the transition occurs at interaction
strengths of the order of the bandwidth which requires the use of a 
non-perturbative
method, and (ii) at $T=0$ the Mott transition is characterized by a
vanishing energy scale, $T^\ast\to 0$, when approached from the
metallic side. Thus, a method is needed that is able to resolve arbitrarily
small energies close to the Fermi level.

The first investigation of the Mott transition with the NRG has been
performed by \textcite{Sakai:1994} 
(see also \textcite{Shimizu:1995}). These
calculations did not use an expression of the self-energy as in 
Eq.~(\ref{eq:self-energy-expression}),
but nevertheless a Mott transition and a hysteresis region 
have been observed, with critical
values very close to the ones reported later in \textcite{Bulla:1999}.

\begin{figure}[htb]
\centerline{
  \includegraphics*[width=3.0in]{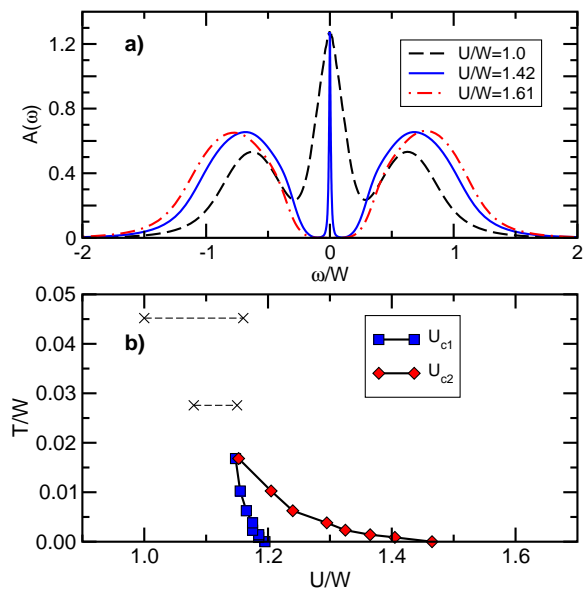}}
\caption{(a) Spectral functions for the half-filled Hubbard model
at $T=0$ for various values of $U$ (similar data as in Fig.~2
in \textcite{Bulla:1999}; (b) phase diagram for the Mott transition.
For a comparison with other methods see the corresponding
Fig.~9 in \textcite{Bulla:2001}.}
\label{fig:mott}
\end{figure}

A detailed discussion of the NRG calculations for the
Hubbard model is given in \textcite{Bulla:1999} for $T=0$
and in \textcite{Bulla:2001} for finite temperatures. The main results
are summarized in Fig.~\ref{fig:mott}:
Spectral functions calculated with NRG for the
half-filled Hubbard model in the paramagnetic regime for different
values of $U$ and $T=0$ are shown in Fig.~\ref{fig:mott}a.
Upon increasing $U$ from the metallic side, the typical 
three-peak structure forms, with upper and lower Hubbard peaks at
$\omega\approx \pm U/2$ and a central quasiparticle peak at
$\omega=0$. The width of this quasiparticle peak goes to zero
when approaching the transition from below, 
$U\nearrow U_{{\rm c},2}\approx 1.47W$, with $W$ the bandwidth
of the noninteracting model. Right at the transition,
when the quasiparticle peak has disappeared, an insulating solution
with a preformed gap is realized.

We should remind the reader at that point that the NRG results for dynamic
quantities have a certain fixed resolution on a {\em logarithmic}
scale (c.f.\ the discussion in Sec.\ \ref{subsec:dynamics}). This means that structures close to $\omega=0$ are
much better resolved than structures at, for example, the band
edges of the Hubbard bands. In contrast, the 
dynamical density-matrix renormalization group
 recently applied to the DMFT for the Hubbard model works with a fixed resolution
on a {\em linear} scale, see for example \textcite{Karski:2005}. 
The structures close to the inner
band edges of the Hubbard bands seen in these calculations (see
Fig.~2 in \textcite{Karski:2005}) cannot be resolved in present 
implementations of the NRG
method.

Figure \ref{fig:mott}b 
shows the $T$-$U$ phase diagram for the Mott transition,
again only considering paramagnetic phases. 
As already observed earlier \cite{Georges:1996},
there are two transition lines, because the insulator-to-metal transition
occurs at a lower critical value ($U_{{\rm c},1}(T)$) as the
metal-to-insulator transition ($U_{{\rm c},2}(T)$). Within this
hysteresis region, both metallic and insulating solutions can be stabilized
within the DMFT self-consistency.
With increasing temperature, the hysteresis region shrinks to zero
at a critical $T_{\rm c}$, above which there is only a crossover
from metallic-like to insulating-like solutions; this crossover region
is indicated by the dashed lines in Fig.~\ref{fig:mott}b.
The NRG values for $U_{{\rm c},1/2}(T)$ have later been
verified by a number of other, non-perturbative techniques 
(see for example \textcite{Tong:2001} or \textcite{Potthoff:2003}).

As discussed above, most of the NRG calculations within DMFT 
have been performed using Eq.~(\ref{eq:self-energy-expression}) 
for the calculation of
the self-energy. This quantity itself shows interesting properties
(see, for example, Fig.~3 in \textcite{Bulla:1999} and Fig.~5 in 
\textcite{Bulla:2001}),
and allows to calculate the $U$-dependence of the
quasiparticle weight, see Fig.~1 in \textcite{Bulla:1999}.

The Mott transition can also be induced by moving away from half-filling,
provided the value of $U$ is larger than the $U_{\rm c}$ for the
half-filled case. Unfortunately, no systematic NRG calculations have
been published for this filling-induced Mott transition, despite
the fact that the NRG can be easily extended to the Hubbard model
away from particle-hole symmetry. Only a few results for the
phase diagram \cite{Ono:2001} and spectral functions 
\cite{Freericks:2003,KrugvonNidda:2003} are available in the literature.


A nice feature of the DMFT is that it also allows for the calculation
of physical quantities other than the single-particle Green function,
in particular susceptibilities and also transport properties, both
static and dynamic. This aspect of the DMFT has been intensively used
already in the early applications (see for example the reviews by
\textcite{Pruschke:1995} and \textcite{Georges:1996}), employing different
methods to solve the effective quantum-impurity problem. However, 
apart from discussions of the $A_{1g}$ Raman response
\cite{Freericks:2001,Freericks:2003} and calculations of the resistivity
\cite{Limelette:2003,Georges:2004} and the local dynamic
susceptibility \cite{KrugvonNidda:2003}, no
detailed studies of such quantities for the paramagnetic phase of the
Hubbard model have been performed yet with the NRG.

\subsubsection{Ordering phenomena}
\label{subsubsec:ordering}

The Mott transition from a paramagnetic metal to a paramagnetic
insulator is merely one of the many features in the rich phase diagram
of the Hubbard model and its generalizations. In addition, 
various types of ordering phenomena occur, such as charge, orbital (in case
of multi-orbital models), and magnetic ordering, and -- possibly --
superconductivity. The NRG has been used in particular 
to study magnetic ordering phenomena in a wide range of parameters.

\begin{figure}[htb]
  \centering
 \includegraphics[width=3in,clip]{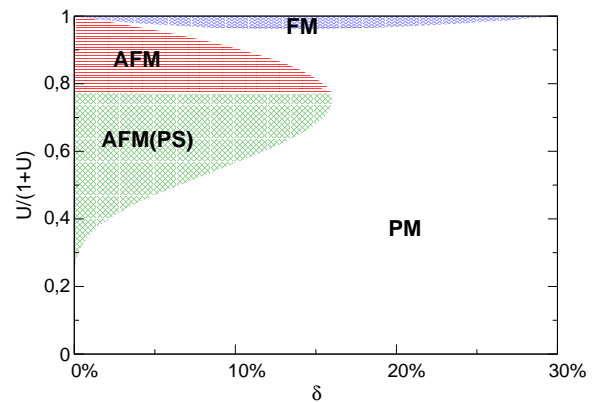} 
  \caption{Ground state magnetic phase diagram for the Hubbard model on a
hypercubic lattice. The phases show antiferromagnetic (AFM) order,
which for smaller $U$ and doping $\delta>0$ also shows phase separation (PS),
and ferromagnetic order at large $U$. To display the whole interval
$[0,\infty)$, the vertical axis was rescaled as $U/(1+U)$ (see also
Fig.~10 in \textcite{Zitzler:2002}). 
}
  \label{fig:hm_magn}
\end{figure}

For the investigation of symmetry broken phases within DMFT, the 
self-consistency equations have to be adapted appropriately
\cite{Georges:1996}. The effective
impurity models still have the structure of the single-impurity
Anderson model so that the application of the NRG is straightforward,
see the discussion in \textcite{Zitzler:2002}.
This work also contains a detailed study of
the magnetic phases of the Hubbard model at $T=0$ both at and away
from half-filling.
Right at half-filling and for a particle-hole symmetric band-structure,
the groundstate is always antiferromagnetically 
ordered. Upon doping, the situation is more complicated as shown in
Fig.~\ref{fig:hm_magn}: For small values of $U$, phase
separation within the antiferromagnetic phase is observed, while
for very large $U$, ferromagnetic solutions can be stabilized.
For intermediate $U$ and finite doping,
magnetic ordering appears to exist
but its type could not be determined yet.

In contrast to this work, the NRG calculations in \textcite{Zitzler:2004}
concentrated on the antiferromagnetic phase at half-filling in
a Hubbard model with frustration. As expected, the antiferromagnetic
region in the $T$-$U$ phase diagram is suppressed upon increasing
frustration. However, the resulting phase diagram turns out to
be significantly different from the one proposed for the
frustrated Hubbard model in \textcite{Georges:1996}, where 
it was claimed that the main
effect of frustration is to suppress $T_N$ such that the
first-order Mott transition is visible above the antiferromagnetic
region. This controversial issue certainly calls for more detailed
calculations (with NRG and other methods); after all, the similarity
of the phase diagram for the Hubbard model in DMFT and the
experimental one for the transition metal oxide V$_2$O$_3$ has been
claimed to be one of the early success of DMFT \cite{Georges:1996}.

The optical conductivity in the antiferromagnetic phase of the
Hubbard model at half-filling and zero temperature has been
studied in detail in \textcite{Pruschke:2003}. For small values of
$U$, the antiferromagnetic phase shows signatures of a Slater
insulator, while for large $U$ a Mott-Heisenberg picture
applies. There is a smooth crossover between these two limiting cases
upon variation of $U$, in contrast to the Mott transition in the
paramagnetic phase. The evidence from the optical data has
been supported by a detailed discussion of the local
dynamical magnetic susceptibility, giving additional insight into the subtle
changes in the physics of charge- and spin-degrees of freedom across
the Mott metal-insulator transition \cite{Pruschke:2003,Pruschke:2005b}.

When the Hubbard model Eq.~(\ref{eq:hub}) is supplemented by
a nearest-neighbor Coulomb repulsion $V$, a transition to
a charge ordered state is observed upon increasing $V$. This
transition has been studied in \textcite{Pietig:1999} for the
quarter-filled case. NRG calculations, together with results from
non-crossing approximation and 
exact diagonalization show a phase diagram with a reentrant charge ordering
transition, a feature which has also been observed in a variety
of transition metal oxides. The NRG results in this work are restricted
to $T=0$, where the transition is of first order. It would be very 
interesting to extend the NRG calculations to a wider range of
parameters, in particular to finite temperatures to study the change of
the character of the transition which is continuous at higher $T$.


\subsubsection{Multi-band Hubbard models}

The application of the NRG to the investigation of
multi-band Hubbard models within DMFT is still in a very early stage. 
This is because (i) the computational effort grows considerably 
with the number of orbitals and (ii) the DMFT requires a very high
accuracy for the calculated dynamic properties. Furthermore,
self-consistent solutions of the DMFT equations have to be
obtained.

The first (and so far the only) DMFT results for a two-band Hubbard
model using the NRG have been presented in \textcite{Pruschke:2005}.
In this work, two different strategies have been used to 
handle the complexity of the problem. The first one is to explicitly
include the orbital quantum numbers in the iterative construction
of the basis states. As for the impurity models discussed
in Sec.~\ref{subsec:orbital}, this additional quantum number significantly
reduces the typical matrix size. However, this approach fails
as soon as the Hamiltonian contains terms which break the orbital
symmetry.

The second strategy is an asymmetric truncation scheme: Instead of
adding both orbital degrees of freedom simultaneously, the
Hilbert space is truncated after adding each orbital individually,
which also leads to a significant reduction of the typical matrix size.
This approach works quite well in a wide range of parameters, but
it appears to violate the orbital symmetry, if present. However, in
the presence of a crystal-field splitting of the orbitals, such a
strategy might be usable.

The focus in \textcite{Pruschke:2005} was on the role of the Hund
exchange coupling $J$ on the Mott transition in the two-band Hubbard model.
It was found that both the position in parameter space and the nature
of the Mott transition depend on the value of $J$ and the precise
form of the coupling. For example, the replacement of a rotationally invariant
Hund exchange by an Ising-like exchange leads to a significant change in the
physics of the Mott transition. Note that such features can be partly
understood already on the level of the corresponding effective
impurity models, which underlines the importance to thoroughly
investigate the properties of the impurity models appearing in the DMFT.

\subsubsection{Other generalizations of the Hubbard model}

Let us conclude this section with a brief
overview of applications of the NRG to various other generalizations
of the Hubbard model.

The influence of correlations in a conduction band (modeled by a
Hubbard model within DMFT) on the physics of the single-impurity
Anderson model has been investigated in \textcite{Hofstetter:2000a}.
The DMFT approach allows to map this model on an effective impurity
model with two coupled correlated sites, the first one
corresponding to the original impurity and the second 
one coming from the DMFT treatment of the Hubbard model. This
two-site cluster couples to a free effective conduction band.
As discussed in \textcite{Hofstetter:2000a}, correlations in the
conduction band have a significant influence on the
low-energy scale and also lead to a suppression of the Kondo resonance.

Within the so-called Anderson-Hubbard model, disorder effects
can be incorporated via a random distribution of the local energies
$\varepsilon_i$. This model has been studied in \textcite{Byczuk:2004}
within DMFT for binary alloy disorder. The application of the NRG
here is standard -- two independent single-impurity Anderson models
have to be considered at each iteration step. Nevertheless, the
physics of this model is already quite interesting, in particular the 
occurrence of a Mott transition at {\em non-integer} filling. The
DMFT treatment in \textcite{Byczuk:2004} does not allow for true Anderson 
localization (as far as disorder is concerned, the DMFT is equivalent
to the coherent potential approximation
 and the main effect of the binary disorder is to split the bands).
This deficiency has been cured in \textcite{Byczuk:2005} where
a generalization of the DMFT approach has been used, based on the
geometrically averaged (typical) local density of states. This allows to study
both Mott insulating and Anderson insulating phases, see Fig.~1 in this
paper. The calculation has been performed using a continuous
probability distribution, approximated by up to 30 different values of
$\varepsilon_i$, so that in each DMFT step a corresponding 
number of independent 
single-impurity Anderson models have to be considered. All the NRG calculations
for the Anderson-Hubbard model have been so far restricted to $T=0$
and to phases without long-range order.

Recently, the NRG has been used within an extension of
the standard DMFT. The `DMFT+$\Sigma_k$' approach as introduced
in \textcite{Sadovskii:2005} and 
\textcite{Kuchinskii:2005} adds to the local self-energy
a $k$-dependent part $\Sigma_k$. Applied to the one-band Hubbard
model, the effective single-impurity Anderson model is still of
the same type as the one appearing in standard DMFT, the only
difference is in the structure of the self-consistency equations.

For details of the NRG calculations and the discussion of the physics
of the Falikov-Kimball model \cite{Anders:2005b} and the
ionic Hubbard model \cite{Jabben:2005} we refer the reader to the 
respective references. Both papers show the usefulness of the NRG approach
to a wide range of correlated lattice models within DMFT, in
particular for the calculation of dynamic quantities at low
temperatures.

  \subsection{Periodic Anderson and Kondo lattice models}
  \label{subsec:pam}

A variety of Lanthanide- and Actinide-based compounds can be characterized 
as heavy fermion systems with a strongly enhanced effective mass of the 
quasiparticles. 
These compounds contain well localized 4$f$ 
or 5$f$ orbitals coupling via a hybridization to a conduction band 
consisting of $s$, $p$ or $d$ orbitals.
The appropriate microscopic model for these materials is the periodic
Anderson model (PAM)
\begin{eqnarray}
  H = \varepsilon_{f} \sum_{i\sigma} f^\dagger_{i\sigma} f_{i\sigma}
      + U \sum_{i}f^\dagger_{i\uparrow} f_{i\uparrow}
                           f^\dagger_{i\downarrow} f_{i\downarrow}
\nonumber \\
     +  \sum_{k\sigma} \varepsilon_k c^\dagger_{k\sigma} c_{k\sigma}
       + \sum_{ij\sigma} V_{ij} \big( f^\dagger_{i\sigma} c_{j\sigma} +
               c^\dagger_{j\sigma}  f_{i\sigma} \big) \ .
\label{eq:pam}
\end{eqnarray}
When charge fluctuations of the $f$ orbitals are negligible, the
low energy physics of the PAM can equally be described by the Kondo
lattice model
\begin{equation}
   H = J \sum_i \vec{S}_i \cdot \vec{s}_i +
\sum_{k\sigma} \varepsilon_k c^\dagger_{k\sigma} c_{k\sigma} \ .
\end{equation}
The large effective mass in these models arises from a strongly
reduced lattice coherence scale $T_0$; this is one of the
reasons why the NRG is very
well suited for the investigation of heavy fermion behavior
when the PAM or the Kondo lattice model are treated within DMFT.
The main difference for the NRG treatment (as compared
to the Hubbard model) lies in the DMFT 
self-consistency.
This means that the structure of the effective impurity model
is changed only via the effective hybridization $\Delta(\omega)$
(which may lead to complications as discussed later).

\begin{figure}[htb]
\centerline{
  \includegraphics*[width=2.9in,clip]{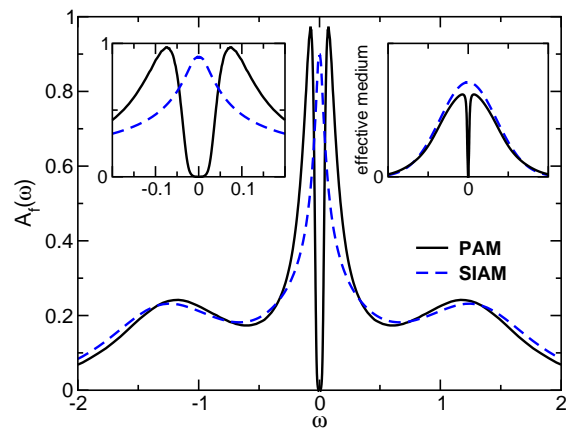}}
\caption{Comparison of dynamic properties for the particle-hole
symmetric periodic Anderson model (solid lines) 
and the corresponding single-impurity Anderson model (dashed lines). 
Main panel: $f$ electron spectral function; left inset: 
enlarged view of the region around the Fermi energy;
right inset: (effective) hybridization function.
(Figure adapted from \textcite{Pruschke:2000}).}
\label{fig:pam1}
\end{figure}

The PAM with on-site hybridization
$V_{ij}=V\delta_{ij}$ and particle-hole symmetry on a hypercubic lattice
has been discussed in
\textcite{Shimizu:1995}, \textcite{Pruschke:2000}, and
\textcite{Shimizu:2000}.
In this case, a hybridization gap at the Fermi
level appears in the spectral functions for both conduction and $f$
electrons.
This effect for the $f$ spectral function is shown 
in Fig.~\ref{fig:pam1} by the full lines in the main panel and left
inset. Apparently, the Kondo resonance 
of the corresponding single-impurity Anderson model 
(dashed curves in the main panel and left inset
in Fig.~\ref{fig:pam1}),
for which the hybridization function is given by the 
bare density
of states of the lattice,
is split in the periodic model. The energy scale of the 
gap in the PAM (proportional to the lattice coherence scale $T_0$)
depends exponentially on the model parameters, similar
as for the width of the Kondo resonance in the impurity model
(proportional to the Kondo temperature $T_{\rm K}$). Further
analysis shows that the lattice coherence scale $T_0$ is enhanced
over the impurity scale $T_{\rm K}$ (for details see 
\textcite{Pruschke:2000}).

The right inset to Fig.~\ref{fig:pam1} shows the hybridization
function of the effective impurity model after self-consistency has
been reached (full line) in comparison to the same quantity entering
the isolated impurity (dashed line). At first sight, the only
difference seems to be the gap at the Fermi level. However, for the
particle-hole symmetric case, one can show that $\Delta(\omega)$ has a
pole at the Fermi level.
At first sight, this pole appears to be a problem for the
NRG as the logarithmic discretization
explicitly excludes the point $\omega=0$, i.e.\ such a pole cannot be
incorporated in the mapping to the semi-infinite chain. The way out is
to take the pole into account via an extra site which couples directly to 
the impurity, thus removing the pole from the hybridization function.

Due to the appearance of the hybridization gap,
the particle-hole symmetric PAM seems rather suitable to describe 
so-called Kondo insulators, but not the metallic heavy
fermion behavior. There are various ways to drive the
PAM into the metallic regime, two of which we will
discuss in the following.
%
One
possibility is to use asymmetric parameters for the $f$ electrons
($\varepsilon_f\!\ne\!-U/2$) and to keep the conduction band symmetric
so that $n_{c}\approx 1$. More interesting (and also physically
more relevant) is the opposite case, namely
keeping $\varepsilon_f = -U/2$ and shifting the conduction band center-of-mass
away from the Fermi level, so that $n_{c}$ is reduced from
1. This situation has been discussed in the context of
Nozi\`eres exhaustion principle \cite{Nozieres:1998} which states that, upon
decreasing $n_{c}$, there will not be enough conduction
electrons available to screen the impurity spins. Collective
screening then becomes effective only at a strongly reduced
lattice coherence scale. We do not want to go into the
details here and refer the reader to the discussion
in Sec.~5.4 of \textcite{Vidhyadhiraja:2004}, which is based
on results for the PAM obtained with the local-moment
approach.

There is one particular feature found in the DMFT calculations which
at first sight seems to support Nozi\`eres idea, namely that,
as shown in Fig.~6b in \textcite{Pruschke:2000},
the effective hybridization function
is strongly reduced in a region close to the Fermi level. Since this
quantity can be interpreted as being proportional to the 
conduction band density of states
effectively seen by the $f$-states, it seems that
indeed  
there are less conduction electrons available to screen
the moments of the $f$ electrons.
Figure 6a in \textcite{Pruschke:2000} shows the
corresponding $f$-spectral function, which consequently now displays
metallic (heavy fermion) behavior. The corresponding lattice coherence
scale is now reduced as compared to 
$T_{\rm K}$, see Fig.~8 in \textcite{Pruschke:2000}
where the dependence of both $T_0$ and $T_{\rm K}$ on
$n_{c}$ is plotted. However, in contrast to Nozi\`eres original claim,
i.e.\ $T_0\propto(T_{\rm K})^2$, a behavior $T_0\propto T_{\rm K}$ is
found, with a prefactor decreasing exponentially with decreasing
$n_{c}$. Again, the ability of the NRG to accurately identify
exponentially small energy scales proves to be of great value here.

Another route to metallic behavior in the PAM is to incorporate
a dispersion of the $f$ electrons of the form
\begin{equation}
    t_f \sum_{<ij>,\sigma} \left(
           f_{i\sigma}^\dagger f_{j\sigma} + f_{j\sigma}^\dagger f_{i\sigma}
        \right) \ .
\end{equation}
The effect of such a dispersion term -- 
in particular the closing of the
gap upon increasing $t_f$ -- 
has been studied in
detail in \textcite{Shimizu:2000} for both the particle-hole
symmetric and asymmetric cases. The authors of this
work also study the relation between 
charge and spin gaps in the dynamical susceptibilities and the hybridization
gap in the spectral function.

A metallic ground state of the particle-hole symmetric
PAM can also be realized when
the hybridization between $f$ electrons and $c$ electrons
is only between nearest neighbors:
\begin{equation}
     V_{ij} = \left\{
       \begin{array}{ccl}
           V & : & i,j {\rm \ nearest\ neighbors} \\
            0 & : & {\rm otherwise}   \ \ \ \ \ \ \ \ \ \ \ \ \ \ \ \ \ \ .
       \end{array}
   \right.   \label{eq:pam-mit}
\end{equation}
For $T=0$, the PAM with nearest-neighbor hybridization shows a
notable difference to the models discussed above: 
the low-energy scale $T_0$ does no longer depend exponentially
on $U$ but vanishes at a {\em finite} 
critical $U_{\rm c}$ \cite{Held:2000}. This behavior
is reminiscent of the physics of the Mott-transition in the Hubbard
model. The difference, however, is that in the PAM defined by 
(\ref{eq:pam}) and (\ref{eq:pam-mit})
the Mott-transition occurs only in the subsystem of the $f$ electrons  
-- a gap
opens in the $f$ electron spectral function
while the $c$ electron part still has finite spectral weight at the Fermi level
(see Fig.~3 in \textcite{Held:2000}) so that the overall system remains
metallic.

The calculations described so far have all been restricted to
$T=0$. Finite temperature calculations for single-particle
and magnetic excitation spectra have been presented in
\textcite{Costi:2002} for the Kondo lattice  model. 
As in \textcite{Pruschke:2000}, one focus
of these studies has been the variation of the spectra with conduction band
filling $n_{c}$. For this case it was found that the spectra
exhibit {\em two} energy scales, one being the Kondo temperature
$T_{\rm K}$
of the corresponding single-impurity Kondo model, the other
one being the Fermi liquid coherence scale $T_0$ which, for
low carrier densities, $n_{c}\ll 1$, is strongly reduced as 
compared to $T_{\rm K}$, similar to the observations made 
in \textcite{Pruschke:2000} for the PAM. A comparison of the 
temperature dependence of the photoemission spectra with 
experimental data on YbInCu$_{4}$ showed good agreement. 
A ferromagnetic version of the Kondo lattice model with Coulomb interactions
in the conduction band was studied by \textcite{Liebsch:2006} in
the context of the  orbitally selective Mott phase of the two-band 
Hubbard model with inequivalent bands. The physics of this model 
is quite different to the usual Kondo lattice model. In particular, one finds,
as in \textcite{Biermann:2005},
non-Fermi liquid or bad metallic behavior, depending on whether the
ferromagnetic exchange is isotropic or anisotropic, respectively.
%

All these results have been obtained for the paramagnetic phases of the
PAM or Kondo lattice model. 
Of course,
the presence of localized moments implies the possibility
for magnetic ordering, which is frequently observed in
heavy fermion compounds, partly in close vicinity to
superconducting phases. These issues have not
been addressed yet with the NRG, but, as has been demonstrated for the
Hubbard model, are of course accessible with this method and are
surely a promising project for future NRG calculations for the PAM or Kondo
lattice
model within the DMFT. 


Magnetic quantum phase transitions in the Kondo lattice model have
been the focus of calculations within the extended DMFT \cite{Si:2001},
for which the effective impurity model includes a coupling to both fermionic
and bosonic baths. The NRG has been generalized to such types
of impurity models, see Sec.~\ref{subsec:bosons},
and recent applications of the NRG within the extended DMFT are
discussed in \textcite{Glossop:2006,Zhu:2006b}.

  \subsection{Lattice models with phonons}
  \label{subsec:phonons}

Let us consider the Hubbard model Eq.~(\ref{eq:hub}) and supplement it by
a local coupling of the electron density to the displacement
of Einstein phonons with frequency $\omega_0$. This results
in the Hubbard-Holstein model with the Hamiltonian
\begin{eqnarray}
  H &=& \varepsilon \sum_{i\sigma} c^\dagger_{i\sigma} c_{i\sigma}
       - t
        \sum_{<ij>\sigma}  c^\dagger_{i\sigma} c_{j\sigma} 
       +U \sum_i c^\dagger_{i\uparrow} c_{i\uparrow}
                 c^\dagger_{i\downarrow} c_{i\downarrow}
   \nonumber \\
      & & + g\sum_{i\sigma}  \left(b_i^\dagger + b_i \right)
              c_{i\sigma}^\dagger c_{i\sigma} 
          +\omega_0 \sum_i  b_i^\dagger  b_i  \ .
\label{eq:hh}
\end{eqnarray}
The limit $U\to 0$ of this Hamiltonian gives the Holstein model,
still a highly non-trivial model as discussed in the following.

Within DMFT, the model Eq.~(\ref{eq:hh}) maps onto the Anderson-Holstein
(impurity) model, to which the NRG method has been first 
applied by \textcite{Hewson:2002}, see Sec.~\ref{subsec:bosons}. 
The self-consistency equations are the
same for both Hubbard-Holstein and the pure Hubbard model, the
only difference lies in the calculation of the self-energy for
the effective impurity model which now contains an additional
contribution from the coupling to the phonons. This contribution
can also be calculated as a ratio of two correlation functions.

From a technical point of view, there is no difference in the
NRG-treatment of the Hubbard-Holstein model with either finite
or zero value of $U$. Historically, the first applications of the
NRG have been for the $U\!=\!0$-case (the Holstein model, see 
\textcite{Meyer:2002} and \textcite{Meyer:2003}) and have already revealed
a number of interesting results. 
As compared to other methods applied to the Holstein-model, the advantage
of the NRG combined with the DMFT is that it is non-perturbative in
both $g$ and $U$ and that it allows to study the case of a 
{\em macroscopic} electron density (in contrast to the few electron
case).

For the half-filled case, the investigations of \textcite{Meyer:2002}
showed some unexpected properties for the transition from a metal
to a bipolaronic insulator at a critical coupling $g_{\rm c}$.
In contrast to the Mott transition in the Hubbard model, no
hysteresis and no preformed gap is observed here (at least for
small values of $\omega_0$) which indicates that the physics
of the transition to the bipolaronic insulator might be completely
different (whether it is connected to locally critical behavior
is an interesting subject for future research).

For large values of $\omega_0$, the physics of the transition is
getting closer to that of the Hubbard model \cite{Meyer:2003}.
This is because in the $\omega_0\!\to\!\infty$-limit the Holstein
model can be mapped onto the attractive Hubbard model which has
the same behavior as the repulsive Hubbard model when charge and
spin degrees of freedom are interchanged.

\begin{figure}[htb]
\centerline{
  \includegraphics*[width=3.0in]{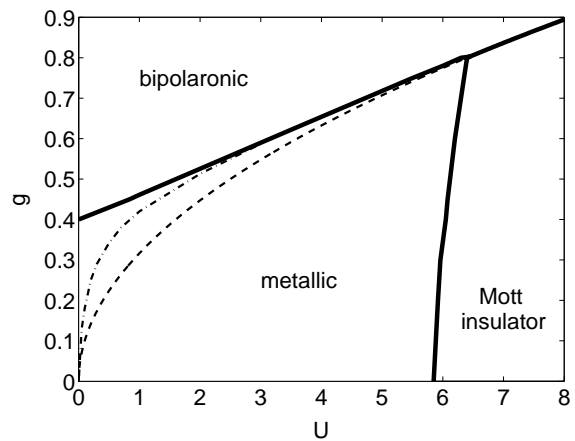}}
\caption{Phase diagram of the half-filled Hubbard-Holstein model
for $T=0$ (figure adapted from \textcite{Koller:2004b}).
The thick solid lines denote the phase boundaries while the
dashed line corresponds to $U_{\rm eff} = U - {2g^2}/\omega_0 = 0$.
The dashed and dot-dashed lines are polaronic lines, 
see \textcite{Koller:2004b}.}
\label{fig:hh}
\end{figure}

The phase diagram of the Hubbard-Holstein model at half-filling,
$T=0$, and neglecting any long-range ordered phases has been
discussed in detail in \textcite{Koller:2004a},
\textcite{Jeon:2004}, and
\textcite{Koller:2004b}.
The main features are summarized in Fig.~\ref{fig:hh} which shows
the position of the phase boundaries between metallic, Mott-insulating,
and bipolaronic insulating phases. The nature of these various transitions,
together with the behavior of dynamic quantities has been discussed
in detail in \textcite{Koller:2004b}. Let us point out here the
behavior of the phonon spectral function which shows a considerable
phonon softening upon approaching the transition to the bipolaronic
insulator. Such a softening is absent in the approach to the Mott
insulator, for the simple reason that close to the Mott transition,
where charge fluctuations are strongly suppressed, the phonons are
effectively decoupled from the electrons. One of the interesting
topics for future research are models which
do show such a phonon softening even close to the Mott transition;
this can possibly be accomplished by considering additional orbital
degrees of freedom.

Recently, the Hubbard-Holstein model has been studied for the
quarter-filled case and large values of $U$ \cite{Koller:2005a}.
In this situation, a strongly renormalized band of polaronic
quasiparticle excitations occurs within the lower Hubbard band
of the electronic spectral function.

All the investigations of lattice models with electron-phonon
coupling described in this
section have been restricted to $T=0$ and to phases without
long-range order. Generalizations to finite temperatures and
ordered phases (such as superconducting and charge ordered
phases) appear to be possible within the DMFT/NRG approach and will certainly
give interesting results and new insights. Other possible generalizations
are different types of lattice models, such as the periodic
Anderson model and multi-orbital models, and models with a 
different type of coupling between electrons and phonons.

\section{Summary}
\label{sec:summary}

Let us first summarize here the main purpose of this
review:
\begin{itemize}
  \item[(i)] To give a general introduction to the basic
        concepts of the numerical renormalization group
        approach (Sec.~\ref{sec:nrg-intro}) and to the
        general strategy for the calculation of physical
        quantities within this method (Sec.~\ref{sec:nrg-calc}).
  \item[(ii)] To cover the whole range of applications over the
        last 25 years, following the seminal work of \textcite{Wilson:1975} 
        on the Kondo problem and the work of \textcite{Krishnamurthy:1980a} on the
        Anderson impurity model (Secs.~\ref{sec:imp} and \ref{sec:dmft}).
%
%
%
\end{itemize}
Apparently, the range of applicability of the NRG widened
considerably, in particular over the last ten years. This
can be easily seen in the list of references in which
more than 50\%  of the entries are from the years starting
with 2000. In physical terms, the NRG is now being used to study
very different phenomena of condensed matter physics:
Typical correlation phenomena such as the Mott transition and 
heavy-fermion behavior, the physics of a two-state system
in a dissipative environment, Kondo correlations in artificial
atoms such as quantum dots, to name but a few.
Of course, we expect that there are still very many problems
to which the NRG will be applied in the future, and we
hope that this review will be helpful as a starting point
for such investigations.

Some of the concepts discussed in Secs.~\ref{sec:nrg-intro} and
\ref{sec:nrg-calc} are fairly recent developments: For example
the generalization of the NRG
to quantum impurities coupled to a bosonic environment (see
also Sec.~\ref{subsec:bosons}) and the calculation of time-dependent
quantities (transient dynamics, see Sec.~\ref{subsubsec:x-ray+transients}). 
As only a few applications of these new
concepts have been considered so far, one line of future research of 
the NRG is their generalization to a broader class of
impurity models.

We already discussed some open issues and ideas for further investigations
in the various subsections of Secs.~\ref{sec:imp} and \ref{sec:dmft}.
Let us mention here a few suggestions for further generalizations and
applications of the NRG:
\begin{itemize}
\item Application of the bosonic NRG to generalizations of the
  spin-boson models such as coupled spins in a dissipative
  environment.
%
%
\item Magnetic, orbital  and charge ordering in lattice
  models within DMFT.
\item Application of multiple-shell techniques (Sec.~\ref{subsec:dynamics})
  to further improvement of the dynamics, particularly at the 
  lowest temperatures.
\end{itemize}

What are the main open issues of the NRG approach? As discussed in
Sec.~\ref{subsec:orbital}, multi-site and multi-orbital models pose 
severe technical problems for the
NRG, because the Hilbert space increases dramatically with
the number of orbitals. This limits, in particular, the accuracy
in the calculation of dynamical quantities which in turn
restricts the applicability of the NRG to multi-band models
within DMFT (see Sec.~\ref{subsec:hubbard}) or its extensions.
Concerning dynamical quantities, another shortcoming of the
present implementations of the NRG is the rather poor
resolution at high frequencies, for example features such
as the band edges of upper and lower Hubbard bands in the
Hubbard model or the sharply peaked and highly asymmetrical 
spin-resolved Kondo resonance at high magnetic fields,
as shown in Fig.~\ref{fig:Kondo-largeB}.

A gradual improvement of the accuracy can, of course, be achieved
by simply increasing the computational effort, but for
a real breakthrough (concerning multi-band models and the
high-energy resolution) we probably need completely new
ideas and concepts.

From a conceptual point of view it will be very interesting to
view the NRG in a broader context. One step in this direction
has been made in \textcite{Verstraete:2005}. The authors of this
work interpret the NRG iteration in terms of matrix product states,
and establish a connection to the widely used density matrix
renormalization group method.

Concerning the future prospects of the numerical renormalization
group, let us conclude with a remark from Wilson's original
paper (\textcite{Wilson:1975}, page 777), about the prospects
of the renormalization group in general:
\begin{quotation}
  {\it ... However, most of the unsolved problems in physics and theoretical
  chemistry are of the kind the renormalization group is intended to
  solve (other kinds of problems do not remain unsolved for long).
  It is likely that there will be a vast extension of the renormalization
  group over the next decade as the methods become more clever
  and powerful; there are very few areas in either elementary
  particle physics, solid state physics, or theoretical chemistry
  that are permanently immune to this infection.}
\end{quotation}

\section*{Acknowledgments} 
It is a pleasure to acknowledge many useful discussions on the topic of this 
review with present and former colleagues, including 
F.B.\ Anders,
J.\ Bon\v{c}a,
L.\ Borda,
S.\ Florens, 
J.\ Freericks,
A.C.\ Hewson, 
W.\ Hofstetter, 
K.\ Ingersent, 
S.\ Kehrein, 
J.\ Kroha,
D.\ Logan,
N.\ Manini,
A.\ Rosch, 
C.\ Varma,
M.\ Vojta, 
D.\ Vollhardt,
J.\ von Delft, 
P.\ W\"{o}lfle,
G.\ Zar\'and, 
A.\ Zawadowski, 
and
V.\ Zlati\'c.
This work was
supported in part by the DFG through the collaborative research
centers SFB 484 and SFB 602.  We acknowledge supercomputer support by 
the Leibniz Computer Center, the Max-Planck Computer Center Garching 
under grant h0301, the {\em Gesellschaft f\"ur
  wissenschaftliche Datenverarbeitung G\"ottingen} (GWDG), the {\em
  Norddeutscher Verbund f\"ur   Hoch- und H\"ochstleistungsrechnen} (HLRN) 
under project nip00015 and the John von Neumann Institute for Computing 
(J\"ulich).


\bibliographystyle{apsrmp}
\bibliography{nrg-review}

\end{document}